\documentclass[english,10pt]{article}

\usepackage[utf8]{inputenc} 
\usepackage[T1]{fontenc} 
\usepackage[a4paper]{geometry} 

\usepackage{xspace} 
\usepackage{dsfont} 
\usepackage{mleftright} 
\usepackage{bropd} 
\usepackage{leftidx}

\usepackage{mathtools} 
\usepackage{amssymb} 
\usepackage{stmaryrd} 
\usepackage{esvect} 
\usepackage{fixdif} 
\usepackage{tdsfrmath} 

\usepackage[amsmath,framed,hyperref,thmmarks]{ntheorem} 
\usepackage{framed} 

\usepackage{array} 
\usepackage{booktabs} 
\usepackage{longtable} 
\usepackage{colortbl} 

\usepackage{fancyhdr}
\usepackage[scr]{rsfso} 
\usepackage{microtype} 
\usepackage{graphicx} 
\usepackage[dvipsnames]{xcolor} 
\usepackage{tikz}
\usetikzlibrary{automata,plotmarks,positioning,decorations.markings,decorations.pathmorphing}
\usepackage{enumitem} 
\usepackage{csquotes} 
\usepackage[style=alphabetic,autocite=inline,backref,giveninits=true,sorting=nyt]{biblatex}
\usepackage{tcolorbox}
\usepackage[multiple,symbol]{footmisc}

\usepackage{babel} 
\usepackage{subcaption} 
\usepackage[breaklinks,colorlinks=true,linkcolor=black,anchorcolor=black,citecolor=black,filecolor=black,menucolor=black,runcolor=black,urlcolor=black,hyperfootnotes=false]{hyperref} 
\usepackage[all]{hypcap} 
\usepackage[nospace]{varioref} 
\usepackage{cleveref} 

\NewDocumentCommand{\posttheorem}{}{%
}

\theoremstyle{plain}
\theoremseparator{~\----}
\theorempostwork{\posttheorem}
\theoremnumbering{Roman}
\newtheorem{Hyp}{Hypothesis}

\theoremstyle{plain}
\theoremseparator{~\----}
\theorempostwork{\posttheorem}
\theoremnumbering{arabic}
\newtheorem{Def}{Definition}
\numberwithin{Def}{subsection}

\theoremheaderfont{\sc\bfseries}
\theoremstyle{break}
\theoremseparator{}
\theorempostwork{\posttheorem}
\newtheorem{Th}[Def]{Theorem}

\theoremheaderfont{\normalfont\bfseries}
\theoremstyle{plain}
\theoremseparator{~\----}
\theorempostwork{\posttheorem}
\newtheorem{Prop}[Def]{Proposition}

\theoremstyle{plain}
\theoremseparator{~\----}
\theorempostwork{\posttheorem}
\newtheorem{Lem}[Def]{Lemma}

\theoremheaderfont{\normalfont\bfseries}
\theoremstyle{plain}
\theoremseparator{~\----}
\theorempostwork{\posttheorem}
\newtheorem{Lemdef}[Def]{Lemma-definition}

\theoremheaderfont{\normalfont\bfseries}
\theoremstyle{plain}
\theoremseparator{~\----}
\theorempostwork{\posttheorem}
\newtheorem{Cor}[Def]{Corollary}

\theoremheaderfont{\bfseries}
\theorembodyfont{\upshape}
\theoremstyle{nonumberbreak}
\theoremnumbering{arabic}
\theorempostwork{\posttheorem}
\theoremseparator{}
\theoremsymbol{\ensuremath{\Box}}
\newtheorem{Demo}{Proof}

\theoremheaderfont{\bfseries}
\theorembodyfont{\upshape}
\theoremstyle{plain}
\theoremnumbering{arabic}
\theorempostwork{\posttheorem}
\theoremseparator{~\----}
\theoremsymbol{}
\newtheorem{Rq}[Def]{Remark}

\theoremheaderfont{\bfseries}
\theorembodyfont{\upshape}
\theoremstyle{plain}
\theoremnumbering{arabic}
\theorempostwork{\posttheorem}
\theoremseparator{~\----}
\theoremsymbol{}
\newtheorem{Ex}[Def]{Example}

\crefname{Hyp}{Hypothesis}{Hypotheses}
\crefname{Def}{Definition}{Definitions}
\crefname{Th}{Theorem}{Theorems}
\crefname{Prop}{Proposition}{Propositions}
\crefname{Lem}{Lemma}{Lemmas}
\crefname{Lemdef}{Lemma-definition}{Lemma-definitions}
\crefname{Cor}{Corollary}{Corollaries}
\crefname{Demo}{Proof}{Proofs}
\crefname{Rq}{Remark}{Remarks}
\crefname{Ex}{Example}{Examples}

\NewDocumentCommand{\dps}{}{\displaystyle}

\NewDocumentCommand{\OptionnalArg}{m}{\prth{#1}}

\NewDocumentCommand{\prth}{m}{\mleftright\br{#1}}
\NewDocumentCommand{\brac}{m}{\mleft(#1\mright)}

\NewDocumentCommand{\ens}{mo}{%
\IfNoValueTF{#2}{\mleft\{#1\mright\}}{\mleft\{#1~\middle/~#2\mright\}}%
}
\NewDocumentCommand{\MathSet}{mo}{%
\IfNoValueTF{#2}{\mleft\{#1\mright\}}{\mleft\{#1~\middle/~#2\mright\}}%
}
\NewDocumentCommand{\abs}{m}{\mleft|#1\mright|}
\NewDocumentCommand{\norm}{O{}m}{\mleft\lVert#2\mright\rVert_{#1}}
\NewDocumentCommand{\ent}{m}{\mleft\lfloor #1\mright\rfloor}
\NewDocumentCommand{\Ent}{m}{\mleft\lceil #1\mright\rceil}
\NewDocumentCommand{\Frac}{mO{}}{\mleft\{#1\mright\}^{#2}}
\NewDocumentCommand{\intgff}{m}{\mleft\llbracket#1\mright\rrbracket}

\NewCommandCopy{\Awitho}{\AA}
\RenewDocumentCommand{\AA}{}{\mathbb{A}}
\NewDocumentCommand{\BB}{}{\mathbb{B}}

\NewDocumentCommand{\EE}{}{\mathbb{E}}

\NewDocumentCommand{\HH}{}{\mathbb{H}}

\NewDocumentCommand{\MM}{}{\mathbb{M}}
\NewDocumentCommand{\NN}{}{\mathbb{N}}

\NewDocumentCommand{\PP}{}{\mathbb{P}}
\NewDocumentCommand{\QQ}{}{\mathbb{Q}}
\NewDocumentCommand{\RR}{}{\mathbb{R}}
\NewCommandCopy{\ligaturedS}{\SS}
\RenewDocumentCommand{\SS}{}{\mathbb{S}}

\NewDocumentCommand{\WW}{}{\mathbb{W}}
\NewDocumentCommand{\XX}{}{\mathbb{X}}

\NewDocumentCommand{\ZZ}{}{\mathbb{Z}}

\NewDocumentCommand{\bB}{}{\mathcal{B}}
\NewDocumentCommand{\cC}{}{\mathcal{C}}

\NewDocumentCommand{\eE}{}{\mathcal{E}}

\NewDocumentCommand{\mM}{}{\mathcal{M}}
\NewDocumentCommand{\nN}{}{\mathcal{N}}
\NewDocumentCommand{\oO}{}{\mathcal{O}}
\NewDocumentCommand{\pP}{}{\mathcal{P}}

\NewDocumentCommand{\Sss}{}{\mathfrack{S}}

\NewDocumentCommand{\nom}{m}{\textsc{#1}}

\NewDocumentCommand{\carac}{m}{\mathds{1}_{#1}}
\NewDocumentCommand{\complementary}{}{^\textnormal{c}}
\NewDocumentCommand{\restrict}{mm}{\mleft.{#1}\!\mright|_{#2}}

\NewDocumentCommand{\Sn}{O{n}}{\Sss_{#1}}

\DeclareMathOperator{\Trace}{Tr}
\NewDocumentCommand{\tr}{O{}s}{\Trace_{#1}\IfBooleanF{#2}{\OptionnalArg}}

\DeclareMathOperator{\GroupeLineaire}{GL}
\NewDocumentCommand{\GL}{O{n}O{\RR}}{\GroupeLineaire_{#1}\brac{#2}}

\NewDocumentCommand{\Mn}{O{n}O{\RR}}{\mM_{#1}\brac{#2}}
\NewDocumentCommand{\On}{O{n}O{\RR}}{\oO_{#1}\brac{#2}}
\DeclareMathOperator{\SpecialOrthogonal}{SO}
\NewDocumentCommand{\SO}{O{n}O{\RR}}{\SpecialOrthogonal_{#1}\brac{#2}}

\NewDocumentCommand{\ZnZ}{O{n}}{\ZZ\mleft/{#1}\ZZ\mright.}
\NewDocumentCommand{\ZnZx}{O{n}}{\prth{\ZZ\mleft/{#1}\ZZ\mright.}^{\times}}

\NewDocumentCommand{\DefaultFunctionName}{}{f}
\NewDocumentCommand{\DefaultStartSpace}{}{\RR}
\NewDocumentCommand{\DefaultVariable}{}{x}

\NewDocumentCommand{\MyFunction}{mmmmmo}{%
\ifthenelse{\equal{#2#3#4#5}{}}{%
	\ifthenelse{\equal{#1}{}}{\DefaultFunctionName}{#1}}{%
\ifthenelse{\equal{#1}{}}{}{#1 : }%
\ifthenelse{\equal{#2#3}{}}{
	\ifthenelse{\equal{#4}{}}{\DefaultVariable\mapsto#5}{%
		\ifthenelse{\equal{#5}{}}{#4\mapsto%
					\ifthenelse{\equal{#1}{}}{\DefaultFunctionName(#4)}{#1(#4)}%
								}{#4\mapsto#5}}%
	}{%
\ifthenelse{\equal{#4#5}{}}{
	\ifthenelse{\equal{#2}{}}{\DefaultStartSpace\to#3}{%
		\ifthenelse{\equal{#3}{}}{#2\to#2}{#2\to#3}}%
	}{%
\IfNoValueTF{#6}{\begin{array}{ccl}}{\begin{array}{#6}}
\ifthenelse{\equal{#2}{}}{\DefaultStartSpace}{#2} & \longrightarrow & \ifthenelse{\equal{#3}{}}{#2}{#3}\\
\ifthenelse{\equal{#4}{}}{\DefaultVariable}{#4} & \longmapsto & \ifthenelse{\equal{#5}{}}{\ifthenelse{\equal{#1}{}}{\DefaultFunctionName(#4)}{#1(#4)}}{#5}
\end{array}%
}}}%
}

\NewDocumentCommand{\MyFunABCDE}{mmmmmo}{%
\IfNoValueTF{#6}{\MyFunction{#1}{#2}{#3}{#4}{#5}}{%
					\MyFunction{#1}{#2}{#3}{#4}{#5}[#6]}%
}

\NewDocumentCommand{\MyFunABCD}{mmmmso}{%
\IfBooleanTF{#5}{\IfNoValueTF{#6}{\MyFunABCDE{#1}{#2}{#3}{#4}{}}{
									\MyFunABCDE{#1}{#2}{#3}{#4}{}[#6]}}{%
				\IfNoValueF{#6}{\MyFunABCDE{#1}{#2}{#3}{#4}{#6}}}%
}

\NewDocumentCommand{\MyFunABC}{mmmso}{%
\IfBooleanTF{#4}{\IfNoValueTF{#5}{\MyFunABCD{#1}{#2}{#3}{}}{%
									\MyFunABCD{#1}{#2}{#3}{}[#5]}}{%
				\IfNoValueF{#5}{\MyFunABCD{#1}{#2}{#3}{#5}}}%
}

\NewDocumentCommand{\MyFunAB}{mmso}{%
\IfBooleanTF{#3}{\IfNoValueTF{#4}{\MyFunABC{#1}{#2}{}}{%
									\MyFunABC{#1}{#2}{}[#4]}}{%
				\IfNoValueF{#4}{\MyFunABC{#1}{#2}{#4}}}%
}

\NewDocumentCommand{\MyFunA}{mso}{%
\IfBooleanTF{#2}{\IfNoValueTF{#3}{\MyFunAB{#1}{}}{\MyFunAB{#1}{}[#2]}}{%
				\IfNoValueF{#3}{\MyFunAB{#1}{#3}}}%
}

\NewDocumentCommand{\MyFun}{so}{%
\IfBooleanTF{#1}{\IfNoValueTF{#2}{\MyFunA{}}{\MyFunA{}[#2]}}{%
				\IfNoValueF{#2}{\MyFunA{#2}}}%
}

\NewDocumentCommand{\tend}{O{}mO{+\infty}}{%
\overset{#1}{\underset{#2\to#3}{\longrightarrow}}}
\NewDocumentCommand{\eqval}{O{+\infty}}{\underset{#1}{\sim}}
\NewDocumentCommand{\Lp}{O{p}}{\text{L}^{#1}}
\NewDocumentCommand{\classeC}{O{\infty}}{\ensuremath{\text{ de classe }\cC^{#1}}}

\NewDocumentCommand{\boule}{O{0,1}}{\BB\prth{#1}}
\NewDocumentCommand{\boulef}{O{0,1}}{\overline{\BB}\prth{#1}}

\DeclareMathOperator{\Probabilite}{\PP}
\NewDocumentCommand{\Pb}{O{}O{}mo}{%
\IfNoValueTF{#4}{\Probabilite_{#1}^{#2}\prth{#3}}{\Probabilite_{#1}^{#2}\brac{#3\middle|#4}}%
}
\DeclareMathOperator{\Esperance}{\EE}
\NewDocumentCommand{\Esp}{O{}O{}mo}{%
\IfNoValueTF{#4}{\Esperance_{#1}^{#2}\prth{#3}%
				}{\Esperance_{#1}^{#2}\brac{#3\middle|#4}}%
}
\DeclareMathOperator{\Variance}{Var}
\NewDocumentCommand{\Var}{O{}mo}{%
\IfNoValueTF{#3}{\Variance_{#1}\prth{#2}%
				}{\Variance_{#1}\brac{#2\middle|#3}}%
}
\DeclareMathOperator{\Covariance}{Cov}
\NewDocumentCommand{\Cov}{O{}mo}{%
\IfNoValueTF{#3}{\Covariance_{#1}\prth{#2}%
								}{\Covariance_{#1}\brac{#2\middle|#3}}%
}

\NewDocumentCommand{\forallpp}{O{\mu}}{\forall_{#1}}
\NewDocumentCommand{\iid}{}{{iid}\xspace}
\NewDocumentCommand{\va}{s}{\IfBooleanTF{#1}{variables aléatoires\xspace}{%
variable aléatoire\xspace}}
\NewDocumentCommand{\fgm}{s}{%
\IfBooleanTF{#1}{fonctions génératrices des moments\xspace}{%
fonction génératrice des moments\xspace}}

\NewDocumentCommand{\dequal}{O{\text{d}}}{\overset{#1}{=}}

\NewDocumentCommand{\LBinomial}{O{n,p}}{\bB\prth{#1}}

\NewDocumentCommand{\LExp}{O{\lambda}}{\eE\prth{#1}}
\NewDocumentCommand{\LNormal}{O{0,1}}{\nN\prth{#1}}
\NewDocumentCommand{\LPoisson}{O{\lambda}}{\pP\prth{#1}}

\NewDocumentCommand{\FK}{}{Feynman-Kac\xspace}

\NewDocumentCommand{\defequal}{}{\vcentcolon=}
\def\overbigdot#1{\overset{\hbox{\tiny$\bullet$}}{#1}}
\NewDocumentCommand{\Thresh}{O{b}s}{%
	\textnormal{Thresh}_{#1}\IfBooleanF{#2}{\OptionnalArg}}

\NewDocumentCommand{\cyc}{}{\delta}

\NewDocumentCommand{\window}{O{L}}{\Lambda_{#1}}
\NewDocumentCommand{\bc}{O{bc}}{%
	(\textnormal{\ifthenelse{\equal{#1}{dir}}{Dir}{#1}})}
\NewDocumentCommand{\bcversion}{O{bc}}{_{\bc[#1],L}}

\NewDocumentCommand{\LoopSymbol}{}{\textnormal{Loop}}

\NewDocumentCommand{\WienerSpace}{O{t}}{\Omega_{#1}}
\NewDocumentCommand{\LoopSpaceRoot}{O{}s}{%
	\LoopSymbol_{\IfBooleanTF{#2}{\infty}{#1}}^{(\textnormal{root})}}
\NewDocumentCommand{\LoopSpace}{sO{bc}}{%
	\LoopSymbol_{\IfBooleanTF{#1}{\infty}{\bc[#2],L}}%
}

\NewDocumentCommand{\WienerTopology}{O{t}}{\mathfrak{W}_{#1}}
\NewDocumentCommand{\Dinf}{s}{\delta\IfBooleanF{#1}{\OptionnalArg}}
\NewDocumentCommand{\WienerTopologyRoot}{O{}}{%
	\mathfrak{W}^{(\textnormal{root})}_{#1}}
\NewDocumentCommand{\dist}{O{bc}s}{%
	\textnormal{dist}_{\bc[#1],L}\IfBooleanF{#2}{\OptionnalArg}}
\NewDocumentCommand{\LoopTopology}{O{bc}}{\mathfrak{W}_{\bc[#1],L}}

\NewDocumentCommand{\WienerAlgebra}{O{t}}{\mathcal{W}_{#1}}
\NewDocumentCommand{\WienerAlgebraRoot}{O{}}{%
	\mathcal{W}^{(\textnormal{root})}_{#1}}
\NewDocumentCommand{\LoopAlgebra}{O{bc}s}{%
	\mathcal{W}_{\bc[#1]\IfBooleanF{#2}{,L}}}

\NewDocumentCommand{\WienerMeasure}{mmm}{%
	W_{#1}^{#2\ifthenelse{\equal{#2}{}\OR\equal{#3}{}}{}{,}%
			\ifthenelse{\equal{#3}{}}{}{(\textnormal{#3})}}%
}
\NewDocumentCommand{\WienerMeasureXYT}{O{0,0}O{1}s}{%
	\WienerMeasure{#1}{#2}{}\IfBooleanF{#3}{\OptionnalArg}%
}
\NewDocumentCommand{\WienerMeasureXYTBC}{O{0,0}O{1}O{bc}s}{%
	\WienerMeasure{#1,\bc[#3],L}{#2}{}\IfBooleanF{#4}{\OptionnalArg}%
}
\NewDocumentCommand{\WienerMeasureNor}{O{1}s}{%
	\WW_{0,0}^{#1}\IfBooleanF{#2}{\OptionnalArg}%
}
\NewDocumentCommand{\WienerMeasureXYTRoot}{O{0,0}O{1}s}{%
	\WienerMeasure{#1}{#2}{root}\IfBooleanF{#3}{\OptionnalArg}%
}
\NewDocumentCommand{\WienerMeasureRootBC}{O{bc}s}{%
	\WienerMeasure{\bc[#1],L}{}{root}\IfBooleanF{#2}{\OptionnalArg}%
}
\NewDocumentCommand{\WienerMeasureBC}{O{bc}s}{%
	\WienerMeasure{\bc[#1],L}{}{}\IfBooleanF{#2}{\OptionnalArg}%
}

\NewDocumentCommand{\Conf}{}{\textnormal{Conf}}
\NewDocumentCommand{\ConfFinite}{}{\textnormal{Conf}_{<\infty}}

\NewDocumentCommand{\ConfSpace}{msO{bc}s}{%
	\textnormal{Conf}^{(\textnormal{#1})}%
		_{\ifthenelse{\equal{#1}{ps}\OR\equal{#1}{rl}}{%
			\IfBooleanTF{#2}{\infty}{\bc[#3]\IfBooleanF{#4}{,L}}}{}}%
}
\NewDocumentCommand{\ConfSpaceAuth}{sO{bc}}{%
\textnormal{ConfAuth}^{(\textnormal{mp})}_{\IfBooleanTF{#1}{\infty}{\bc[#2],L}}}
\NewDocumentCommand{\ConfSpacePerm}{msO{bc}}{%
	\textnormal{ConfPerm}^{(\textnormal{#1})}%
	_{\ifthenelse{\equal{#1}{FK}}{}{\IfBooleanTF{#2}{\infty}{\bc[#3],L}}}%
}

\NewDocumentCommand{\ConfAlgebra}{smsO{bc}}{%
	\mathcal{C}\IfBooleanTF{#1}{_{<\infty}}{%
	^{(\textnormal{#2})}%
		_{\ifthenelse{\equal{#2}{ps}\OR\equal{#2}{rl}}{%
			\IfBooleanTF{#3}{\infty}{\bc[#4],L}}{}}}%
}

\NewDocumentCommand{\finiteconfig}{}{\xi}
\NewDocumentCommand{\mpconfig}{}{\gamma}
\NewDocumentCommand{\FKconfig}{}{\gamma}

\NewDocumentCommand{\rlconfig}{}{\eta}

\NewDocumentCommand{\inward}{}{%
	\partial^{\textnormal{in}}_{\Delta}\FKconfig}
\NewDocumentCommand{\outward}{}{%
	\partial^{\textnormal{out}}_{\Delta}\FKconfig}

\NewDocumentCommand{\interior}{}{^{\textnormal{int}}_{\Delta}}
\NewDocumentCommand{\exterior}{O{\Delta}}{^{\textnormal{ext}}_{#1}}

\NewDocumentCommand{\SpatialComponent}{sO{bc}}{%
	\overbigdot{\mpconfig}\IfBooleanF{#1}{_{\bc[#2],L}}%
}
\NewDocumentCommand{\brown}{}{\omega}
\NewDocumentCommand{\sect}{}{\textnormal{w}}
\NewDocumentCommand{\loopath}{}{w}

\NewDocumentCommand{\loopEquiv}{sO{bc}}{%
	\equiv_{\IfBooleanTF{#1}{\infty}{\bc[#2],L}}%
}

\NewDocumentCommand{\tame}{m}{$(\textnormal{#1})$-tame\xspace}

\NewDocumentCommand{\Proj}{mO{\Delta}s}{%
	\textnormal{Proj}_{#1#2}\IfBooleanF{#3}{\OptionnalArg}%
}

\NewDocumentCommand{\versionbc}{O{bc}}{\FKconfig_{\bc[#1],L}}

\NewDocumentCommand{\Sausage}{O{\delta}m}{\textnormal{Saus}_{#1}\prth{#2}}
\NewDocumentCommand{\Neighbour}{mO{\secconfig}}{\textnormal{Neigh}\prth{#2,#1}}
\NewDocumentCommand{\Predecessor}{mO{\secconfig}}{%
	\textnormal{Pred}\prth{#2,#1}}
\NewDocumentCommand{\Successor}{mO{\secconfig}}{\textnormal{Succ}\prth{#2,#1}}

\NewDocumentCommand{\TimeReversal}{ms}{%
	R^{(\textnormal{#1})}\IfBooleanF{#2}{\OptionnalArg}}

\NewDocumentCommand{\length}{s}{\ell\IfBooleanF{#1}{\OptionnalArg}}
\NewDocumentCommand{\Length}{s}{L\IfBooleanF{#1}{\OptionnalArg}}
\NewDocumentCommand{\TimeTranslation}{O{s}s}{%
	T_{#1}^{(\textnormal{ps})}\IfBooleanF{#2}{\OptionnalArg}%
}

\NewDocumentCommand{\case}{O{r}}{\Lambda_{#1}}
\NewDocumentCommand{\CardCase}{sO{bc}O{r}mO{\SpatialComponent*}}{%
	N^{#3}_{\IfBooleanTF{#1}{\infty}{\bc[#2],L}}\prth{#5,#4}%
}

\NewDocumentCommand{\Quotient}{O{bc}s}{%
	\pi_{\bc[#1],L}\IfBooleanF{#2}{\OptionnalArg}%
}

\NewDocumentCommand{\PermutationSpace}{m}{\SS\brac{#1}}
\NewDocumentCommand{\PermutationSymbol}{}{\sigma}

\NewDocumentCommand{\Permutationmp}{sO{bc}mO{\mpconfig}}{%
	\PermutationSymbol^{(\textnormal{mp})}%
		_{\IfBooleanTF{#1}{\infty}{\bc[#2],L}}\prth{#4,#3}%
}
\NewDocumentCommand{\PermutationFK}{mO{\FKconfig}}{%
	\PermutationSymbol^{(\textnormal{FK})}\prth{#2,#1}%
}

\NewDocumentCommand{\ViewChangeDelta}{s}{%
	\varphi_{\Delta}^*\IfBooleanF{#1}{\OptionnalArg}%
}

\NewDocumentCommand{\ViewChange}{sO{bc}O{L}mms}{%
	\phantom{\overset{(\textnormal{#4})}{}}%
	\overset{\mathclap{(\textnormal{#4})\leftarrow(\textnormal{#5})}}{%
		\varphi}_{\ifthenelse{\equal{#5}{ptp}\and\equal{#4}{sec}}{%
			\IfBooleanF{#1}{\phantom{<\infty}}}{%
				\IfBooleanTF{#1}{\infty}{\bc[#2],#3}}}%
	\ifthenelse{\equal{#5}{ptp}\and\equal{#4}{sec}}{%
			\IfBooleanTF{#1}{%
				\phantom{\overset{(\textnormal{#4})}{}}}{\phantom{p}}}{%
		\IfBooleanT{#1}{\phantom{\overset{\textnormal{m}}{}}}}%
	\IfBooleanF{#6}{\OptionnalArg}%
}

\NewDocumentCommand{\Leb}{O{d}}{\textnormal{Leb}^{(#1)}}
\NewDocumentCommand{\pMeasure}{O{}s}{\nu_{#1}\IfBooleanF{#2}{\OptionnalArg}}

\NewDocumentCommand{\mpMeasure}{}{\mathfrak{m}}

\NewDocumentCommand{\PoissonLeb}{O{\window[L]}s}{%
	\Pi_{#1}\IfBooleanF{#2}{\OptionnalArg}%
}

\NewDocumentCommand{\Poisson}{mO{bc}s}{%
	\Pi^{\ifthenelse{\equal{#1}{}}{}{(\textnormal{#1})}}%
	_{\ifthenelse{\equal{#1}{rl}\OR\equal{#1}{ps}}{\bc[#2],}{}L}%
	\IfBooleanF{#3}{\OptionnalArg}%
}

\NewDocumentCommand{\PoissonRef}{O{\lambda}s}{%
\Pi^{#1\mpMeasure}_{L}\IfBooleanF{#2}{\OptionnalArg}%
}

\NewDocumentCommand{\Entropy}{mm}{\textnormal{I}\brac{#1\middle|#2}}
\NewDocumentCommand{\SpecificEntropy}{O{}m}{\textnormal{I}_{#1}\brac{#2}}

\NewDocumentCommand{\PairPotentiel}{O{}s}{%
						\Phi_{#1}\IfBooleanF{#2}{\OptionnalArg}}
\NewDocumentCommand{\InteractionFree}{O{}s}{%
	U_{#1}\IfBooleanF{#2}{\OptionnalArg}}
\NewDocumentCommand{\Interaction}{O{bc}s}{%
	U_{\bc[#1],L}\IfBooleanF{#2}{\OptionnalArg}}
\NewDocumentCommand{\InteractionLocal}{O{\Delta}s}{%
	U_{#1}\IfBooleanF{#2}{\OptionnalArg}%
}

\NewDocumentCommand{\Ham}{mO{bc}s}{%
	H^{(\textnormal{#1})}_{\bc[#2],L}\IfBooleanF{#3}{\OptionnalArg}%
}

\NewDocumentCommand{\Hamconditional}{O{\Delta}s}{%
	H^{(\textnormal{FK})}_{#1}\IfBooleanF{#2}{\OptionnalArg}%
}

\NewDocumentCommand{\Hamexterior}{s}{%
	H^{(\textnormal{FK})}_{\Delta\complementary,\bc,L}%
		\IfBooleanF{#1}{\OptionnalArg}%
}

\NewDocumentCommand{\Zgrandcanonic}{sO{bc}}{%
	Z_{\IfBooleanTF{#1}{\infty}{\bc[#2],L}}%
}

\NewDocumentCommand{\Zconditional}{O{\Delta}s}{%
	Z_{#1}\IfBooleanF{#2}{\OptionnalArg}%
}

\NewDocumentCommand{\Model}{smO{bc}s}{%
	\PP_{\bc[#3],\IfBooleanTF{#1}{\infty}{L}}%
		^{\ifthenelse{\equal{#2}{FK}}{(\textnormal{#2})}{(\textnormal{#2})}}%
	\IfBooleanF{#4}{\OptionnalArg}%
}

\NewDocumentCommand{\EmpiricalField}{mO{bc}O{L}s}{%
	\widetilde{\PP}_{\bc[#2],#3}%
	^{\ifthenelse{\equal{#1}{FK}}{(\textnormal{#1})}{(\textnormal{#1})}}%
	\IfBooleanF{#4}{\OptionnalArg}%
}

\NewDocumentCommand{\Modelbar}{smO{bc}s}{%
	\overline{\PP}_{\bc[#3],\IfBooleanTF{#1}{\infty}{L}}%
		^{\ifthenelse{\equal{#2}{FK}}{(\textnormal{#2})}{(\textnormal{#2})}}%
	\IfBooleanF{#4}{\OptionnalArg}%
}

\NewDocumentCommand{\ConditionnalModel}{O{\Delta}mo}{%
	\PP_{#1}^{(\textnormal{FK})}%
		\brac{#2\middle|\IfNoValueTF{#3}{\FKconfig}{#3}}%
}

\NewDocumentCommand{\diffsym}{}{\ominus}

\NewDocumentCommand{\bound}{m}{\textnormal{bound}\prth{#1}}

\NewDocumentCommand{\ConfAlgebraPerm}{m}{%
	\mathcal{C}^{(\textnormal{#1})}_{\textnormal{perm}}}

\RenewDocumentCommand{\ConfSpace}{msO{}s}{%
	\textnormal{Conf}^{(\textnormal{#1})}%
}
\RenewDocumentCommand{\ConfSpaceAuth}{sO{}}{%
	\textnormal{ConfAuth}^{(\textnormal{mp})}%
}
\RenewDocumentCommand{\ConfSpacePerm}{msO{}}{%
	\textnormal{ConfPerm}^{(\textnormal{#1})}%
}

\RenewDocumentCommand{\SpatialComponent}{sO{}}{%
	\overbigdot{\mpconfig}%
}

\RenewDocumentCommand{\CardCase}{sO{bc}O{r}mO{\SpatialComponent*}}{N^{#3}\prth{#5,#4}%
}

\RenewDocumentCommand{\Permutationmp}{sO{bc}mO{\mpconfig}}{%
	\PermutationSymbol^{(\textnormal{mp})}\prth{#4,#3}%
}

\RenewDocumentCommand{\ViewChange}{sO{}O{}mms}{%
	\overset{(\textnormal{#4})\leftarrow(\textnormal{#5})}{\varphi}%
	\IfBooleanF{#6}{\OptionnalArg}%
}

\RenewDocumentCommand{\bc}{O{}}{}
\RenewDocumentCommand{\bcversion}{O{}}{}

\RenewDocumentCommand{\WienerMeasureXYTBC}{O{0,0}O{1}O{bc}s}{%
	\WienerMeasure{#1}{#2}{}\IfBooleanF{#4}{\OptionnalArg}%
}

\RenewDocumentCommand{\Interaction}{O{bc}s}{%
	U_{L}\IfBooleanF{#2}{\OptionnalArg}}
\RenewDocumentCommand{\Ham}{mO{bc}s}{%
	H^{(\textnormal{#1})}_{L}\IfBooleanF{#3}{\OptionnalArg}%
}
\RenewDocumentCommand{\Zgrandcanonic}{sO{bc}}{%
	Z_{\IfBooleanTF{#1}{\infty}{L}}%
}
\RenewDocumentCommand{\Model}{smO{bc}s}{%
	\PP_{\IfBooleanTF{#1}{\infty}{L}}%
	^{\ifthenelse{\equal{#2}{FK}}{(\textnormal{#2})}{(\textnormal{#2})}}%
	\IfBooleanF{#4}{\OptionnalArg}%
}
\RenewDocumentCommand{\EmpiricalField}{mO{bc}O{L}s}{%
	\widetilde{\PP}_{\bc[#2]#3}%
	^{\ifthenelse{\equal{#1}{FK}}{(\textnormal{#1})}{(\textnormal{#1})}}%
	\IfBooleanF{#4}{\OptionnalArg}%
}
\RenewDocumentCommand{\Modelbar}{smO{bc}s}{%
	\overline{\PP}_{\bc[#3]\IfBooleanTF{#1}{\infty}{L}}%
	^{\ifthenelse{\equal{#2}{FK}}{(\textnormal{#2})}{(\textnormal{#2})}}%
	\IfBooleanF{#4}{\OptionnalArg}%
}

\RenewDocumentCommand{\TimeTranslation}{O{s}s}{%
	T_{#1}^{(\textnormal{rl})}\IfBooleanF{#2}{\OptionnalArg}%
}

\RenewDocumentCommand{\Hamexterior}{s}{%
	H^{(\textnormal{FK})}_{\Delta\complementary,L}%
	\IfBooleanF{#1}{\OptionnalArg}%
}

\RenewDocumentCommand{\ConfAlgebra}{smsO{bc}}{%
	\mathcal{C}\IfBooleanTF{#1}{_{<\infty}}{%
		^{(\textnormal{#2})}%
		_{\ifthenelse{\equal{#2}{ps}\OR\equal{#2}{rl}}{%
				\IfBooleanTF{#3}{}{}}{}}}%
}

\NewDocumentCommand{\WienerSpaceExtended}{m}{%
	\widetilde{\Omega}_{#1}%
}
\NewDocumentCommand{\WienerSpaceLoop}{m}{%
	\Omega^{(\textnormal{#1})}%
}
\NewDocumentCommand{\WienerAlgebraExtended}{m}{%
	\widetilde{\mathcal{W}}_{#1}%
}
\NewDocumentCommand{\WienerAlgebraLoop}{m}{%
	\mathcal{W}^{(\textnormal{#1})}%
}
\NewDocumentCommand{\WienerTopologyExtended}{m}{%
	\widetilde{\mathfrak{W}}_{#1}%
}
\NewDocumentCommand{\WienerTopologyLoop}{m}{%
	\mathfrak{W}^{(\textnormal{#1})}%
}
\RenewDocumentCommand{\WienerMeasureRootBC}{O{bc}s}{%
	\WienerMeasure{L}{}{rl}\IfBooleanF{#2}{\OptionnalArg}%
}

\RenewDocumentCommand{\Poisson}{mO{bc}s}{%
	\Pi^{\ifthenelse{\equal{#1}{}}{}{(\textnormal{#1})}}%
	_{\ifthenelse{\equal{#1}{rl}\OR\equal{#1}{ps}}{}{}L}%
	\IfBooleanF{#3}{\OptionnalArg}%
}

\RenewDocumentCommand{\WienerMeasureNor}{O{1}O{0,0}s}{%
	\WW_{#2}^{#1}\IfBooleanF{#3}{\OptionnalArg}%
}

\fancypagestyle{plain}{}
\NewDocumentCommand{\stoptime}{O{i}sO{1}}{\tau_{#1}\IfBooleanT{#2}{^{(#3)}}}

\NewDocumentCommand{\mailGuillaume}{}{guiblt.pro@gmail.com}
\NewDocumentCommand{\mailDavid}{}{david.dereudre@univ-lille.fr}
\NewDocumentCommand{\mailMylene}{}{mylene.maida@univ-lille.fr}

\let\oldFootnote\footnote
\newcommand\nextToken\relax

\renewcommand\footnote[1]{%
	\oldFootnote{#1}\futurelet\nextToken\isFootnote}

\newcommand\isFootnote{%
	\ifx\footnote\nextToken\textsuperscript{,}\fi}

\begin{document}

\title{DLR Equations for the Superstable Bose Gas at any Temperature and Activity}
\author{\nom{G. Bellot}\footnotemark[1]\footnotemark[2] ,
\nom{D. Dereudre}\footnotemark[1]\footnotemark[3] { and}
\nom{M. Ma\"ida}\footnotemark[1]\footnotemark[4]}
\maketitle
\fancyhead[C]{}

\footnotetext[1]{Univ. Lille, CNRS, UMR 8524 - Laboratoire Paul Painlevé, F-59000 Lille, France}
\footnotetext[2]{\href{mailto:guiblt.pro@gmail.com}{\texttt{\mailGuillaume}}}
\footnotetext[3]{\href{mailto:david.dereudre@univ-lille.fr}{\texttt{\mailDavid}}}
\footnotetext[4]{\href{mailto:mylene.maida@univ-lille.fr}{\texttt{\mailMylene}}}

\begin{abstract}
We construct a thermodynamic limit for the grand canonical Bose gas in dimension $d\geqslant1$ (in its Feynman-Kac representation) with superstable interaction at any inverse temperature $\beta>0$ and any chemical potential $\mu\in\RR$. Our infinite volume model is naturally a distribution over configurations of finite loops and possibly interlacements. We prove the limiting process to solve a new class of DLR equations involving random permutations and Brownian paths.
\\

\end{abstract}

\tableofcontents

\section{Introduction and Results}

Since the celebrated lesson of Ginibre in Les Houches \cite{Gin70}, it has been formally established that, under broad hypotheses, a canonical ensemble of $N$ indistinguishable bosons in a bounded domain $\Lambda\subset\RR^d$ at thermal equilibrium at inverse temperature $\beta>0$ interacting through a potential $\MyFunction{\InteractionFree*}{\Lambda^N}{\RR\cup\{+\infty\}}{}{}$ can be modeled as a point process. More precisely, if we denote as $\SS_N$ the $N$'th permutation group and $\WW_{x,y}^{\beta}$ the normalized Wiener measure over Brownian bridges $\MyFunction{\sect}{\interff{0;\beta}}{\RR^d}{}{}$ going from $x$ to $y$ then the square modulus of the bosons' wave function at $(x_1\dots x_N)\in(\RR^d)^N$ is proportional to
\begin{equation*}
	\sum_{\sigma\in\SS_N}\int 
	\prth{\bigotimes_{i=1}^n\carac{\sect_i\subset\Lambda}\WW_{x_i,x_{\sigma(i)}}^{\beta}
		(\d\sect_i)}~
	\exp\prth{-\frac{1}{2\beta}\sum_{i=1}^{n}\norm{x_{\sigma(i)}-x_i}^2-\int_0^{\beta}\InteractionFree{\sect_1\prth{s}\dots\sect_N\prth{s}}\d s}.
\end{equation*}
This probability density is called the Feynman-Kac (FK) representation of the Bose gas, as it is obtained through  the Feynman-Kac formula for bosonic solutions of the Schrödinger equation in the case of Dirichlet boundary condition. From the mathematical point of view, this model can be interpreted as a random spatial permutation, in which the presence or absence of macroscopic cycles is related to physical phenomena as explained below. 

For physical reasons, as in \autocite{AFY21,Vog23}, \emph{etc}, we study the grand canonical ensemble, where the number of points in the point process is random, with average density controlled by the parameter $\mu\in\RR,$ known as the chemical potential: high densities are attained through high chemical potentials.

A standard first step to the study of the Bose gas from point processes consists in taking its thermodynamic limit, which means inflating the domain to infinity $\Lambda\uparrow\RR^d$. One may then hope to deduce facts about the physical Bose gas from the properties of this infinite volume model. In particular, the community is most interested in proving when and how much the Bose gas undergoes Bose-Einstein condensation (BEC). The mathematical literature provides a good picture of the non-interacting case ($U=0$), known as the free Bose gas. \citeauthor{Sut93} (\autocite{Sut93,Sut02}) has proven Feynman's conjecture \autocite{Fey53} claiming that the emergence of infinite cycles (interlacements) in the sampled permutation $\sigma$ in infinite volume is equivalent to BEC. Later, it has then been proven in various ways and frameworks (\autocite{BU09,AFY21,Vog23}) that these interlacements appear through a saturation effect, beyond some critical density $\rho_c>0$. The link between interlacements and BEC is still to be clarified for interacting Bose gas. Note that interlacements do not only appear in Bose gas models but also in other probabilistic models \autocite{Szn09}.

For the interacting Bose gas, very few mathematical results are available and even constructing the thermodynamic limit rigorously is a challenge. This is one of the main result of the paper and our first main theorem can be summarized as follows:
\begin{center}
	\begin{tcolorbox}[width=12cm,colframe=blue!8,colback=blue!2]
		The thermodynamic limit of a superstable interacting Bose gas exists in a suitable (FK) path space at any inverse temperature $\beta$ and  chemical potential $\mu,$ or activity $e^{\beta\mu}.$
	\end{tcolorbox}
\end{center}

This problem has been previously addressed in the literature (\autocite{Pa84,Pa85,ACK11,NPZ13,SKS20,BV23,DV24}) but none of
these results simultaneously deal with large densities/chemical potentials and
keep track of the Feynman-Kac paths in the limiting structure. This achievement could be crucial, because it is precisely the setting where it is expected to show BEC via the presence of infinite cycles.

Under additional assumptions on the interaction, we provide a local description of the limiting distribution:
\begin{center}
	\begin{tcolorbox}[width=12cm,colframe=blue!8,colback=blue!2]
		When the interaction is finite range, the infinite volume interacting Bose gas satisfies a new class of Dobrushin-Lanford-Ruelle (DLR) equations, compatible with the (FK) formalism.
	\end{tcolorbox}
\end{center}
The local distributions of infinite volume Gibbs measures are never tractable in practice. The DLR equations are the standard way to describe the classical equilibrium states by providing local \emph{conditional} distribution  in any bounded set given the configuration outside \autocite{Geo11, Vel18}. 

We believe it is particularly relevant to use this DLR formalism in the present setting as the FK representation of the Bose gas  is a reformulation of  a quantum model through classical probabilistic objects.
Moreover, DLR equations  are a powerful tool to prove probabilistic properties of infinite volume models. For instance, in percolation theory, DLR equations are equivalent to the finite-energy property which is crucial for local surgery to study the emergence of infinite branches, as in Burton and Keane-type arguments \cite{BuKe89}. This paves the way  to use our DLR equations in the future to study the emergence of interlacements in the interacting Bose gas.
On a slightly different line of research, one can mention the use of DLR equations to obtain central limit theorems 
for linear statistics or number rigidity of some point processes \autocite{Leb21, Leb24}.

Let us note that we provide an infinite volume equilibrium state, in the sense that it is a specific solution of DLR equations. The next step would be to study all solutions of such DLR equations and compare them to equilibrium states in the quantum formalism, such as KMS states, in the spirit of \autocite{DrvV23}. These are  important and difficult questions that we keep  for future investigation. We refer to \autocites{DLL25} for recent work on bosonic KMS states in the lattice case.

To conclude this introduction, let us say a few words about our technical contributions.
In the previous results relying on the FK model, the cycles of the random permutations induced by the  model were encoded as a loop soup. The thermodynamical limit was obtained through tightness in a suitable space of loops. Unfortunately, for large activity, the size of some cycles is expected to grow to infinity, becoming therefore invisible in any rooted loop model. In particular, this prevents loop soup approaches (\autocite{SKS20,BV23}) to go beyond a critical density or activity. A key point of our approach is that this phenomenon does not occur here, since our model encompasses directly the rooted bridges, allowing us to reach any activity.
More precisely, our main contribution was to construct a sophisticated marked point version of the model, where tightness can be analyzed through the powerful entropy machinery developed in 
\autocite{GZ93}. Technical issues arise from establishing a suitable  encoding and decoding correspondance between the initial FK model and our marked point process.

The paper is organized as follows. We begin by stating rigorously the FK framework, our hypotheses and our main results. In the second section, we introduce the two other frameworks we will need in this paper, ae the usual loop soup and our new marked point definition of the Bose gas. Most technical arguments are contained in the proof section. An extensive notation table can be found at the end of the paper.

The paper is organized as follows. In \Cref{sec_thermodynamic_limit,sec_dlr_equations}, we state our main results (existence of the thermodynamic limit and DLR equations -- \Cref{th_limit_FK,th_DLR}) in the FK setting. In \Cref{sec_mp}, we describe in detail our new marked point model and show its equivalence with the (FK) model. We also recall in \Cref{subsection_rl} the rooted loop soup model, used in finite volume for its simple properties. The rest of the paper is devoted to the proofs. An extensive table of notations is available in \Cref{sec_table}.

\subsection{Interaction and Assumptions}

We begin by introducing the family of interactions we are considering in this paper.

\begin{Def}\label{def_confFinite}
Let us denote the set of finite point configurations in $\RR^d$ ($d\geqslant1$) as
\[
\ConfFinite\defequal\ens{\finiteconfig\subset\RR^d}[\#\finiteconfig<+\infty]
\]
where $\#\finiteconfig$ is the cardinal of $\finiteconfig$. We equip this space with the smallest $\sigma$-algebra $\ConfAlgebra*{}$ making measurable the maps $\MyFunction{}{}{}{\finiteconfig}{\#\prth{\finiteconfig\cap E}}$ for every Borel set $E\subseteq\RR^d$.
\end{Def}

Let $\MyFunction{\InteractionFree*}{\ConfFinite}{\RR\cup\{+\infty\}}{}{}$ measurable be an interaction.\label{lab_InteractionFree}

\begin{Hyp}
We assume the interaction to be \emph{non-degenerate} $\InteractionFree{\emptyset}<+\infty.$
\end{Hyp}

\begin{Hyp}\label{hyp_superstable}
We assume the interaction to be \emph{superstable \autocite{Rue70}}, that is to say there exist $A\geqslant 0$, $B>0$ and $r>0$ such that
\begin{equation*}
\forall\finiteconfig\in\ConfFinite,~\InteractionFree{\finiteconfig}\geqslant -A\#\finiteconfig+B\sum_{z\in r\ZZ^d}\#\prth{\finiteconfig\cap\prth{z+\case}}^{2}
\end{equation*}
where $\case\defequal\interfo{-r/2;r/2}^d$.
\end{Hyp}

\begin{Rq}\label{rq_superstability}
Superstability is a very standard assumption for an interaction. We refer to Proposition 1.2 from \autocite{Rue70} for criteria for a pairwise interaction to be superstable.

The hypotheses of this subsection are always considered throughout the following pages and not stated anymore. Similarly, the numbers $A\geqslant0$, $B>0$ and $r>0$ always refer to the superstability constants of \Cref{hyp_superstable}. Other assumptions are be added when needed.
\end{Rq}

We are investigating the Bose gas in \emph{Dirichlet boundary conditions}. This has consequences on the interaction we do calculations with: we need to add an infinite exterior potential to $\InteractionFree*$, restraining the particles inside the domain $\window$.

\begin{Def}\label{def_interaction}
	Let $L>0$. We define $\MyFunction{\Interaction*}{\ConfFinite}{\RR\cup\{+\infty\}}{}{}$ by
	\begin{align*}
		\Interaction[dir]{\finiteconfig}&\defequal
		\begin{cases}
			\InteractionFree{\finiteconfig} & \text{ if }\finiteconfig\subset\window\\
			+\infty & \text{ otherwise.}
		\end{cases}
	\end{align*}
\end{Def}

Our conclusions still hold for several other boundary conditions, including periodic and Neumann's. But we think presenting our results in full generality would have damaged readability without enriching significantly the theorems. We refer to a previous arXiv version \autocite{BDM24} for more details.

\subsection{Feynman-Kac State Space}

Since we are not sampling usual point configurations, we need to take some time to introduce the adapted state space of configurations of trajectories in $\RR^d$.

\begin{Def}\label{def_Wiener}
For any $t>0$, we denote as $\WienerSpace[t]$ the set of continuous functions from $\interff{0;t}$ to $\RR^d$. This set is equipped with the topology $\WienerTopology[t]$ associated to the uniform norm $\norm[\infty]{\cdot}$. We denote as $\WienerAlgebra$ the associated Borel $\sigma$-algebra.

On the set $\WienerSpace[t]$, we consider the Wiener measure $\WienerMeasureXYT[x,y][t]*$ weighing Brownian bridges going from $x$ to $y$ in time $t$, whose finite-dimensional distributions are given by
\[
\d\WienerMeasureXYT[x,y][t]{\brown\prth{s_1}=z_1\dots\brown\prth{s_n}=z_n}=
\prod_{i=0}^{n}{\prth{2\pi\prth{s_{i+1}-s_i}}^{-d/2}}\exp\prth{-\frac{\norm{z_{i+1}-z_i}^2}{2\prth{s_{i+1}-s_i}}}\d z_1\dots\d z_n
\]
with $0=s_0<\dots<s_{n+1}=t$ and the convention $z_0=x$ and $z_{n+1}=y$. This measure is \emph{un-normalized}, as
\[
\WienerMeasureXYT[x,y][t]{\WienerSpace[t]}=\prth{2\pi t}^{-d/2}\exp\prth{-\frac{1}{2t}\norm{y-x}^2}.
\]
\end{Def}

Since we are studying the Bose gas at thermal equilibrium at inverse temperature $\beta>0$, we are focusing on the $t=\beta$ case. The parameter $\beta>0$ is fixed in the sequel and omitted in most notations.

\begin{Def}\label{def_confspace_FK}
We denote the set of \emph{\FK} configurations $(\textnormal{FK})$ as
\begin{align*}
\ConfSpace{FK}&\defequal
\ens{\FKconfig\subset\WienerSpace[\beta]}[\FKconfig\text{ is locally finite for }\WienerTopology[\beta]]
\end{align*}
and we equip this configuration set with the smallest $\sigma$-algebra $\ConfAlgebra{FK}$ making measurable the maps $\MyFunction{}{}{}{\FKconfig}{\#\prth{\FKconfig\cap E}}$ for every bounded $E\in\WienerAlgebra[\beta]$.
\end{Def}

\begin{Def}\label{def_confspaceperm_FK}
A \FK configuration $\FKconfig\in\ConfSpace{FK}$ is said to be \emph{permutation-wise} if the following condition is satisfied
\[
\forall\sect\in\FKconfig,~
\begin{cases}
\exists!\sect'\in\FKconfig,~\sect'(0)=\sect(\beta)\\
\exists!\sect''\in\FKconfig,~\sect(0)=\sect''(\beta).
\end{cases}
\]
We denote
\[
\ConfSpacePerm{FK}\defequal\ens{\FKconfig\in\ConfSpace{FK}}[\FKconfig\text{ is permutation-wise}].
\]
and equip this set with the subset $\sigma$-algebra $\ConfAlgebraPerm{FK}$ induced by $\ConfAlgebra{FK}$.

For any $\FKconfig\in\ConfSpacePerm{FK}$, we define the permutation
\[
\MyFunction{\PermutationFK{\cdot}}{\FKconfig}{\FKconfig}{\sect}{\sect'\text{ such that }\sect'(0)=\sect(\beta).}
\]
\end{Def}

In finite volume, all probability measures are defined so that they are supported on permutation-wise configurations. We later prove the thermodynamic limits to have the same property.

\subsection{Locality and Tameness}

According to \autocite{GZ93}, the functionals whose integral is compatible with the thermodynamic limit are \emph{local} functionals. For marked point processes, locality just means the function can only depend on points inside some compact. But in our setting this is not so simple.

\begin{Def}\label{def_proj}
	Let $\Delta\subset\RR^d$ be a compact. We define the following projections $\MyFunction{}{\ConfSpace{FK}}{\ConfSpace{FK}}{}{}$
	\begin{align*}
		&\Proj{\in}{\FKconfig}\defequal\MathSet{\sect\in\FKconfig}[\sect(0)\in\Delta]
		\\
		&\Proj{\cap}{\FKconfig}\defequal\MathSet{\sect\in\FKconfig}[\sect\cap\Delta\neq\emptyset]
	\end{align*}
	if we accept the abuse of notation \enquote{$\sect\cap\Delta$} to mean $\sect\prth{\interff{0;\beta}}\cap\Delta$.
	
	For any $n\geqslant0$, we also define the maps\footnote{The maps $\Proj{\cap^n}*$ are not \emph{per se} projections but we keep this notation by coherence with the other maps. Technically, we could have defined $\Proj{\cap^n}*$ on the whole configuration space but it would have been unhelpful.} $\MyFunction{}{\ConfSpacePerm{FK}}{\ConfSpace{FK}}{}{}$
	\begin{align*}
		\Proj{\cap^n}{\FKconfig}=
		\MathSet{\sect\in\FKconfig}[
		\exists k\in\intgff{-n;n},~
		\prth{\PermutationFK{\cdot}}^k(\sect)\cap\Delta\neq\emptyset
		].
	\end{align*}
\end{Def}

All those sets are illustrated in \Cref{fig_local}.

\begin{figure}[h]
	\centering
	\begin{tikzpicture}[scale=1.25,decoration={markings}]
		\tikzstyle{fleche} = [rounded corners,->,>=latex]
		\tikzstyle{out} = [red]
		\tikzstyle{point_in} = [thick,dash dot dot,blue]
		\tikzstyle{inter} = [thick,dashed,blue]
		\tikzstyle{inter_plus} = [
		blue]
		\tikzstyle{inter_moins} = [
		blue]
		\draw (1,1) rectangle (3,3);
		\draw (1,1) node[left] {$\Delta$};
		\draw (1.3,1.1) node (A) {$\bullet$} ;
		\draw (3.3,0.5) node (B) {$\bullet$} ;
		\draw (3.4,3) node (C) {$\bullet$} ;
		\draw (2.8,1.7) node (D) {$\bullet$} ;
		\draw (0.7,3.6) node (E) {$\bullet$} ;
		\draw (4.5,1.5) node (F) {$\bullet$} ;
		\draw (0.5,2.6) node (G) {$\bullet$} ;
		\draw (-0.2,1.7) node (H) {$\bullet$} ;
		\draw (2.3,3.2) node (I) {$\bullet$};
		\draw (-0.5,3) node (M) {};
		\draw (4.7,2.7) node (N) {};
		
		\draw[fleche,point_in] (1.3,1.1) to[bend right, looseness=1] (D) ;
		\draw[fleche,point_in] (2.8,1.7) to[bend left, looseness=1.5] (C);
		\draw[fleche,inter] (0.7,3.6) to[bend right, looseness=2] (1.5,2.5) to[bend right, looseness=1] (E);
		\draw[fleche,inter_plus] (3.4,3) to[bend left, looseness=1] (B);
		\draw[fleche,inter_plus] (3.3,0.5) to[bend right,looseness=1] (F);
		\draw[fleche,inter] (0.5,2.6) to[bend left,looseness=1] (A);
		\draw[fleche,inter_moins] (-0.2,1.7) to[bend left,looseness=1.6] (G);
		\draw[fleche,inter_moins] (-0.5,3) to[bend right,looseness=1.6] (H);
		\draw[fleche,red] (4.5,1.5) to[bend left,looseness=1.6] (N);
		
		\draw[fleche,red] (2.3,3.2) to[bend right, looseness=1.8] (2.9,3.8) to[bend right, looseness=1] (I);
		
		\draw[fleche,point_in] (6,3.2) to (7,3.2);
		\draw (7,3.2) node[right] {$\Proj{\in}{\FKconfig}$};
		\draw[fleche,inter] (6,2.675) to (7,2.675);
		\draw (7,2.675) node[right] {$\Proj{\cap}{\FKconfig}\setminus\Proj{\in}{\FKconfig}$};
		\draw[fleche,inter_plus] (6,2.05) to (7,2.05);
		\draw (7,2.05) node[right] {$\Proj{\cap^2}{\FKconfig}\setminus\Proj{\cap}{\FKconfig}$};
		\draw[fleche,red] (6,1.475) to (7,1.475);
		\draw (7,1.475) node[right] {$\FKconfig\setminus\Proj{\cap^2}{\FKconfig}$};

		\draw (1.3,1.1) node {$\bullet$} ;
		\draw (3.3,0.5) node {$\bullet$} ;
		\draw (3.4,3) node {$\bullet$} ;
		\draw (2.8,1.7) node {$\bullet$} ;
		\draw (0.7,3.6) node {$\bullet$} ;
		\draw (4.5,1.5) node {$\bullet$} ;
		\draw (0.5,2.6) node {$\bullet$} ;
		\draw (-0.2,1.7) node {$\bullet$} ;
		\draw (2.3,3.2) node {$\bullet$};
	\end{tikzpicture}
	\caption{Illustration of various projection maps}
	\label{fig_local}
\end{figure}

The following notions of locality ensue.

\begin{Def}\label{def_local}
	A function $\MyFunction{f}{}{}{}{}$ defined over $\ConfSpacePerm{FK}$ is said to be \emph{$\in$-local} if there exists a compact $\Delta\subset\RR^d$ such that
	\[
	\Proj{\in}{\FKconfig}=\Proj{\in}{\FKconfig'}\implies f(\FKconfig)=f(\FKconfig').
	\]

	Let $n\geqslant 0$. Similarly, a function $\MyFunction{f}{}{}{}{}$ defined over $\ConfSpacePerm{FK}$ is said to be \emph{$\cap^n$-local} if there exists a compact $\Delta\subset\RR^d$ such that
	\[
	\Proj{\cap^n}{\FKconfig}=\Proj{\cap^n}{\FKconfig'}\implies f(\FKconfig)=f(\FKconfig').
	\]
\end{Def}

We also need to define some notions of tameness which are compatible with our respective definitions of locality.

\begin{Def}\label{def_tame}
	For any trajectory $\sect\in\WienerSpace[\beta]$ and $\delta>0$, we define its \emph{Wiener sausage} of thickness $\delta$ as
	\[
	\Sausage[\delta]{\sect}\defequal\ens{x\in\RR^d}[\exists s\in\interff{0;\beta},~\norm{x-\sect(s)}\leqslant\delta]
	\]
	whose volume we denote $\abs{\Sausage[\delta]{\sect}}$.
	
	A function $\MyFunction{f}{\ConfSpacePerm{FK}}{\RR}{}{}$ is said to be \emph{$\in$-tame} if there exists $a,\delta>0$, $\alpha\in\interfo{0;2}$, and a compact $\Delta\subset\RR^d$ such that for any $\FKconfig\in\ConfSpacePerm{FK}$,
	\[
	a\abs{f(\FKconfig)}\leqslant1+\sum_{\sect\in\Proj{\in}{\FKconfig}}\abs{\Sausage[\delta]{\sect}}^{\alpha}.
	\]
	
	Let $n\geqslant 0$. A function $\MyFunction{f}{\ConfSpacePerm{FK}}{\RR}{}{}$ is said to be \emph{$\cap$-tame} if there exists $a,\delta>0$, $\alpha\in\interfo{0;1}$, and a compact $\Delta\subset\RR^d$ such that for any $\FKconfig\in\ConfSpacePerm{FK}$,
	\begin{align*}
		a\abs{f(\FKconfig)}\leqslant
		1+\sum_{\sect\in\Proj{\in}{\FKconfig}}\abs{\Sausage[\delta]{\sect}}^{1+\alpha}+\sum_{\sect\in\Proj{\cap}{\FKconfig}}\abs{\Sausage[\delta]{\sect}}^{\alpha}.
	\end{align*}

	A function $\MyFunction{f}{\ConfSpacePerm{FK}}{\RR}{}{}$ is said to be \emph{$\cap^n$-Lipschitz} if there exists $a,\delta>0$, $\alpha\in\interfo{0;1}$, and a compact $\Delta\subset\RR^d$ such that for any $\FKconfig,\FKconfig'\in\ConfSpacePerm{FK}$,
	\begin{align*}
		a\abs{f(\FKconfig)-f\prth{\FKconfig'}}
		\leqslant&
		\sum_{\sect\in\Proj{\in}{\FKconfig\diffsym\FKconfig'}}\abs{\Sausage{\sect}}^{1+\alpha}
		+\sum_{\sect\in\Proj{\cap}{\FKconfig\diffsym\FKconfig'}}\abs{\Sausage{\sect}}^{\alpha}
		\\
		&+\sum_{\sect\in\Proj{\cap}{\FKconfig}}\carac{\exists k\in\intgff{-n;n},~\prth{\PermutationFK{\cdot}}^{k}(\sect)\notin\FKconfig'}
		\\
		&+\sum_{\sect\in\Proj{\cap}{\FKconfig'}}\carac{\exists k\in\intgff{-n;n},~\prth{\PermutationFK{\cdot}[\FKconfig']}^{k}(\sect)\notin\FKconfig}
	\end{align*}
	where $\diffsym$ denotes the symmetric difference between two sets.
\end{Def}

\begin{Rq}\label{rq_inter_n_lipschitz}
	The Lipschitz property can be explained as follows: if the configuration $\FKconfig$ is slightly modified then the value $f(\FKconfig)$ only marginally changes.

	If a function is $\cap^n$-Lipschitz, then it also is $\cap^n$-local and $\cap$-tame. Conversely, if a function is $\cap^n$-local and bounded, then it also is $\cap^n$-Lipschitz.
\end{Rq}

\begin{Prop}\label{Prop_lipschitz}
	Let $\MyFunction{f}{\ConfSpacePerm{FK}}{\RR}{}{}$, $\Delta\subset\RR^d$ a compact and $a>0$ be such that for any $\FKconfig,\FKconfig'\in\ConfSpacePerm{FK}$,
	\begin{align*}
		a\abs{f(\FKconfig)-f\prth{\FKconfig'}}
		\leqslant
		\#\Proj{\cap}{\FKconfig\diffsym\FKconfig'}
		+\#\prth{\textnormal{cyc}\prth{\FKconfig,\Delta,n}\diffsym \textnormal{cyc}\prth{\FKconfig',\Delta,n}}
	\end{align*}
	where $\textnormal{cyc}\prth{\FKconfig,\Delta,n}$ denotes the set of cycles of $\PermutationFK{\cdot}$ of length at most $n$ whose trajectories intersect $\Delta$.
	
	Then $f$ is $\cap^n$-Lipschitz.
\end{Prop}

\begin{Demo}[\Cref{Prop_lipschitz}]
	We can upper bound the assumed condition on $f$ by
	\begin{align*}
		a\abs{f(\FKconfig)-f\prth{\FKconfig'}}
		\leqslant
		\sum_{\sect\in\Proj{\cap}{\FKconfig\diffsym\FKconfig'}}1~+&\sum_{\sect\in\Proj{\cap}{\FKconfig}}\carac{\exists k\in\intgff{-n;n},~\prth{\PermutationFK{\cdot}}^{k}(\sect)\notin\FKconfig'}
		\\
		+&\sum_{\sect\in\Proj{\cap}{\FKconfig'}}\carac{\exists k\in\intgff{-n;n},~\prth{\PermutationFK{\cdot}[\FKconfig']}^{k}(\sect)\notin\FKconfig}
	\end{align*}
	The sausage with thickness $1$ of any trajectory is bounded from below by the volume of a $d$ dimensional unit ball. So this is enough to conclude.
\end{Demo}

\begin{Ex}\label{ex_local_tame_Lipschitz}
	We provide a few examples of functions which are local, tame or Lipschitz in various ways.
	\begin{itemize}
		\item $\MyFunction{f_1}{\ConfSpacePerm{FK}}{\RR}{}{}$
		defined by
		\[
		f_1(\FKconfig)=\sum_{\sect\in\FKconfig,~\sect(0)\in\interff{0;1}^d}\norm{\sect(\beta)-\sect(0)}
		\]
		is $\in$-local and $\in$-tame.
		
		Locality is clear. Tameness comes from the following fact: a cylinder whose axis goes from $\sect(0)$ to $\sect(\beta)$ with radius $\delta$ has a smaller volume than $\Sausage{\sect}$ (see \Cref{prop_saucisse}).

		\item $\MyFunction{f_2}{\ConfSpacePerm{FK}}{\RR}{}{}$ defined by
		\[
		f_2(\FKconfig)=\sum_{\sect\in\FKconfig,~\sect\subset\interff{0;1}^d}\frac{1}{\inf\MathSet{j\geqslant1}[\prth{\PermutationFK{\cdot}}^j(\sect)=\sect]}\carac{\forall j\in\ZZ,~\prth{\PermutationFK{\cdot}}^j(\sect)\subset\interff{0;1}^d}
		\]
		is $\in$-local and $\in$-tame.
		
		Indeed, the function $f_2$ counts the number of disjoint cycles in the cycle structure of $\FKconfig$ which are completely included inside $\interff{0;1}^d$. Locality is then intuitive. Tameness is clear because $f_2(\FKconfig)\leqslant\#\Proj{\in}[\interff{0;1}^d]{\FKconfig}$.

		\item $\MyFunction{f_3}{\ConfSpacePerm{FK}}{\RR}{}{}$ defined by
		\[
		f_3(\FKconfig)=\#\MathSet{\sect\in\FKconfig}[\sect\cap\interff{0;1}^d\neq\emptyset]\cdot\carac{\#\MathSet{\sect\in\FKconfig}[\sect\cap\interff{0;1}^d\neq\emptyset]\text{ is even}}
		\]
		is $\cap^0$-local and $\cap$-tame, but not $\cap^n$-Lipschitz for any $n\geqslant0$.
		
		There is no hope for $f_3$ to be $\cap^n$-Lipschitz because its variations can be arbitrarily large.

		\item$\MyFunction{f_4}{\ConfSpacePerm{FK}}{\RR}{}{}$ defined by
		\[
		f_4(\FKconfig)=\#\MathSet{\sect\in\Proj{\in}[\interff{0;1}^d]{\FKconfig}}[\prth{\PermutationFK{\cdot}}^2(\sect)=\sect]
		\]
		is $\cap^1$-local, $\in$-tame and $\cap^1$-Lipschitz.
		
		Locality is straightforward. Tameness is clear because $f_4(\FKconfig)\leqslant\#\Proj{\in}[\interff{0;1}^d]{\FKconfig}$.
		
		Adding or removing a number $n$ of bridges from $\Proj{\cap^1}{\FKconfig}$ increases or decreases by at most $n$ the counted number of closed cycles of length $1$ or $2$ in the configuration $\FKconfig$. Then \Cref{Prop_lipschitz} ensures $f_4$ is $\cap^1$-Lipschitz.
	\end{itemize}
	
	None of the examples above are bounded. So we state the thermodynamic limit in the most general possible way, without restricting ourselves to bounded functions. This allows us to state \Cref{cor_interlacements}.
\end{Ex}

\subsection{Thermodynamic Limit}\label{sec_thermodynamic_limit}

We now properly define the Bose gas in finite volume with Dirichlet boundary conditions and state our first major result.

\begin{Def}\label{def_ham_FK}
We define a Hamiltonian over finite configurations $\FKconfig\in\ConfSpace{FK}$ with
\[
\Ham{FK}{\FKconfig}\defequal\int_0^{\beta}\Interaction{\ens{\sect\prth{s},~\sect\in\FKconfig}}\d s\in\RR\cup\{+\infty\}.
\]
\end{Def}

We justify in \Cref{rq_ham} this integral is well defined.

\begin{Def}\label{def_poisson}
For any compact $\Delta\subset\RR^d$, we denote as $\PoissonLeb[\Delta]*$ the Poisson point process over $\Delta$ with intensity $1$.
\end{Def}

We introduce the Feynman-Kac representation of the grand canonical Bose gas interacting through the potential $\InteractionFree*$ at inverse temperature $\beta>0$ and chemical potential $\mu\in\RR$. The parameter $\mu$ controls the average density of the Bose gas at some given temperature. Just like $\beta$, the parameter $\mu$ is fixed in the sequel and omitted in the notations.

\begin{Lemdef}[proof: see \Cref{rq_demo}]\label{lemdef_FK}
Let $L>0$, $\beta>0$ and $\mu\in\RR$.

We define the probability measure $\Model{FK}*$ over $\ConfSpace{FK}$ by
\[
\Model{FK}{\d\FKconfig}
\defequal
\frac{1}{\Zgrandcanonic}\int \E^{\beta\mu\#\finiteconfig}~\PoissonLeb{\d\finiteconfig}\sum_{\PermutationSymbol\in\PermutationSpace{\finiteconfig}} \E^{-\Ham{FK}{\FKconfig}}
\prth{\bigotimes_{x\in\finiteconfig}\WienerMeasureXYTBC[x,\PermutationSymbol(x)][\beta]*}(\d\FKconfig)
\]
where $\PermutationSpace{\finiteconfig}$ is the set of permutations over $\finiteconfig$ and
\[
\Zgrandcanonic\defequal\int \E^{\beta\mu\#\finiteconfig}~\PoissonLeb{\d\finiteconfig}\sum_{\PermutationSymbol\in\PermutationSpace{\finiteconfig}}\int\E^{-\Ham{FK}{\FKconfig}}~
\prth{\bigotimes_{x\in\finiteconfig}\WienerMeasureXYTBC[x,\PermutationSymbol(x)][\beta]*}(\d\FKconfig).
\]
\end{Lemdef}

We illustrate the kind of configurations sampled by $\Model{FK}*$ in \Cref{fig_model_FK}. The points of $\finiteconfig$ are indicated with \enquote{$\bullet$} and we represent smooth trajectories instead of Brownian bridges for readability purposes. We also did not represent bridges that intersect each others because, although common, this situation could make the drawings less clear.

\begin{figure}[h]
\centering
\begin{tikzpicture}[scale=1.25]
\tikzstyle{fleche} = [rounded corners,->,>=latex]
\draw[loosely dashed] (0,0) rectangle (4,4);
\draw[<->] (-0.2,0) -- (-0.2,4);
\draw (-0.4,2) node {$L$};
\draw (1.3,1) node (A) {$\bullet$} ;
\draw (3.3,0.5) node (B) {$\bullet$} ;
\draw (3.4,3) node (C) {$\bullet$} ;
\draw (2.8,1.7) node (D) {$\bullet$} ;
\draw (0.7,3.6) node (E) {$\bullet$} ;
\draw[fleche] (1.3,1) to[bend left, looseness=1] (0.3,0.5) to[bend left, looseness=1] (0.7,2) to[bend left, looseness=1] (2,2) to[bend left, looseness=1] (A);
\draw[fleche] (3.3,0.5) to[bend left, looseness=1] (2.5,0.7) to[bend left, looseness=1] (D);
\draw[fleche] (2.8,1.7) to[bend left, looseness=1.5] (C);
\draw[fleche] (3.4,3) to[bend left, looseness=1] (B);
\draw[fleche] (0.7,3.6) to[bend right, looseness=2] (1.5,2.5) to[bend right, looseness=1] (E);
\end{tikzpicture}
\caption{Example of (FK) configuration}
\label{fig_model_FK}
\end{figure}

\begin{Rq}
By construction, it is clear that $\Model{FK}{\ConfSpacePerm{FK}}=1$.
\end{Rq}

\begin{Def}\label{def_empirical_field_FK}
For any $L>0$, we define the \emph{empirical field} $\EmpiricalField{FK}*$ over $\ConfSpace{FK}$ by
\[
\int f\d\EmpiricalField{FK}*\defequal
\frac{1}{L^d}\int_{\window}\d v\int f(\FKconfig+v)~\Model{FK}{\d\FKconfig}
\]
for any measurable $\MyFunction{f}{\ConfSpace{FK}}{\RR^+}{}{}$.
\end{Def}

\begin{Rq}
One should think the empirical field as a partially stationarized version of the probability $\Model{FK}*$.
\end{Rq}

\begin{Th}[Thermodynamic limit, proof pp. \pageref{demo_th_limit_FK_1},\pageref{demo_th_limit_FK_2}]\label{th_limit_FK}
Under hypotheses I and II, for any inverse temperature $\beta>0$ and chemical potential $\mu\in\RR$, there exists a stationary probability measure $\Model*{FK}*$ over $\ConfSpacePerm{FK}$ and an increasing sequence $L_m\tend{m}+\infty$ such that,
\begin{itemize}
\item for any measurable $\MyFunction{f}{\ConfSpacePerm{FK}}{\RR}{}{}$ which is $\in$-local and $\in$-tame
\item for any measurable $\MyFunction{f}{\ConfSpacePerm{FK}}{\RR}{}{}$ which is $\cap^n$-Lipschitz for some $n\geqslant 0$
\end{itemize}
then
\[
\lim_{m\to+\infty}\int f\d\EmpiricalField{FK}[bc][L_m]*=\int f\d\Model*{FK}*.
\]
\end{Th}

In the following, we abbreviate this fact as
\[
\lim_{L\to+\infty}\int f\d\EmpiricalField{FK}[bc]*=\int f\d\Model*{FK}*.
\]

\begin{Rq}
Our \Cref{th_limit_FK} concerns functions defined on $\ConfSpacePerm{FK}$. This is not a limitation. Any function defined on the whole $\ConfSpace{FK}$ can be restricted to $\ConfSpacePerm{FK}$. But the reverse is not as trivial: extending $f_2$ from \Cref{ex_local_tame_Lipschitz} to the whole configuration space would be an unpleasant exercise.
\end{Rq}

\begin{Def}
Let $\FKconfig\in\ConfSpacePerm{FK}$. We call an \emph{infinite cycle} (or \emph{interlacement}) a family $\prth{\sect_n}_{n\in\ZZ}$ of distinct bridges of $\FKconfig$ such that
\[
\forall n\in\ZZ,~\PermutationFK{\sect_n}=\sect_{n+1}.
\]
\end{Def}

\begin{Rq}
We were not able to prove the induced permutation $\PermutationFK{\cdot}$ to comprise infinite cycles at low enough temperature (or high enough chemical potential) with positive probability under $\Model*{FK}*$. But we believe our construction naturally includes this possibility, because it is a local to global construction, rather than a cycle-wise definition like in loop soup models.
\end{Rq}

\begin{Cor}\label{cor_interlacements}
The proportion of Brownian bridges which are part of an infinite cycle in the limiting process is the limiting proportion of bridges which are part of an arbitrarily large cycle in finite volume. More precisely,
\begin{align*}
&\int\#\MathSet{\sect\in\Proj{\in}[\interff{0;1}^d]{\FKconfig}}[\forall j\geqslant2,~\prth{\PermutationFK{\cdot}}^j(\sect)\neq\sect]~\Model*{FK}{\d\FKconfig}
\\
=&\lim_{n\to+\infty}\lim_{L\to+\infty}\int\#\MathSet{\sect\in\Proj{\in}[\interff{0;1}^d]{\FKconfig}}[\forall j\in\intgff{2;n},~\prth{\PermutationFK{\cdot}}^j(\sect)\neq\sect]~\EmpiricalField{FK}{\d\FKconfig}.
\end{align*}
\end{Cor}

\begin{Demo}[\Cref{cor_interlacements}]
We apply the second half of \Cref{th_limit_FK} to the functions $\MyFunction{f_n}{\ConfSpacePerm{FK}}{\RR}{}{}$, $n\geqslant2$, defined by
\begin{align*}
	f_n(\FKconfig)=\#\MathSet{\sect\in\Proj{\in}[\interff{0;1}^d]{\FKconfig}}[
	\exists k\in\intgff{2;n},~\prth{\PermutationFK{\cdot}}^k(\sect)=\sect
	]
\end{align*}
which makes it $\cap^n$-Lipschitz (see explanations in the 4'th point of \Cref{ex_local_tame_Lipschitz}).

Then we pass to the limit on $n\to+\infty$ to go from the density of cycles of length at most $n$ to finite cycles of any arbitrary length.
\end{Demo}

\subsection{DLR Equations}\label{sec_dlr_equations}

In the previous section, we stated the existence of an infinite volume model $\Model*{FK}*$. Since this probability measure is a thermodynamic limit, we can hope to compute the probability of some events as limits, but we did not provide any information on this infinite volume distribution itself yet.

The standard way to describe an infinite model like $\Model*{FK}*$ is to write DLR (Dobrushin-Lanford-Ruelle) equations, that is to say, write the conditional distribution of the infinite configuration inside some compact $\Delta$, given the configuration outside $\Delta$. This description is the best we can hope for. It is not possible to describe the probability $\Model*{FK}*$ as simply as we did in finite volume in \Cref{lemdef_FK}, because an infinite volume Hamiltonian would always value any infinite configuration to an infinite energetic cost.

\begin{Hyp}
We assume the interaction to be \emph{stationary}
\[
\forall\finiteconfig\in\ConfFinite,~\forall v\in\RR^d,~\InteractionFree{\finiteconfig+v}=\InteractionFree{\finiteconfig}.
\]
\end{Hyp}

\begin{Hyp}\label{hyp_finite_range}
We assume the interaction to be \emph{finite range with range $R>0$}, that is to say for any compact $\Delta\subset\RR^d$, there exists a local interaction $\MyFunction{\InteractionLocal*}{\ConfFinite}{\RR\cup\{+\infty\}}{}{}$ such that
\[
\forall\finiteconfig\in\ConfFinite,~\InteractionFree{\finiteconfig}=\InteractionLocal{\finiteconfig\cap\prth{\Delta+B_R}}+\InteractionFree{\finiteconfig\cap\Delta\complementary}
\]
where $B_R$ is the closed ball of radius $R$ and $\Delta+B_R$ is the Minkowski sum of those two sets.
\end{Hyp}

\begin{Rq}
The local interaction $\MyFunction{\InteractionLocal*}{\ConfFinite}{\RR\cup\{+\infty\}}{}{}$ does not need to be uniquely defined. More precisely, the value $\InteractionLocal{\finiteconfig}$ is not uniquely characterized if and only if $\InteractionFree{\finiteconfig}=+\infty$ or $\finiteconfig\not\subset\Delta+B_R$. Yet, Lebesgue almost everywhere (over $s\in\interff{0;\beta}$), the interaction term $\InteractionFree{\MathSet{\sect\prth{s},~\sect\in\FKconfig}}$ is $\Model{FK}*$ almost surely finite. Furthermore, the model $\ConditionnalModel{\cdot}$ (see \Cref{lemdef_DLR}) only depends on the sets $\MathSet{\sect(s),~\sect\in\FKconfig}\cap(\Delta+B_R),~s\in\interff{0;\beta}$. So the measure $\ConditionnalModel{\cdot}$ is $\Model{FK}*$ almost surely invariant under the choice of local interaction. It is also true $\Model*{FK}*$ almost surely.

The existence of those $\InteractionLocal*$ implies heredity of the original interaction $\InteractionFree*$.
\end{Rq}

\begin{Prop}
Let $\InteractionFree*$ be a pairwise stationary interaction, that is to say there exists a potential $\MyFunction{\Phi}{\RR^d}{\RR\cup\{+\infty\}}{}{}$ such that
\[
\forall\finiteconfig\in\ConfFinite,~
\InteractionFree{\finiteconfig}=\frac{1}{2}\sum_{\underset{x\neq y}{x,y\in\finiteconfig}}\Phi(x-y).
\]
We assume that
\[
\forall x\in\RR^d,~\norm{x}>R\implies\Phi(x)=0.
\]
Then the interaction is finite range with range $R$ and for any compact $\Delta\subset\RR^d$,
\[
\forall\finiteconfig\in\ConfFinite,~
\InteractionLocal{\finiteconfig}=
\frac{1}{2}\sum_{\underset{x\neq y}{x,y\in\finiteconfig\cap\Delta}}\Phi(x-y)
+\sum_{x\in\finiteconfig\cap\Delta}~\sum_{y\in\finiteconfig\cap\prth{\prth{\Delta+B_R}\setminus\Delta}}\Phi(x-y).
\]
\end{Prop}

\begin{Hyp}\label{hyp_boundedly_attractive}
We assume the interaction to be \emph{uniformly regular from below}, that is to say, for any compact $\Delta\subset\RR^d$ and integer $N\geqslant0$, there exists $C_{\Delta,N}\in\RR$ such that
\[
\forall\finiteconfig\in\ConfFinite,~\#\prth{\finiteconfig\cap\prth{\Delta+B_R}}=N\implies\InteractionFree{\finiteconfig}\geqslant\InteractionFree{\finiteconfig\cap\Delta\complementary}+C_{\Delta,N}.
\]
\end{Hyp}

In other words, we assume the energetic value of a configuration $\finiteconfig$ (with a given number of points close to $\Delta$) not to decrease by an arbitrarily large value when adding a given number of points inside $\Delta$.

This could probably be guaranteed by some weak regularity criteria (hence the name of \Cref{hyp_boundedly_attractive}) because we were not able to find an example of interaction which would check hypotheses I to IV but not the V'th. A finite range pairwise interaction satisfies the hypothesis as long as the potential is bounded from below, which is guaranteed by superstability.

\begin{Rq}[proof p. \pageref{demo_rq}]\label{rq_ham_local}
Thanks to \Cref{hyp_boundedly_attractive}, we can assume without any loss of generality that for any compact $\Delta\subset\RR^d$ and integer $N\geqslant0$,
\[
\forall\finiteconfig\in\ConfFinite,~\#\prth{\finiteconfig\cap\prth{\Delta+B_R}}=N\implies\InteractionLocal{\finiteconfig\cap\prth{\Delta+B_R}}\geqslant C_{\Delta,N}.
\]
\end{Rq}

\begin{Def}\label{def_ham_local}
For any compact $\Delta\subset\RR^d$, we define the local Hamiltonian $\Hamconditional*$ over
\[
\MathSet{\FKconfig\in\ConfSpace{FK}}[\sup_{s\in\interff{0;\beta}}\#\prth{\MathSet{\sect(s),~\sect\in\FKconfig}\cap\prth{\Delta+B_R}}<+\infty]
\]
by
\[
\Hamconditional{\FKconfig}=\int_0^{\beta}\InteractionLocal{\MathSet{\sect(s),~\sect\in\FKconfig}\cap\prth{\Delta+B_R}}\d s\in\RR\cup\{+\infty\}.
\]
\end{Def}

According to \Cref{rq_ham_local}, the integrand is bounded from below by $\min_{0\leqslant k\leqslant N}C_{\Delta,k}$ where
\[
N=\sup_{s\in\interff{0;\beta}}\#\prth{\MathSet{\sect(s),~\sect\in\FKconfig}\cap\prth{\Delta+B_R}}
\]
so the integral is well defined.

\begin{Lemdef}[proof pp. \pageref{demo_lemdef_DLR_1},\pageref{demo_lemdef_DLR_2}]\label{lemdef_DLR}
Let $\Delta\subset\RR^d$ be a compact and $\FKconfig\in\ConfSpace{FK}$.

We define the \emph{exterior} configuration relatively to $\Delta$ as
\[
\FKconfig\exterior\defequal\MathSet{\sect\in\FKconfig}[\sect\not\subset\Delta].
\]

If $\sup_{s\in\interff{0;\beta}}\#\prth{\MathSet{\sect(s),~\sect\in\FKconfig\exterior}\cap\prth{\Delta+B_R}}$ is finite, then one can define the constant
\begin{align*}
\Zconditional{\FKconfig}\defequal
\int\PoissonLeb[\Delta]{\d\zeta}~\E^{\beta\mu\#\zeta}\sum_{\PermutationSymbol\interior\in\PermutationSpace{\inward\cup\zeta\to\outward\cup\zeta}}
\int\prth{\bigotimes_{x\in\inward\cup\zeta}\WienerMeasureXYT[x,\PermutationSymbol\interior(x)][\beta,\subset\Delta]*}\!\prth{\d\eta}~\E^{-\Hamconditional{\eta\cup\FKconfig\exterior}}
\end{align*}
where we denote
\[
\PermutationSpace{X\to Y}\defequal\MathSet{\MyFunction{\PermutationSymbol}{X}{Y}{}{}}[\PermutationSymbol\text{ is bijective}]
\]
the inward and outward boundaries are the finite point configurations
\begin{align*}
\inward&\defequal\MathSet{x\in\Delta}[
\begin{array}{l}
\exists\sect\in\FKconfig\exterior,~x=\sect(\beta)\\
\forall\sect\in\FKconfig\exterior,~x\neq\sect(0)
\end{array}
]
\\
\outward&\defequal\MathSet{x\in\Delta}[
\begin{array}{l}
\exists\sect\in\FKconfig\exterior,~x=\sect(0)\\
\forall\sect\in\FKconfig\exterior,~x\neq\sect(\beta)
\end{array}
]
\end{align*}
and we define the measure
\[
\WienerMeasureXYT[x,y][\beta,\subset\Delta]{\d\sect}\defequal\carac{\sect\subset\Delta}\WienerMeasureXYT[x,y][\beta]{\d\sect}.
\]

The normalization constant $\Zconditional*$ is well defined, positive and finite $\Model{FK}*$ almost surely for any $L>0$. It is also true $\Model*{FK}*$ almost surely.

If the quantity $\Zconditional{\FKconfig}$ is well defined and $\Zconditional{\FKconfig}\in\interoo{0;+\infty}$ then we define the probability measure $\ConditionnalModel{~\cdot~}$ over $\ConfSpace{FK}$ by
\begin{align*}
\ConditionnalModel{\d\eta}
\defequal
\frac{1}{\Zconditional{\FKconfig}}\int&\PoissonLeb[\Delta]{\d\finiteconfig_{\Delta}}~\E^{\beta\mu\#\finiteconfig_{\Delta}}\sum_{\PermutationSymbol\interior\in\PermutationSpace{\inward\cup\finiteconfig_{\Delta}\to\outward\cup\finiteconfig_{\Delta}}}
\\
&\prth{\bigotimes_{x\in\inward\cup\finiteconfig_{\Delta}}\WienerMeasureXYT[x,\PermutationSymbol\interior(x)][\beta,\subset\Delta]*}\prth{\d\eta}~\E^{-\Hamconditional{\eta\cup\FKconfig\exterior}}.
\end{align*}
Otherwise we define the measure by $\ConditionnalModel{~\cdot~}\defequal0$.
\end{Lemdef}

We illustrate in \Cref{fig_int_ext} the detail of exterior and interior configurations.

Only points inside $\Delta$ and bridges intersecting the compact are represented.

\begin{figure}[h]
\centering
\begin{tikzpicture}[scale=1]
\tikzstyle{fleche_int} = [rounded corners,->,>=latex,color=red]
\tikzstyle{fleche_ext} = [rounded corners,->,>=latex,color=blue,dashed]
\draw[dotted] (0,0) rectangle (4,4);
\draw (-0.3,1) node {$\Delta$};
\draw (1.3,1) node (A) {} ;
\draw (3.3,0.5) node (B) {} ;
\draw (3.4,3) node (C) {} ;
\draw (2.8,1.7) node (D) {} ;
\draw (0.7,3.6) node (E) {} ;
\draw (4.9,3.4) node (F) {};
\draw (2.3,3.7) node (G) {};
\draw (0.2,3.85) node (H) {};
\draw[fleche_int] (1.3,1) to[bend left, looseness=1] (0.3,0.5) to[bend left, looseness=1] (0.7,2) to[bend left, looseness=1] (2,2) to[bend left, looseness=1] (A);
\draw[fleche_int] (3.3,0.5) to[bend left, looseness=1] (2.5,0.7) to[bend left, looseness=1] (D);
\draw[fleche_int] (2.8,1.7) to[bend left, looseness=1.5] (C);
\draw[fleche_int] (0.7,3.6) to[bend left, looseness=2] (1.5,2.5) to[bend left, looseness=1] (H);
\draw (4.7,1.1) node (Z) {};
\draw[fleche_ext] (4.7,1.1) to[bend right, looseness=1.7] (B);
\draw[fleche_ext] (3.4,3) to[bend right, looseness=1.2] (F);
\draw[fleche_ext] (2.3,3.7) to[bend right, looseness=1.1] (2.4,4.6) to[bend right, looseness=1.1] (G);
\draw[fleche_ext] (0.2,3.85) to[bend left, looseness=0.9] (-1.3,3.5);
\draw[fleche_ext] (0.9,4.9) to[bend left, looseness=0.6] (E);
\draw (1.3,1) node {$\bullet$} ;
\draw (3.3,0.5) node {\tiny$\blacksquare$} ;
\draw (3.4,3) node {\small$\blacktriangle$} ;
\draw (2.8,1.7) node {$\bullet$} ;
\draw (0.7,3.6) node {\tiny$\blacksquare$} ;
\draw (2.3,3.7) node {\small$\blacklozenge$};
\draw (0.2,3.85) node {\small$\blacktriangle$};
\draw[fleche_int] (6,4) to (7,4);
\draw (7,4) node[right] {$\FKconfig\interior$};
\draw[fleche_ext] (6,3) to (7,3);
\draw (7,3) node[right] {$\FKconfig\exterior$};
\draw (10,4) node {\tiny$\blacksquare$};
\draw (10.5,4) node[right] {$\inward$};
\draw (10,3) node {\small$\blacktriangle$};
\draw (10.5,3) node[right] {$\outward$};
\draw (6.5,2) node {$\bullet$};
\draw (7,2) node[right] {Start and end of a bridge from $\textcolor{red}{\FKconfig\interior}$};
\draw (6.5,1) node {\small$\blacklozenge$};
\draw (7,1) node[right] {Start and end of a bridge from $\textcolor{blue}{\FKconfig\exterior}$};
\end{tikzpicture}
\caption{Exterior and interior configurations relatively to $\Delta$}
\label{fig_int_ext}
\end{figure}

\begin{Rq}
If the exterior configuration $\FKconfig\exterior$ is far enough from the compact $\Delta$
\[
\forall\sect\in\FKconfig\exterior,~\sect\cap(\Delta+B_R)=\emptyset
\]
then the conditional distribution does not depend on the exterior configuration :
\[
\ConditionnalModel{~\cdot~}=\ConditionnalModel{~\cdot~}[\emptyset].
\]
In fact, it coincides with a finite volume model on $\Delta$ with Dirichlet boundary conditions.
\end{Rq}

\begin{Th}[DLR equations, proof p. \pageref{demo_th_DLR}]\label{th_DLR}
Let $\Delta\subset\RR^d$ be a compact. Under hypotheses I to V, for any inverse temperature $\beta>0$ and chemical potential $\mu\in\RR$, for any measurable $\MyFunction{f}{\ConfSpacePerm{FK}}{\RR^+}{}{}$,
\[
\int f(\FKconfig)~\Model*{FK}{\d\FKconfig}=\int \prth{\int f\prth{\eta\cup\FKconfig\exterior}~\ConditionnalModel{\d\eta}}~\Model*{FK}{\d\FKconfig}.
\]
\end{Th}

The resampling of the interior configuration in the DLR equations consists in the following: we erase $\bullet$ points and \textcolor{red}{solid bridges}, then sample new interior $\bullet$ points, a new interior bijection $\PermutationSymbol\interior\in\SS(${\tiny$\blacksquare$}$\cup\bullet\to${\small$\blacktriangle$}$\cup\bullet)$ and the associated \textcolor{red}{solid bridges}. This is illustrated in \Cref{fig_DLR}.

\begin{figure}[!h]
\centering
\begin{subfigure}[c]{0.45\linewidth}
\centering
\begin{tikzpicture}[scale=1]
\tikzstyle{fleche_int} = [rounded corners,->,>=latex,color=red]
\tikzstyle{fleche_ext} = [rounded corners,->,>=latex,color=blue,dashed]
\draw[dotted] (0,0) rectangle (4,4);
\draw (-0.3,1) node {$\Delta$};
\draw (1.3,1) node (A) {} ;
\draw (3.3,0.5) node (B) {} ;
\draw (3.4,3) node (C) {} ;
\draw (2.8,1.7) node (D) {} ;
\draw (0.7,3.6) node (E) {} ;
\draw (4.9,3.4) node (F) {};
\draw (2.3,3.7) node (G) {};
\draw (0.2,3.85) node (H) {};
\draw[fleche_int] (1.3,1) to[bend left, looseness=1] (0.3,0.5) to[bend left, looseness=1] (0.7,2) to[bend left, looseness=1] (2,2) to[bend left, looseness=1] (A);
\draw[fleche_int] (3.3,0.5) to[bend left, looseness=1] (2.5,0.7) to[bend left, looseness=1] (D);
\draw[fleche_int] (2.8,1.7) to[bend left, looseness=1.5] (C);
\draw[fleche_int] (0.7,3.6) to[bend left, looseness=2] (1.5,2.5) to[bend left, looseness=1] (H);
\draw (4.7,1.1) node (Z) {};
\draw[fleche_ext] (4.7,1.1) to[bend right, looseness=1.7] (B);
\draw[fleche_ext] (3.4,3) to[bend right, looseness=1.2] (F);
\draw[fleche_ext] (2.3,3.7) to[bend right, looseness=1.1] (2.4,4.6) to[bend right, looseness=1.1] (G);
\draw[fleche_ext] (0.2,3.85) to[bend left, looseness=0.9] (-1.3,3.5);
\draw[fleche_ext] (0.9,4.9) to[bend left, looseness=0.6] (E);
\draw (1.3,1) node {$\bullet$} ;
\draw (3.3,0.5) node {\tiny$\blacksquare$} ;
\draw (3.4,3) node {\small$\blacktriangle$} ;
\draw (2.8,1.7) node {$\bullet$} ;
\draw (0.7,3.6) node {\tiny$\blacksquare$} ;
\draw (2.3,3.7) node {\small$\blacklozenge$};
\draw (0.2,3.85) node {\small$\blacktriangle$};
\end{tikzpicture}
\caption{}
\label{fig_sample1}
\end{subfigure}
$\Longrightarrow$
\begin{subfigure}[c]{0.45\linewidth}
\centering
\begin{tikzpicture}[scale=1]
\tikzstyle{fleche_int} = [rounded corners,->,>=latex,color=red]
\tikzstyle{fleche_ext} = [rounded corners,->,>=latex,color=blue,dashed]
\draw[dotted] (0,0) rectangle (4,4);
\draw (-0.3,1) node {$\Delta$};
\draw (3.3,0.5) node (B) {} ;
\draw (3.4,3) node (C) {} ;
\draw (0.7,3.6) node (E) {} ;
\draw (4.9,3.4) node (F) {};
\draw (2.3,3.7) node (G) {};
\draw (0.2,3.85) node (H) {};
\draw (2,3) node (I) {};
\draw (0.5,2) node (J) {};
\draw (1.5,0.9) node (K) {};
\draw[fleche_int] (0.7,3.6) to[bend right, looseness=1.6] (I);
\draw[fleche_int] (2,3) to[bend left, looseness=0.7] (C);
\draw[fleche_int] (3.3,0.5) to[bend left, looseness=0.5] (K);
\draw[fleche_int] (1.5,0.9) to[bend right, looseness=1.4] (J);
\draw[fleche_int] (0.5,2) to[bend right, looseness=0.9] (H);
\draw (4.7,1.1) node (Z) {};
\draw[fleche_ext] (4.7,1.1) to[bend right, looseness=1.7] (B);
\draw[fleche_ext] (3.4,3) to[bend right, looseness=1.2] (F);
\draw[fleche_ext] (2.3,3.7) to[bend right, looseness=1.1] (2.4,4.6) to[bend right, looseness=1.1] (G);
\draw[fleche_ext] (0.2,3.85) to[bend left, looseness=0.9] (-1.3,3.5);
\draw[fleche_ext] (0.9,4.9) to[bend left, looseness=0.6] (E);
\draw (3.3,0.5) node {\tiny$\blacksquare$} ;
\draw (3.4,3) node {\small$\blacktriangle$} ;
\draw (0.7,3.6) node {\tiny$\blacksquare$} ;
\draw (2.3,3.7) node {\small$\blacklozenge$};
\draw (0.2,3.85) node {\small$\blacktriangle$};
\draw (2,3) node {$\bullet$};
\draw (0.5,2) node {$\bullet$};
\draw (1.5,0.9) node {$\bullet$};
\end{tikzpicture}
\caption{}
\label{fig_sample2}
\end{subfigure}
\caption{Resampling in the DLR equations}
\label{fig_DLR}
\end{figure}

\section{Equivalent Models}

As we mentioned previously, there is more than one formulation of the Bose gas. We properly introduce two other equivalent models. The \emph{marked point} (mp) framework is essential to establish the thermodynamic limit. The \emph{rooted loops} (rl) framework is necessary because of \Cref{Prop_time-shift} which states the invariance of the Bose gas under a time-shift of the Brownian bridges in the configuration. This is useful in the proof of the (mp) entropic bound. It is also worth examining how incompatible the topologies of (rl) and (FK) thermodynamic limits are.

In the light of the previous description, one might question the necessity of introducing the (FK) framework, instead of using natively the (rl) one. We chose to express our main results in the (FK) setting for two main reasons. Of course, this writing is the closest to the physics. But more importantly, the DLR equations from \Cref{th_DLR} necessitate to cut down the cycles of $\PermutationFK{\cdot}$ (or equivalently to cut the loops in a loop soup configuration) in \emph{interior} and \emph{exterior} parts relatively to a given compact. The (FK) framework is naturally adapted to this operation, unlike the (rl) one.

\subsection{Rooted Loops (rl)}\label{subsection_rl}

A usual formulation of the Bose gas is the loop soup one (\autocite{BV23}, \autocite{DV24}, \emph{etc}). Instead of sampling separately $N$ different Brownian bridges which happen to draw a permutation $\PermutationSymbol\in\PermutationSpace{N}$, this model samples directly the cycles of $\PermutationSymbol$ as Brownian loops $\MyFunction{\loopath}{\RR}{\RR^d}{}{}$. This model can be naturally interpreted as a process sampling marked points: the position is the root $\loopath(0)$ of the loop, and the mark is the loop itself. This is why we call this representation the \emph{rooted loop} (rl) model of the Bose gas. In \Cref{rq_pas_equivalence}, we discuss the flaws this representation has that prevent its thermodynamic limit from sampling interlacements.

\begin{Def}\label{def_Loop_t}
For any $\brown\in\WienerSpace[t]$, we define the continuation $\MyFunction{\widetilde{\brown}}{\RR}{\RR^d}{}{}$ of the trajectory
\[
\widetilde{\brown}(s+tq)\defequal\brown(s)+q\prth{\brown\prth{t}-\brown\prth{0}}
\]
for $q\in\ZZ$ and $s\in\interfo{0;t}$.

We introduce the set
\[
\WienerSpaceExtended{t}\defequal
\ens{\widetilde{\brown},~\brown\in\WienerSpace[t]}.
\]
The map $\MyFunction{}{}{}{\brown}{\widetilde{\brown}}$ induces the respective topology $\WienerTopologyExtended{t}$ and $\sigma$-algebra $\WienerAlgebraExtended{t}$.
\end{Def}

\begin{Def}\label{def_LoopRoot}
We define the \emph{length} $\length{\loopath}$ of a given $\loopath\in\bigcup_{j\geqslant 1}\WienerSpaceExtended{\beta j}$ to be the integer
\[
\length{\loopath}\defequal\inf\ens{j\geqslant 1}[\loopath\in\WienerSpaceExtended{\beta j}].
\]

We define
\begin{align*}
\WienerSpaceLoop{rl}&\defequal
\bigcup_{j\geqslant 1}\ens{\loopath\in\WienerSpaceExtended{\beta j}}[
\begin{array}{l}
\length{\loopath}=j\\
\loopath(\beta j)=\loopath(0)
\end{array}
].
\end{align*}
Trajectories in a set $\WienerSpaceLoop{rl}$ are called \emph{rooted loops}.
\end{Def}

\begin{Lemdef}\label{lemdef_LoopRoot_alg_top}
Defining $\WienerSpaceLoop{rl}$ as a discrete disjoint union of subsets induces a topology and $\sigma$-algebra over this space, which we denote $\WienerTopologyLoop{rl}$ and $\WienerAlgebraLoop{rl}$ respectively.

The topological space $\prth{\WienerSpaceLoop{rl},\WienerTopologyLoop{rl}}$ is Polish for the distance
\[
\Dinf{\loopath,\loopath'}\defequal
\begin{cases}
\sup_{s\in\interff{0;\beta j}}\norm{\loopath(s)-\loopath'(s)}
& \text{ if }j=\length{\loopath}=\length{\loopath'}
\\
+\infty & \text{ otherwise}.
\end{cases}
\]
\end{Lemdef}

It is clear $\MyFunction{\Dinf*}{\WienerSpaceLoop{rl}\times\WienerSpaceLoop{rl}}{\RR\cup\{+\infty\}}{}{}$ metrizes the topology of the subsets of $\WienerSpaceExtended{\beta j}$, $j\geqslant1$, described in the definition of $\WienerSpaceLoop{rl}$. Moreover, $\Dinf*$ sets a $+\infty$ distance between those subsets, which is coherent with a discrete disjoint union. The completeness and separability of $\WienerSpaceLoop{rl}$ come from the completeness and separability of each $\WienerSpaceExtended{\beta j}$, $j\geqslant1$.

The property of being Polish is sufficient to ensure that a Poisson point process can be defined on this space.

\begin{Def}\label{def_confspace_rl}
	We define a set of \emph{rooted loops} configurations as
	\[
	\ConfSpace{rl}*\defequal
	\ens{\rlconfig\subset\WienerSpaceLoop{rl}}[\rlconfig\text{ is locally finite in }\WienerTopologyLoop{rl}]
	\]
	and we equip this space with the smallest $\sigma$-algebra $\ConfAlgebra{rl}*$ making measurable the maps $\MyFunction{}{}{}{\rlconfig}{\#\prth{\rlconfig\cap E}}$ for every bounded $E\in\WienerAlgebraLoop{rl}$.
	
	We also define the transition map $\MyFunction{\ViewChange[dir][\infty]{FK}{rl}*}{\ConfSpace{rl}*}{\ConfSpace{FK}}{}{}$ by
	\[
	\ViewChange[dir][\infty]{FK}{rl}{\rlconfig}\defequal
	\ens{\MyFunction{}{\interff{0;\beta}}{\RR^d}{s}{\loopath(\beta j+s)}\!,~
		0\leqslant j<\length{\loopath},~\loopath\in\rlconfig}
	\]
	which cuts down each loop of length $j\geqslant1$ into $j$ distinct pieces.
\end{Def}

\begin{Def}\label{def_W_root}
On the set of rooted loops, we introduce the following measure
\begin{align*}
\WienerMeasureRootBC[Dir]*
&\defequal\int_{\Lambda_L}\d x\sum_{j\geqslant 1}\frac{1}{j}\WienerMeasureXYT[x,x][\beta j]*
\end{align*}
We denote by $\Poisson{rl}*$ the Poisson point process over $\ConfSpace{rl}$ with intensity measure $\WienerMeasureRootBC*$.
\end{Def}

\begin{Def}\label{def_ham_rl}
We define a Hamiltonian for rooted loops configurations
\[
\Ham{rl}{\rlconfig}\defequal
\int_0^{\beta}\Interaction{\ens{\loopath(\beta j+s),~0\leqslant j<\length{\loopath},~\loopath\in\rlconfig}}\d s.
\]
\end{Def}

\begin{Lemdef}[proof: see \Cref{rq_demo}]\label{lemdef_rl}
Let $L>0$. The probability measure over $\ConfSpace{rl}$
\[
\Model{rl}{\d\rlconfig}\defequal\frac{\E^{-L^d+\WienerMeasureRootBC{\WienerSpaceLoop{rl}}}}{\Zgrandcanonic} \exp\prth{\beta\mu\sum_{\loopath\in\rlconfig}\length{\loopath}-\Ham{rl}{\rlconfig}}~\Poisson{rl}{\d\rlconfig}.
\]
where $\Zgrandcanonic$ is the same normalization constant as in \Cref{lemdef_FK}, is well defined. We call this probability the \emph{rooted loops model} of the Bose gas.
\end{Lemdef}

The following proposition states the Feynman-Kac and rooted loops models are equivalent.

\begin{Prop}[proof: see \Cref{rq_demo}]\label{Prop_equiv_FK_rl}
For any measurable $\MyFunction{f}{\ConfSpace{FK}}{\RR^+}{}{}$,
\[
\int f\circ\ViewChange{FK}{rl}*\d\Model{rl}*=\int f\d\Model{FK}*
\]
\end{Prop}

As noted in in \Cref{rq_pas_equivalence}, this result may only be true in finite volume.

\begin{Prop}[proof p. \pageref{demo_prop_time-shift}]\label{Prop_time-shift}
	The rooted loops model with Dirichlet boundary conditions is time-shift invariant. More precisely, for any $s\in\RR$ and measurable $\MyFunction{f}{\ConfSpace{rl}}{\RR^+}{}{}$,
	\[
	\int f\circ\TimeTranslation*\d\Model{rl}*=\int f\d\Model{rl}*
	\]
	with the time-shift operator $\TimeTranslation*$ defined as
	\[
	\MyFunction{\TimeTranslation*}{\ConfSpace{rl}}{\ConfSpace{rl}}{\rlconfig}{\MathSet{\loopath\prth{\cdot+s},~\loopath\in\rlconfig}.}
	\]
\end{Prop}

\begin{Rq}
We could also have formulated time-shift invariance in the (FK) framework but it was useless here and the proof was much easier in the (rl) model, due to the time-shift invariance of the reference measure $\WienerMeasureRootBC*$ itself. This property does not translate well in (mp).
\end{Rq}

\begin{Def}
For any $L>0$, we define the \emph{empirical field} $\EmpiricalField{rl}[Dir]*$ over $\ConfSpace{rl}*$ by
\[
\int f\d\EmpiricalField{rl}[Dir]*\defequal
\frac{1}{L^d}\int_{\window}\d v\int f(\FKconfig+v)~\Model{rl}[Dir]{\d\FKconfig}
\]
for any measurable $\MyFunction{f}{\ConfSpace{rl}*}{\RR}{}{}$.
\end{Def}

\begin{Def}
A function $\MyFunction{f}{\ConfSpace{rl}*}{\RR}{}{}$ is said to be \emph{\tame{rl} local} if there exists $a>0$ and a compact $\Delta\subset\RR^d$ such that
\begin{equation*}
\forall\rlconfig\in\ConfSpace{rl}*,~
f(\rlconfig)=f\prth{\ens{\loopath\in\rlconfig}[\loopath(0)\in\Delta]}
\end{equation*}
and
\[
\forall\rlconfig\in\ConfSpace{rl}*,~a\abs{f(\rlconfig)}\leqslant 1+\#\ens{\loopath\in\rlconfig}[\loopath(0)\in\Delta].
\]
\end{Def}

\begin{Th}[proof p. \pageref{demo_th_limit_rl}]\label{th_limit_rl}
There exists a stationary probability measure $\Model*{rl}[Dir]*$ over $\ConfSpace{rl}*$ and an increasing sequence $L_m\tend{m}+\infty$ such that, for any \tame{rl} local measurable $\MyFunction{f}{\ConfSpace{rl}*}{\RR}{}{}$,
\[
\lim_{m\to+\infty}\int f\d\EmpiricalField{rl}[Dir][L_m]*
=\int f\d\Model*{rl}[Dir]*.
\]
\end{Th}

\begin{Cor}\label{cor_rl_no_interlacement}
By construction, $\Model*{rl}[Dir]{\rlconfig\subset\WienerSpaceLoop{rl}}=1$.
\end{Cor}

\begin{Rq}\label{rq_pas_equivalence}
With \Cref{cor_rl_no_interlacement}, one may conclude that the Feynman-Kac model $\Model*{FK}[Dir]*$, just as $\Model*{rl}[Dir]*$, only produces finite cycles. But this is not so trivial. The infinite volume models $\Model*{FK}[Dir]*$ and $\Model*{rl}[Dir]*$ may \emph{not} be equivalent, as we conjecture :
\[
\Model*{rl}[dir]{\prth{\!\ViewChange[dir][\infty]{FK}{rl}*}^{-1}(E)}
\neq\Model*{FK}[dir]{E}
\]
for some event $E\in\ConfAlgebraPerm{FK}$, because the topologies of convergence for \FK and rooted loops models are fundamentally incompatible. At first glance the class of functions on which rooted loops models converge seems strictly larger. Indeed, (rl)-local functions can depend on whole loops, whereas $\cap^n$-local functions can only see $n$ Brownian bridges beyond the boundary of some compact $\Delta$. However (rl)-local functions are limited as they can only see loops whose root is inside $\Delta$, contrary to $\cap^n$-local functions. Those two modes of convergence can not be compared and we believe this is not just a technicality.

We could have chosen a more general notion of tameness for the (rl) framework but this would not have changed the discussion above because the problem comes from locality.
\end{Rq}

\subsection{Marked Points (mp)}\label{sec_mp}

We present now a framework in which a lot of the work has been done. But it unfortunately also is the least elegant one. We encode the Feynman Kac representation into a configuration of \emph{marked points} (mp). The goal is to localize the global information of the permutation $\PermutationSymbol$ into marks, so that the mark of each point $x$ is enough to reconstruct the Brownian bridge starting in $x$ and ending at some $y=\PermutationSymbol(x)$.

Let us explain with words and picture what the mathematics describe more precisely later in \Cref{def_confspace_mp}. Each point $x\in\finiteconfig$ of the point configuration is now equipped with a 3-part mark $(p,u,\brown)\in\ZZ^d\times\interff{0;1}\times\WienerSpace[1]$. To identify the position of $\PermutationSymbol(x)$, we first consider a square lattice of size $r$ centered in $x$ (the point $x$ is at the center of a cell). The image of $x$ is in the cell whose center is $x+rp$.

\begin{figure}[!h]
\centering
\begin{tikzpicture}
\draw[<->] (-1.75,0) to (-1.75,1);
\draw (-2,0.5) node{$r$};

\draw (0.5,0.5) node{$\bullet$};
\draw (0.5,0.25) node{$x$};
\draw (3,0.75) node[right]{$p=(1,1)$};
\draw (3,0.25) node[right]{$u=0.383\dots$};
\draw (2.1,1.3) node{$\PermutationSymbol(x)$};
\draw[->,thick] (0.5,0.5) to (1.5,1.5);
\draw (1,1) rectangle (2,2);

\draw (-1.5+0.019,2.202) node{$\bullet$};
\draw (-1+0.065,2.129) node{$\bullet$};
\draw (0.060,2.236) node{$\bullet$};
\draw (1.264,1.711) node{$\bullet$};
\draw (1.758,1.513) node{$\bullet$};
\draw (1.884,1.832) node{$\bullet$};
\draw (-1+0.571,0.880) node{$\bullet$};
\draw (-1+0.035,0.883) node{$\bullet$};
\draw (+0.872,-1+0.954) node{$\bullet$};
\draw (+0.671,-1+0.212) node{$\bullet$};
\draw (2+0.365,-1+0.547) node{$\bullet$};
\draw (-1.5+0.119,-1.5+0.240) node{$\bullet$};
\draw (1+0.205,-1.5+0.137) node{$\bullet$};

\clip (-1.5,-1.5) rectangle (2.5,2.5);
\draw[dashed] (-2,-2) grid (3,3);
\end{tikzpicture}
\caption{Use of the $p$ and $u$ marks in the (mp) encoding}
\label{fig_encoding}
\end{figure}

It may be the case that the cell $x+rp+\case$ does not contain any point of $\finiteconfig$, but let us ignore this pathological case for now. Another apparently problematic possibility is the target cell may contain more than one point of $\finiteconfig$ (as represented in \Cref{fig_encoding}). How does one choose $\PermutationSymbol(x)$ among all those possibilities? The solution we present here is to order all points $y\in\finiteconfig\cap(x+rp+\case)$ lexicographically and choose one of those depending on the value of $u$. If there are $n$ possibilities and $u\in\interff{0;\frac{1}{n}}$, we choose the first point. If $u\in\interof{\frac{1}{n};\frac{2}{n}}$, we choose the second one, \emph{etc}. The knowledge of the point configuration in $x+rp+\case$ is necessary to perform this reconstruction, in addition to the marks of $x$ of course. Finally, to reconstruct the bridge between $x$ and $\PermutationSymbol(x)$, we unfold the $\brown$ mark of $x$ along the straight line from $x$ to $\PermutationSymbol(x)$:
\[
\MyFunction{}{\interff{0;\beta}}{\RR^d}{s}{\dps x+\frac{s}{\beta}\prth{\PermutationSymbol\prth{x}-x}+\sqrt{\beta}~\brown\prth{\frac{s}{\beta}}.}
\]

This decoding protocol does not natively guarantee the reconstructed map $\MyFunction{\PermutationSymbol}{\finiteconfig}{\finiteconfig}{}{}$ to be bijective. Actually we have seen previously that, if a target cell is empty, the reconstruction protocol can not even be performed. We take care of those edge cases by limiting ourselves to a specific subset of marked points configurations.

\begin{Def}\label{def_confspace_mp}
We define
\begin{align*}
\ConfSpace{mp}&\defequal\MathSet{\mpconfig=\prth{x,p_x,u_x,\brown_x}_{x\in\finiteconfig}\subset\RR^d\times\ZZ^d\times\interff{0;1}\times\WienerSpace[1]}[\finiteconfig\text{ is locally finite in }\RR^d].
\end{align*}
and for any $\mpconfig\in\ConfSpace{mp}$ we denote its spatial component
\[
\SpatialComponent*\defequal\MathSet{x,~(x,p,u,\brown)\in\mpconfig}\subset\RR^d.
\]
For any $\mpconfig\in\ConfSpace{mp}$, we also denote
\[
\CardCase{z}\defequal\#\prth{\SpatialComponent\cap\prth{z+\case}}.
\]

We define the set of \emph{authorized} configurations as
\[
\ConfSpaceAuth\defequal\MathSet{\mpconfig\in\ConfSpace{mp}}[\SpatialComponent*\text{ is simple and }\forall(x,p,u,\brown)\in\mpconfig,~\CardCase{x+rp}\geqslant 1].
\]
For any authorized configuration $\mpconfig\in\ConfSpaceAuth$, we define the map $\MyFunction{\Permutationmp{\cdot}}{\SpatialComponent*}{\SpatialComponent}{}{}$ such that for any $(x,p,u,\brown)\in\mpconfig$,
\[
\Permutationmp{x}\defequal\Ent{\CardCase{x+rp}\cdot u}\text{'th element of }\SpatialComponent\cap\prth{x+rp+\case}
\]
if we order them lexicographically\footnote{It is an abuse of notation to only refer to the position $x$ in the writing $\Permutationmp{x}$ since it also depends on the marks of this point. But given the fact the point configuration is assumed to be simple, there is no ambiguity in which marks are being used to define $\Permutationmp{x}$.}.

Finally we call an authorized configuration \emph{permutation-wise} if it satisfies
\[
\forall y\in\SpatialComponent*,~\exists!x\in\SpatialComponent*,~\Permutationmp{x}=y
\]
and we denote
\[
\ConfSpacePerm{mp}\defequal\MathSet{\mpconfig\in\ConfSpaceAuth}[\mpconfig\text{ is permutation-wise}].
\]
\end{Def}

\begin{Def}\label{def_vc_mp_FK}
We define the transition map $\MyFunction{\ViewChange{FK}{mp}*}{\ConfSpacePerm{mp}}{\ConfSpacePerm{FK}}{}{}$ by
\begin{align*}
\ViewChange[dir]{FK}{mp}{\mpconfig}&\defequal
\MathSet{\MyFunction{}{\interff{0;\beta}}{\RR^d}{s}{x+\frac{s}{\beta}\prth{\Permutationmp[dir]{x}-x}+\sqrt{\beta}~\brown\prth{\frac{s}{\beta}}},~(x,p,u,\brown)\in\mpconfig}.
\end{align*}
\end{Def}

\begin{Def}\label{def_Poisson_mp}
Let $\pMeasure*$ be a probability measure over $\ZZ^d$ such that $\forall p\in\ZZ^d,~\pMeasure{p}>0.$

We also denote as $\WienerMeasureNor*$ the probability measure
$
\WienerMeasureNor*\defequal\frac{\WienerMeasureXYT[0,0][1]*}{\WienerMeasureXYT[0,0][1]{\WienerSpace[1]}}.
$

We denote as $\Poisson{mp}*$ the Poisson point process with intensity measure $\Leb[d]\otimes\pMeasure*\otimes\Leb[1]_{\interff{0;1}}\otimes\WienerMeasureNor*$ over configurations $\mpconfig\in\ConfSpace{mp}$ such that $\SpatialComponent*\subset\window$.
\end{Def}

We chose the reference measure $\WienerMeasureNor*$ so that the state space does not depend on $\beta$, unlike the (rl) framework, and chose a normalized measure for mathematical elegance (the intensity measure's mass is exactly $L^d$).

\begin{Lemdef}[proof: see \Cref{rq_demo_mp}]\label{lemdef_mp}
We define a Hamiltonian over finite marked points configurations as follows

if $\mpconfig\in\ConfSpacePerm{mp}$,
\begin{align*}
\Ham{mp}{\mpconfig}\defequal
&\sum_{(x,p,u,\brown)\in\mpconfig}\prth{\frac{d}{2}\log\prth{2\pi\beta}+\frac{1}{2\beta}\norm{\Permutationmp{x}-x}^2+\log\prth{\pMeasure{p}}-\log\prth{\CardCase{x+rp}}}
\\
&+\int_0^{\beta}\Interaction{\MathSet{x+\frac{s}{\beta}\brac{\Permutationmp{x}-x}+\sqrt{\beta}~\brown\prth{\frac{s}{\beta}},~\prth{x,p,u,\brown}\in\mpconfig}}\d s
\end{align*}
else $\Ham{mp}{\mpconfig}\defequal+\infty.$

Then for any $L>0$, the probability measure over $\ConfSpace{mp}$
\[
\Model{mp}{\d\mpconfig}\defequal\frac{1}{\Zgrandcanonic}\exp\prth{\beta\mu\#\mpconfig-\Ham{mp}{\mpconfig}}~\Poisson{mp}{\d\mpconfig}
\]
where $\Zgrandcanonic$ is the same normalization constant as in \Cref{lemdef_FK}, is well defined.
\end{Lemdef}

\begin{Rq}
The model $\Model{mp}*$ does not depend on the chosen density $\MyFunction{\pMeasure*}{\ZZ^d}{\RR}{}{}$.
\end{Rq}

This horrendous formulation of the Bose gas is equivalent to the previously defined models.

\begin{Prop}[proof: see \Cref{rq_demo_mp}]\label{Prop_equiv_mp_FK}
For any measurable $\MyFunction{f}{\ConfSpace{FK}}{\RR^+}{}{}$,
\[
\int f\circ\ViewChange{FK}{mp}*\d\Model{mp}*=\int f\d\Model{FK}*.
\]
\end{Prop}

\begin{Def}\label{def_local_tame_mp}
A function $\MyFunction{f}{\ConfSpace{mp}}{\RR}{}{}$ is said to be \emph{\tame{mp} local} if there exists $a,\delta>0$, $\alpha\in\interfo{0;2}$ and a compact $\Delta\subset\RR^d$ such that
\[
\forall\mpconfig\in\ConfSpace{mp},~
f(\mpconfig)=f\prth{\MathSet{\prth{x,p,u,\brown}\in\mpconfig}[x\in\Delta]}
\]
and
\[
\forall\mpconfig\in\ConfSpace{mp},~
a\abs{f(\mpconfig)}\leqslant 1+\sum_{(x,p,u,\brown)\in\mpconfig}\abs{\Sausage[\delta]{\MyFunction{}{\interff{0;\beta}}{\RR^d}{s}{\frac{s}{\beta}rp+\sqrt{\beta}~\brown\brac{\frac{s}{\beta}}}}}^{\alpha}.
\]
\end{Def}

\begin{Def}
For any $L>0$, we define the empirical field $\EmpiricalField{mp}*$ over $\ConfSpace{mp}$ by
\[
\int f(\mpconfig)~\EmpiricalField{mp}{\d\mpconfig}=
\frac{1}{L^d}\int_{\window}\d v\int f(\mpconfig+v)~\Model{mp}{\d\mpconfig}.
\]
\end{Def}

\begin{Th}[proof p. \pageref{demo_th_limit_mp}]\label{th_limit_mp}
There exists a stationary probability measure $\Model*{mp}*$ over $\ConfSpace{mp}$ and an increasing sequence $L_m\tend{m}+\infty$ such that, for any \tame{mp} local measurable $\MyFunction{f}{\ConfSpace{mp}}{\RR}{}{}$,
\[
\lim_{m\to+\infty}\int f\d\EmpiricalField{mp}[bc][L_m]*
=\int f\d\Model*{mp}*.
\]
\end{Th}

This theorem is the one unlocking everything. The probability $\Model*{mp}*$ is used to directly define $\Model*{FK}*$ through the transition map $\ViewChange*{FK}{mp}*$.

\subsection{Proof of Equivalences}

In this section, we prove the various models to be well-defined (\Cref{lemdef_FK,lemdef_rl,lemdef_mp}), equivalent (\Cref{Prop_equiv_FK_rl,Prop_equiv_mp_FK}) and possibly time-shift invariant (\Cref{Prop_time-shift}).

The following proposition establishes a link between (FK) and (rl) frameworks. It proves that, if the models (FK) and (rl) are well defined, then they are equivalent.

\begin{Prop}\label{Prop_proof_equiv_FK_rl}
	For any measurable $\MyFunction{f}{\ConfSpace{FK}}{\RR^+}{}{}$,
	\begin{align*}
		&\E^{L^d}\int\PoissonLeb{\d\finiteconfig}\sum_{\PermutationSymbol\in\PermutationSpace{\finiteconfig}}\int f\prth{\FKconfig\bcversion}~\E^{-\Ham{FK}{\FKconfig}}~\prth{\bigotimes_{x\in\finiteconfig}\WienerMeasureXYTBC[x,\PermutationSymbol(x)][\beta]*}\prth{\d\FKconfig}
		\\
		=&\E^{\WienerMeasureRootBC{\WienerSpaceLoop{rl}}}\int\prth{f\circ\ViewChange{FK}{rl}*}(\rlconfig)~\E^{-\Ham{rl}{\rlconfig}}~\Poisson{rl}{\d\rlconfig}.
	\end{align*}
\end{Prop}

\begin{Demo}[\Cref{Prop_proof_equiv_FK_rl}]
	The following proof is simply a rewriting of a special case of the contents from \autocite{Gin70}, approximately from the bottom of page 358 to page 360 (the "QS case").
	
	By definition of the standard Poisson point process,
	\begin{align*}
		&\int\PoissonLeb{\d\finiteconfig}\sum_{\PermutationSymbol\in\PermutationSpace{\finiteconfig}}\int f\prth{\FKconfig\bcversion}~\E^{-\Ham{FK}{\FKconfig}}~\prth{\bigotimes_{x\in\finiteconfig}\WienerMeasureXYTBC[x,\PermutationSymbol(x)][\beta]*}\prth{\d\FKconfig}
		\\
		&=\E^{-L^d}\sum_{N=0}^{+\infty}\frac{1}{N!}\int_{\window}\d x^{\otimes N}\sum_{\PermutationSymbol\in\PermutationSpace{N}}\int f\prth{\FKconfig\bcversion}~\E^{-\Ham{FK}{\FKconfig}}~\prth{\bigotimes_{i=1}^N\WienerMeasureXYTBC[x_i,x_{\PermutationSymbol(i)}][\beta]*}\prth{\d\FKconfig}.
	\end{align*}
	Let $N\geqslant 0$ and $\PermutationSymbol\in\PermutationSpace{N}$. We denote as $\ell_i,~1\leqslant i\leqslant n$ the respective cycle lengths of $\PermutationSymbol$.
	
	We can re-order the variables $\prth{x_i}_{1\leqslant i\leqslant N}$ into $\prth{x_{i,j}}_{0\leqslant j<\ell_i,~1\leqslant i\leqslant n}$ such that
	\[
	\PermutationSymbol\prth{x_{i,j}}=x_{i,j+1}
	\]
	with the convention $x_{i,\ell_i}=x_{i,0}$. Then
	\begin{align*}
		&\int_{\window}\d x^{\otimes N}\int f\prth{\FKconfig\bcversion}~\E^{-\Ham{FK}{\FKconfig}}~\prth{\bigotimes_{i=1}^N\WienerMeasureXYTBC[x_i,x_{\PermutationSymbol(i)}][\beta]*}\prth{\d\FKconfig}
		\\
		=&\int_{\window}\bigotimes_{i=1}^n\d x_i^{\otimes\ell_i}\int f\prth{\FKconfig\bcversion}~\E^{-\Ham{FK}{\FKconfig}}~\prth{\bigotimes_{i=1}^n\bigotimes_{j=0}^{\ell_i-1}\WienerMeasureXYTBC[x_{i,j},x_{i,j+1}][\beta]*}\prth{\d\FKconfig}
	\end{align*}
	
	We can observe that for any $\MyFunction{g}{\WienerSpace[\beta]\times\WienerSpace[\beta]}{\RR^+}{}{}$,
	\[
	\int_{\window}\d y~g\prth{\sect_1,\sect_2}~\carac{\sect_1\subset\window}\WienerMeasureXYT[x,y][\beta]{\d\sect_1}
	~\carac{\sect_2\subset\window}\WienerMeasureXYT[y,z][\beta]{\d\sect_2}
	=g\prth{\restrict{\sect}{\interff{0;\beta}},\restrict{\sect}{\interff{\beta;2\beta}}}~\carac{\sect\subset\window}\WienerMeasureXYT[x,z][2\beta]{\d\sect}.
	\]
	
	Therefore, by integrating over $\window$, we can assemble the respective Brownian bridges into Brownian loops, along the cycle structure of $\PermutationSymbol$
	\begin{align*}
		&\int_{\window}\bigotimes_{i=1}^n\d x_i^{\otimes\ell_i}\int f\prth{\FKconfig\bcversion}~\E^{-\Ham{FK}{\FKconfig}}~\prth{\bigotimes_{i=1}^n\bigotimes_{j=0}^{\ell_i-1}\WienerMeasureXYTBC[x_{i,j},x_{i,j+1}][\beta]*}\prth{\d\FKconfig}
		\\
		&=\int_{\window}\d x^{\otimes n}\int \prth{f\circ\ViewChange{FK}{rl}*}\prth{\rlconfig}~\E^{-\Ham{rl}{\rlconfig}}~\prth{\bigotimes_{i=1}^n\WienerMeasureXYTBC[x_{i},x_{i}][\beta\ell_i]*}\prth{\d\rlconfig}.
	\end{align*}
	
	For any given sequence $\cyc\in\NN^{\NN}$ such that $\sum_{j\geqslant1}j\cyc_j=N$, there are $N!\cdot\prod_{j\geqslant1}\frac{1}{\cyc_j!j^{\cyc_j}}$ permutations $\PermutationSymbol\in\PermutationSpace{N}$ which have exactly $\delta_j$ cycles of length $j$, for any $j\geqslant1$.
	
	Furthermore, for any $\MyFunction{g}{\NN^{\NN}}{\RR^+}{}{}$,
	\[
	\sum_{(\delta_j)_{j\geqslant1}}\prth{\prod_{j\geqslant1}\frac{1}{\delta_j!}}
	g\prth{\delta}
	=\sum_{n\geqslant0}\frac{1}{n!}\sum_{j_1\geqslant1}\dots\sum_{j_n\geqslant1}g\prth{\MyFunction{}{\NN}{\NN}{q}{\#\MathSet{i\in\intgff{1;n}}[j_i=q]}}.
	\]
	where $\sum_{(\delta_j)_{j\geqslant 1}}$ is a summation over sequences of integers with a finite number of non-zero entries.
	
	We conclude
	\begin{align*}
		&\sum_{N=0}^{+\infty}\frac{1}{N!}\int_{\window}\d x^{\otimes N}\sum_{\PermutationSymbol\in\PermutationSpace{N}}\int f\prth{\FKconfig\bcversion}~\E^{-\Ham{FK}{\FKconfig}}~\prth{\bigotimes_{i=1}^N\WienerMeasureXYTBC[x_i,x_{\PermutationSymbol(i)}][\beta]*}\prth{\d\FKconfig}
		\\
		&=\sum_{n\geqslant0}\frac{1}{n!}\sum_{j_1\geqslant1}\frac{1}{j_1}\dots\sum_{j_n\geqslant1}\frac{1}{j_n}\int_{\window}\d x^{\otimes n}
		\int\prth{f\circ\ViewChange{FK}{rl}*}\prth{\rlconfig}~\E^{-\Ham{rl}{\rlconfig}}~\prth{\bigotimes_{i=1}^n\WienerMeasureXYTBC[x_{i},x_{i}][\beta j_i]*}\prth{\d\rlconfig}
		\\
		&=\sum_{n\geqslant0}\frac{1}{n!}\prth{f\circ\ViewChange{FK}{rl}*}\prth{\rlconfig}~\E^{-\Ham{rl}{\rlconfig}}~\prth{\bigotimes_{i=1}^n\int_{\window}\d x_i\sum_{j_i\geqslant 1}\frac{\E^{-\beta j_i}}{j_i}\WienerMeasureXYTBC[x_{i},x_{i}][\beta j_i]*}\prth{\d\rlconfig}
		\\
		&=\E^{\WienerMeasureRootBC{\WienerSpaceLoop{rl}}}\int\prth{f\circ\ViewChange{FK}{rl}*}\prth{\rlconfig}~\E^{-\Ham{rl}{\rlconfig}}~\Poisson{rl}{\d\rlconfig}
	\end{align*}
	because the measure $\WienerMeasureRootBC*$ has finite mass.
\end{Demo}

As mentionned previously, \Cref{Prop_proof_equiv_FK_rl} proved neither (FK) nor (rl) models to be well defined. This is what \Cref{lem_Z} is for.

\begin{Lem}\label{lem_Z}
	For all $L>0$, $\beta>0$ and $\mu\in\RR$,
	\[
	\Zgrandcanonic\defequal\int \E^{\beta\mu\#\finiteconfig}~\PoissonLeb{\d\finiteconfig}\sum_{\PermutationSymbol\in\PermutationSpace{\finiteconfig}}\int\E^{-\Ham{FK}{\FKconfig}}~
	\prth{\bigotimes_{x\in\finiteconfig}\WienerMeasureXYTBC[x,\PermutationSymbol(x)][\beta]*}(\d\FKconfig)\in\interoo{0;+\infty}.
	\]
\end{Lem}

\begin{Demo}[\Cref{lem_Z}]
	According to \Cref{Prop_proof_equiv_FK_rl},
	\[
	\Zgrandcanonic=\E^{-L^d+\WienerMeasureRootBC{\WienerSpaceLoop{rl}}}
	\int\exp\prth{\beta\mu\sum_{\loopath\in\rlconfig}\length{\loopath}-\Ham{rl}{\rlconfig}}~\Poisson{rl}{\d\rlconfig}.
	\]
	We deduce immediately from \Cref{hyp_superstable,def_interaction} that for any $\finiteconfig\in\ConfFinite$,
	\[
	\Interaction{\finiteconfig}\geqslant -A\#\finiteconfig+B\sum_{\underset{\prth{z+\case}\cap\window\neq\emptyset}{z\in r\ZZ^d}}\CardCase{z}[\finiteconfig]^2.
	\]
	In particular,
	\[
	\mu\#\finiteconfig-\Interaction{\finiteconfig}
	\leqslant
	\sum_{\underset{\prth{z+\case}\cap\window\neq\emptyset}{z\in r\ZZ^d}}
	\prth{A+\mu}\CardCase{z}[\finiteconfig]
	-B\cdot\CardCase{z}[\finiteconfig]^2.
	\]
	There are at most $\prth{\frac{L}{r}+1}^d$ such $z\in r\ZZ^d$ so
	\begin{equation}\label{eq_U_bounded}
		\mu\#\finiteconfig-\Interaction{\finiteconfig}
		\leqslant\prth{\frac{L}{r}+1}^d\frac{\prth{A+\mu}^2}{4B}.
	\end{equation}
	Let $\rlconfig\in\ConfSpace{rl}$. By applying inequality (\ref{eq_U_bounded}) to sets $\MathSet{\loopath(\beta j+s),~0\leqslant j<\length{\loopath},~\loopath\in\rlconfig}$ for any $s\in\interff{0;\beta}$ and integrating it, we conclude
	\begin{equation}\label{eq_ham_bounded}
		\beta\mu\sum_{\loopath\in\rlconfig}\length{\loopath}-\Ham{rl}{\rlconfig}\leqslant\beta\prth{\frac{L}{r}+1}^d\frac{\prth{A+\mu}^2}{4B}.
	\end{equation}
	Since the measure $\WienerMeasureRootBC*$ has finite mass,
	\[
	\Zgrandcanonic\leqslant
	\exp\prth{-L^d+\WienerMeasureRootBC{\WienerSpaceLoop{rl}}+\beta\prth{\frac{L}{r}+1}^d\frac{\prth{A+\mu}^2}{4B}}
	<+\infty.
	\]
	Since the interaction is non-degenerate,
	\begin{equation}\label{eq_Z_lower_bound}
		\Zgrandcanonic\geqslant\exp\prth{-L^d-\beta\InteractionFree{\emptyset}}>0.
	\end{equation}
\end{Demo}

\begin{Rq}\label{rq_ham}
	If we apply inequality (\ref{eq_U_bounded}) for $\mu=0$, it is clear the interaction $\Interaction*$ is bounded from below by a finite quantity. The integral from the formula of $\Ham{FK}*$ is then well defined.
\end{Rq}

\begin{Rq}\label{rq_demo}
	Together, \Cref{Prop_proof_equiv_FK_rl,lem_Z} are enough to prove \Cref{lemdef_FK,lemdef_rl,Prop_equiv_FK_rl}.
\end{Rq}

Now we quickly justify the time-shift invariance property of the (rl) model.

\begin{Demo}[\Cref{Prop_time-shift}]\label{demo_prop_time-shift}
	First, it is clear that the measure $\carac{\loopath\subset\window}\WienerMeasureRootBC[dir]{\d\loopath}$ is time-shift invariant:
	
	for any measurable $\MyFunction{f}{\WienerSpaceLoop{rl}}{\RR^+}{}{}$,
	\[
	\forall s\in\RR,~\int f\prth{\loopath\prth{\cdot+s}}~\carac{\loopath\subset\window}\WienerMeasureRootBC[dir]{\d\loopath}=\int f\prth{\loopath}~\carac{\loopath\subset\window}\WienerMeasureRootBC[dir]{\d\loopath}.
	\]
	
	We deduce immediately that the process $\carac{\rlconfig\subset\window}\Poisson{ps}[dir]{\d\rlconfig}$ is time-shift invariant too:
	
	for any measurable $\MyFunction{f}{\ConfSpace{rl}[dir]}{\RR^+}{}{}$,
	\[
	\forall s\in\RR,~\int \prth{f\circ\TimeTranslation*}(\rlconfig)~\carac{\rlconfig\subset\window}\Poisson{rl}[dir]{\d\rlconfig}
	=\int f\prth{\rlconfig}~\carac{\rlconfig\subset\window}\Poisson{rl}[dir]{\d\rlconfig}.
	\]
	
	Since the Hamiltonian $\Ham{rl}*$ is also time-shift invariant:
	\[
	\forall s\in\RR,~\Ham{rl}*\circ\TimeTranslation*=\Ham{rl}*
	\]
	we conclude on the probability measure $\Model{rl}*$.
\end{Demo}

\begin{Rq}\label{rq_demo_mp}
	Proving the (mp) model to be both well defined (\Cref{lemdef_mp}) and equivalent to the (FK) model (\Cref{Prop_equiv_mp_FK}) can be done with only the proposition below.
\end{Rq}

\begin{Prop}\label{Prop_proof_equiv_mp_FK}
	For any measurable $\MyFunction{f}{\ConfSpace{FK}}{\RR^+}{}{}$,
	\begin{align*}
		&\int\PoissonLeb{\d\finiteconfig}\sum_{\PermutationSymbol\in\PermutationSpace{\finiteconfig}}\int f\prth{\FKconfig\bcversion}~\E^{-\Ham{FK}{\FKconfig}}~\prth{\bigotimes_{x\in\finiteconfig}\WienerMeasureXYTBC[x,\PermutationSymbol(x)][\beta]*}\prth{\d\FKconfig}
		\\
		&=\int\prth{f\circ\ViewChange{FK}{mp}*}(\mpconfig)~\E^{-\Ham{mp}{\mpconfig}}~\Poisson{mp}{\d\mpconfig}.
	\end{align*}
\end{Prop}

\begin{Demo}[\Cref{Prop_proof_equiv_mp_FK}]
	Let $\finiteconfig=\MathSet{x_1\dots x_N}\subset\window$ be a finite simple configuration.
	
	Sampling uniformly a permutation $\PermutationSymbol\in\PermutationSpace{\finiteconfig}$ is the same as sampling independently and uniformly the image of each $x\in\finiteconfig$ among all possible images, conditioned to the whole map being indeed a permutation over $\finiteconfig$. Therefore
	\begin{align*}
		&\sum_{\PermutationSymbol\in\PermutationSpace{\finiteconfig}}\int f\prth{\FKconfig\bcversion}~\E^{-\Ham{FK}{\FKconfig}}~\prth{\bigotimes_{i=1}^N\WienerMeasureXYTBC[x_i,\PermutationSymbol(x_{i})][\beta]*}\prth{\d\FKconfig}
		\\
		&=\sum_{y_1\in\finiteconfig}\dots\sum_{y_N\in\finiteconfig}\carac{(x_i\mapsto y_i)\in\PermutationSpace{\finiteconfig}}\int f\prth{\FKconfig\bcversion}~\E^{-\Ham{FK}{\FKconfig}}~\prth{\bigotimes_{i=1}^N\WienerMeasureXYTBC[x_i,y_{i}][\beta]*}\prth{\d\FKconfig}
	\end{align*}

	Then we restrict the choice of $y_i\in\finiteconfig$ along the constraint of a $p_i\in\ZZ^d$ mark:
	\begin{align*}
		&=\sum_{p_1\in\ZZ^d}\carac{\finiteconfig\bcversion\cap\prth{x_1+rp_1+\case}\neq\emptyset}\sum_{y_1\in\finiteconfig\bcversion\cap\prth{x_1+rp_1+\case}}\dots\sum_{p_N\in\ZZ^d}\carac{\finiteconfig\bcversion\cap\prth{x_N+rp_N+\case}\neq\emptyset}\sum_{y_N\in\finiteconfig\bcversion\cap\prth{x_N+rp_N+\case}}
		\\
		&\qquad
		\carac{\prth{x_i\mapsto{y_i}}\in\PermutationSpace{\finiteconfig}}
		\int f\prth{\FKconfig\bcversion}~\E^{-\Ham{FK}{\FKconfig}}
		~\prth{\bigotimes_{i=1}^N\WienerMeasureXYT[x_i,y_{i}][\beta]*}\prth{\d\FKconfig}.
	\end{align*}
	Next, we choose the $y_i\in\finiteconfig\cap(x_i+rp_i+\case)$ based on the value of a $u_i\in\interff{0;1}$ mark. Specifically, $y_i$ is the $\Ent{\CardCase{x_i+rp_i}[\finiteconfig]\cdot u_i}$'th element of $\finiteconfig\cap(x_i+rp_i+\case)$ in the lexicographic order.
	\begin{align*}
		&=\sum_{p_1\in\ZZ^d}\CardCase{x_1+rp_1}[\finiteconfig]\int_0^1\d u_1\dots\sum_{p_N\in\ZZ^d}\CardCase{x_N+rp_N}[\finiteconfig]\int_0^1\d u_N
		\\
		&\qquad
		\carac{\prth{x_i\mapsto\Ent{\CardCase{x_i+rp_i}[\finiteconfig]\cdot u_i}\textnormal{'th element of }\finiteconfig\cap\prth{x_i+rp_i+\case}}\in\PermutationSpace{\finiteconfig}}\int f\prth{\FKconfig\bcversion}~\E^{-\Ham{FK}{\FKconfig}}
		~\prth{\bigotimes_{i=1}^N\WienerMeasureXYT[x_i,\Permutationmp{x_i}][\beta]*}\prth{\d\FKconfig}.
	\end{align*}
	This procedure corresponds to the definition of the map $\Permutationmp{\cdot}$ so this equals
	\begin{align*}
		&=\sum_{p_1\in\ZZ^d}\CardCase{x_1+rp_1}[\finiteconfig]\int_0^1\d u_1\dots\sum_{p_N\in\ZZ^d}\CardCase{x_N+rp_N}[\finiteconfig]\int_0^1\d u_N
		\\
		&\qquad
		\carac{\Permutationmp{\cdot}\in\PermutationSpace{\finiteconfig}}\int f\prth{\FKconfig\bcversion}~\E^{-\Ham{FK}{\FKconfig}}
		~\prth{\bigotimes_{i=1}^N\WienerMeasureXYT[x_i,\Permutationmp{x_i}][\beta]*}\prth{\d\FKconfig}.
	\end{align*}
	Finally, all the additional factors are assembled into the (mp) Hamiltonian
	\begin{align*}
		=\sum_{p_1\in\ZZ^d}\pMeasure{p_1}\int_0^1\d u_1\int\WienerMeasureNor{\d\brown_1}\dots\sum_{p_N\in\ZZ^d}\pMeasure{p_N}\int_0^1\d u_N
		\int\WienerMeasureNor{\d\brown_N}~\prth{f\circ\ViewChange{FK}{mp}*}\!\prth{\mpconfig}~\E^{-\Ham{mp}{\mpconfig}}
	\end{align*}
	where $\mpconfig=\MathSet{(x_i,p_i,u_i,\brown_i),~1\leqslant i\leqslant N}$.
\end{Demo}

\section{Proofs}
\subsection{Entropic Method}\label{sec_entropic_method}

To the best of our knowledge, there is no reference in the literature fully exposing the entropic method as we are using it here. We state here the plan we follow to establish the thermodynamic limits of the (mp) and (rl) models and quickly justify its validity. None of the results we state in this section are new and  we refer to the famous works of Georgii \cite{Geo11}, Zessin \cite{GZ93} and the mini-course \cite{Der19} for more details. 
\\

Let $\MM$ be a Polish space, called the mark space. We denote $\XX\defequal\RR^d\times\MM$. We define the configuration space
\[
\Conf(\XX)\defequal\MathSet{\mpconfig=\prth{x,y_x}_{x\in\finiteconfig}}[\finiteconfig\subset\RR^d\text{ is locally finite}]
\]
which we equip with the smallest $\sigma$-algebra $\mathcal{C}(\XX)$ making measurable the maps $\MyFunction{}{}{}{\mpconfig}{\#\prth{\mpconfig\cap E}}$ for all bounded Borel set $E\subset\XX$.

Let $\theta$ be a \emph{finite} measure over $\MM$. We denote by $\Leb$ the $d$ dimensional Lebesgue measure and by $\Pi^{\Leb\otimes\theta}_L$ the Poisson point process over $\window\times\MM$ with intensity measure $\Leb\otimes\theta$.

\begin{Def}\label{def_entropy}
Let $P$ and $Q$ be probability measures over $\Conf(\XX)$. We define the relative entropy of $P$ over $Q$ by\footnote{
	The integral defining the relative entropy is $\int f\log(f)\d Q$ where $f=\frac{\d P}{\d Q}$. Since the function $\MyFunction{}{}{}{x}{x\log(x)}$ is bounded from below, the integral of $f\log(f)$ along a probability measure is always well defined, although it may be $+\infty$.\label{rq_relative_entropy}
}
\begin{equation*}
\Entropy{P}{Q}\defequal
\begin{cases}
\int\log\prth{\frac{\d P}{\d Q}}\d P & \text{ if }P\ll Q
\\
+\infty & \text{ otherwise.}
\end{cases}
\end{equation*}

We also define the specific entropy of $P$ with reference measure $\theta$ by
\[
\SpecificEntropy[\theta]{P}=\limsup_{L\to+\infty}\frac{1}{L^d}\Entropy{P\big|_{\window}}{\Pi^{\Leb\otimes\theta}_L}
\]
where $P\big|_{\window}$ is the restriction of $P$ to $\window$ defined by
\[
\int f\d P\big|_{\window}\defequal\int f\prth{\MathSet{\prth{x,y}\in\mpconfig}[x\in\window]}~P\prth{\d\mpconfig}.
\]
\end{Def}

\begin{Def}\label{def_local_tame_general}
Let $\MyFunction{\psi}{\MM}{\RR^+}{}{}$ be such that
\[
\forall\lambda>0,~\int\E^{\lambda\psi\prth{y}}\theta\prth{\d y}<+\infty.
\]
We call a function $\MyFunction{f}{\Conf(\XX)}{\RR}{}{}$ local and tame if there exists $a>0$ and a compact $\Delta\subset\RR^d$ such that
\[
\forall\mpconfig\in\Conf(\XX),~f(\mpconfig)=f\prth{\MathSet{\prth{x,y}\in\mpconfig}[x\in\Delta]}
\]
and
\[
\forall\mpconfig\in\Conf(\XX),~a\abs{f(\mpconfig)}\leqslant1+\sum_{(x,y)\in\mpconfig,~x\in\Delta}\psi(y).
\]
\end{Def}

\begin{Rq}
It is clear from the previous definition that the faster $\psi$ grows, the bigger the class of tame functions turns out to be. But we are restricted by the constraint of $\psi$ having all its exponential moments finite along the reference measure $\theta$. Choosing $\psi=1$ is always possible and leads to a less general notion of tameness, namely
\[
a\abs{f(\mpconfig)}\leqslant1+\#\MathSet{\prth{x,y}\in\mpconfig}[x\in\Delta].
\]
\end{Rq}

Let $\prth{P_L}_{L\geqslant1}$ be a family of probabilities over the spaces $\MathSet{\mpconfig\in\Conf(\XX)}[\SpatialComponent*\subset\window]$ respectively, such that
\[
\sup_{L\geqslant1}\Entropy{P_L}{\Pi^{\Leb\otimes\theta}_L}<+\infty
\]
We define for all $L\geqslant1$ the associated empirical field $\widetilde{P}_L$ by
\[
\int f\d\widetilde{P}_L\defequal\frac{1}{L^d}\int_{\window}\d v\int f\prth{\mpconfig+v}~P_L\prth{\d\mpconfig}
\]
for any measurable $\MyFunction{f}{\Conf(\XX)}{\RR^+}{}{}$.

We also define the associated stationary field $\overline{P}_L$ by
\[
\int f\d\overline{P}_L\defequal\frac{1}{L^d}\int_{\window}\d v\int f\prth{\bigcup_{k\in\ZZ^d}\mpconfig_k+Lk+v}~\bigotimes_{k\in\ZZ^d}P_L\prth{\d\mpconfig_k}
\]
for any measurable $\MyFunction{f}{\Conf(\XX)}{\RR^+}{}{}$.

The next two results are used extensively throughout the proofs of the following pages.

\begin{Lem}\label{lem_entropic_method}
For any $V,\varepsilon>0$, there exists some $m\geqslant 1$ such that for any $L\geqslant1$ and compact $\Delta$ of volume $V$,
\[
\int\Thresh[m]{\sum_{(x,y)\in\mpconfig,~x\in\Delta}\psi\prth{y}}~\widetilde{P}_L(\d\mpconfig)\leqslant\varepsilon
\]
where we define the threshold function
\[
\Thresh[m]{x}\defequal
\begin{cases}
0 & \text{ if }x\leqslant m\\
x & \text{ if }x>m.
\end{cases}
\]
\end{Lem}

\begin{Demo}[\Cref{lem_entropic_method}]
First let us justify inequality
\[
\int\Thresh[m]{\sum_{(x,y)\in\mpconfig,~x\in\Delta}\psi\prth{y}}~\overline{P}_L(\d\mpconfig)\leqslant\varepsilon.
\]
We are merely restating Lemma 5.2 from \autocite{GZ93} with one subtle difference: the compact $\Delta$ can depend on $L$, as long as its volume remains constant. This slight generalization is not exclusive to our model and could be stated for any point process of marked points. Indeed the proof of Lemma 5.2 from \autocite{GZ93} never involves the shape of $\Delta$ and exclusively uses its volume.

From there, it is clear
\begin{align*}
\int\Thresh[m]{\sum_{(x,y)\in\mpconfig,~x\in\Delta}\psi\prth{y}}~\widetilde{P}_L(\d\mpconfig)
\leqslant
\int\Thresh[m]{\sum_{(x,y)\in\mpconfig,~x\in\Delta}\psi\prth{y}}~\overline{P}_L(\d\mpconfig)\leqslant\varepsilon.
\end{align*}
\end{Demo}

\begin{Th}\label{th_limit_thermo}
The family $\prth{\widetilde{P}_L}_{L\geqslant1}$ is sequentially compact. More precisely, there exists a sequence $\prth{L_n}_{n\in\NN}$ and a probability distribution $P_{\infty}$ over $\Conf(\XX)$ such that for any local tame $\MyFunction{f}{\Conf(\XX)}{\RR}{}{}$,
\[
\lim_{n\to+\infty}\int f\d\widetilde{P}_{L_n}=\int f\d P_{\infty}.
\]
Furthermore, for any compact $\Delta\subset\RR^d$,
\[
\int\sum_{(x,y)\in\mpconfig,~x\in\Delta}\psi\prth{y}~P_{\infty}\prth{\d\mpconfig}<+\infty.
\]
\end{Th}

\begin{Demo}[\Cref{th_limit_thermo}]
According to Proposition 15.52 (p. 330) from \autocite{Geo11},
\[
\SpecificEntropy[\theta]{\overline{P}_L}=\frac{1}{L^d}\Entropy{P_L}{\Pi^{\Leb\otimes\theta}_L}.
\]
The setting in \autocite{Geo11} is discrete, but the result is still valid in a continuous setting. We deduce immediately
\[
\sup_{L\geqslant1}\SpecificEntropy[\theta]{\overline{P}_L}<+\infty.
\]
According to Proposition 2.6 from \autocite{GZ93}, this is enough to prove sequential compactness of the family $\prth{\overline{P}_L}_{L\geqslant1}$, which we abusively write as
\[
\lim_{L\to+\infty}\int f\d\overline{P}_L=\int f\d P_{\infty}
\]
for all local tame functions $f$. Furthermore, according to Lemma 5.2 from \autocite{GZ93} again,
\[
\int\sum_{(x,y)\in\mpconfig,~x\in\Delta}\psi(y)~P_{\infty}\prth{\d\mpconfig}<+\infty.
\]

According to sections 2.1 and 2.2 from \autocite{Der19}, the empirical field has the same limit
\[
\lim_{L\to+\infty}\int f\d\widetilde{P}_L=\int f\d P_{\infty}
\]
for all local \emph{bounded} functions $f$.

Thanks to \Cref{lem_entropic_method}, we can extend this limit to any local \emph{tame} function.
\end{Demo}

\subsection{Entropic Bounds}\label{sec_entropy}

In this section, we prove the thermodynamic limits of \Cref{th_limit_rl,th_limit_mp}. We begin with the (rl) model because it is both simpler than and necessary for proving the (mp) limit. This is done by uniformly bounding the relative entropy of the finite volume probabilities.

\begin{Demo}[\Cref{th_limit_rl}]\label{demo_th_limit_rl}
The goal of this proof is to establish
\[
\sup_{L\geqslant1}\frac{1}{L^d}\Entropy{\Model{rl}*}{\Poisson{rl}*}<+\infty.
\]
According to inequations (\ref{eq_ham_bounded}) and (\ref{eq_Z_lower_bound}) from the proof of \Cref{lem_Z},
\[
\Entropy{\Model{rl}*}{\Poisson{rl}*}\leqslant\WienerMeasureRootBC{\WienerSpaceLoop{rl}}+\beta\InteractionFree{\emptyset}+\beta\prth{\frac{L}{r}+1}^d\frac{\prth{A+\mu}^2}{4B}.
\]
Furthermore,
\begin{align*}
\WienerMeasureRootBC[Dir]{\WienerSpaceLoop{rl}}
= L^d\frac{1}{(2\pi\beta)^{d/2}}\zeta\prth{\frac{d}{2}+1}.
\end{align*}
Therefore
\[
\sup_{L\geqslant1}\frac{1}{L^d}\Entropy{\Model{rl}*}{\Poisson{rl}*}
\leqslant \frac{1}{(2\pi\beta)^{d/2}}\zeta\prth{\frac{d}{2}+1}+\beta\prth{\frac{1}{r}+1}^d\frac{\prth{A+\mu}^2}{4B}<+\infty.
\]

According to \Cref{th_limit_thermo}, this is enough to prove the thermodynamic limit. The class of tame functions is defined by the choice $\psi=1$.
\end{Demo}

To prove the thermodynamic limit of the (mp) model, we first need the following technical result.

\begin{Lem}\label{lem_xlogx}
Let $L>0$. For any $\mpconfig\in\ConfSpacePerm{mp}$ such that $\SpatialComponent*\subset\window$,
\[
2^{-d}\sum_{(x,p,u,\brown)\in\mpconfig}\log\prth{\CardCase{x+rp}}\leqslant
\log\prth{2^d}\#\mpconfig+
\sum_{z\in r\ZZ^d}
\CardCase{z}\cdot\log\prth{\CardCase{z}}.
\]
\end{Lem}

\begin{Demo}[\Cref{lem_xlogx}]
Let $(x,p,u,\brown)\in\mpconfig$. If for any $y\in\RR^d$ we denote as $\ent{y}$ the closest point to $y$ in the lattice $r\ZZ^d+\frac{r}{2}(1\dots 1)$, then
\[
\CardCase{y}\leqslant\CardCase[bc][2r]{\ent{y}}
\]
Furthermore, for any $z\in r\ZZ^d+\frac{r}{2}(1\dots1)$, we know $z$ is the closest point of that lattice to some $\Permutationmp{x}$ at most $\CardCase[bc][r]{z}$ times. So $z$ is the closest point of the lattice to some ${x+rp}$ at most $\CardCase[bc][2r]{z}$ times. Thus
\begin{align*}
\sum_{(x,p,u,\brown)\in\mpconfig}\log\prth{\CardCase{x+rp}}
&\leqslant\sum_{(x,p,u,\brown)\in\mpconfig}
\log\prth{\CardCase[bc][2r]{\ent{x+rp}}}
\\
&\leqslant\sum_{z\in r\ZZ^d+\frac{r}{2}(1\dots1)}\CardCase[bc][2r]{z}\cdot\log\prth{\CardCase[bc][2r]{z}}.
\end{align*}
Since for any $z\in r\ZZ^d+\frac{r}{2}(1\dots1)$,
\[
\frac{1}{2^d}\CardCase[bc][2r]{z}=\frac{1}{2^d}
\sum_{\varepsilon\in\{\pm 1\}^d}\CardCase[bc][r]{z+\frac{r}{2}\varepsilon}
\]
by convexity of $\MyFunction{}{}{}{x}{x\log(x)}$ then
\begin{align*}
&\sum_{z\in r\ZZ^d+\frac{r}{2}(1\dots1)}\CardCase[bc][2r]{z}\cdot\log\prth{\CardCase[bc][2r]{z}}
\\
\leqslant&
\sum_{z\in r\ZZ^d+\frac{r}{2}(1\dots1)}
\prth{\log\prth{2^d}\CardCase[bc][2r]{z}+\sum_{\varepsilon\in\{\pm 1\}^d}\CardCase{z+\frac{r}{2}\varepsilon}\cdot\log\prth{\CardCase{z+\frac{r}{2}\varepsilon}}}.
\end{align*}

We take care of the first term
\begin{align*}
\sum_{z\in r\ZZ^d+\frac{r}{2}(1\dots1)}\CardCase[bc][2r]{z}
\leqslant
\sum_{z\in r\ZZ^d+\frac{r}{2}(1\dots1)}~\sum_{\varepsilon\in\{\pm 1\}^d}\CardCase{z+\frac{r}{2}\varepsilon}.
\end{align*}
As illustrated in \Cref{fig_xlogx}, for each $z\in r\ZZ^d$, the term $\CardCase{z}$ appears in the summation at most $2^d$ times.

\begin{figure}[h]
\centering
\begin{tikzpicture}[scale=1]
\draw[<->] (-3.1,-0.5) -- (-3.1,0.5);
\draw (-3.3,0) node {$r$};
\draw (5,1) node {%
\textcolor{red}{$\bullet$} $\in r\ZZ^d+\frac{r}{2}(1\dots1)$
};
\draw (6.85,0.5) node {%
{\tiny\textcolor{red}{$=\!\!=\!\!=$}} limits of a cell of size $2r$ centered in a \textcolor{red}{$\bullet$}
};
\draw (4.185,-0.5) node {\textcolor{blue}{$+$} $\in r\ZZ^d$};
\draw (6.775,-1) node {{\tiny\textcolor{blue}{$-~-$}} limits of a cell of size $r$ centered in a \textcolor{blue}{$+$}};
\clip (-2.9,-2.9) rectangle (2.9,2.9);
\draw[color=red,double] (-4,4) grid[step=1] (4,-4);
\draw[color=blue,dashed] (-4,4) grid (4,-4);
\foreach \kx in {-3,...,3}{
	\foreach \ky in {-4,...,4}{
		\draw[color=red] (\kx,\ky) node {$\bullet$};
		\draw[color=blue] (\kx+0.5,\ky+0.5) node {$+$};
};};
\end{tikzpicture}
\caption{Counting the $\CardCase{z},~z\in r\ZZ^d$}
\label{fig_xlogx}
\end{figure}

Thus
\begin{align*}
\sum_{z\in r\ZZ^d+\frac{r}{2}(1\dots1)}\CardCase[bc][2r]{z}
&\leqslant
2^d\sum_{z\in r\ZZ^d}\CardCase{z}
\leqslant
2^d\#\mpconfig.
\end{align*}

We control similarly the second term
\begin{align*}
&\sum_{z\in r\ZZ^d+\frac{r}{2}(1\dots1)}~\sum_{\varepsilon\in\{\pm 1\}^d}\CardCase{z+\frac{r}{2}\varepsilon}\cdot\log\prth{\CardCase{z+\frac{r}{2}\varepsilon}}
\\
\leqslant&
~2^d\sum_{z\in r\ZZ^d}\CardCase{z}\cdot\log\prth{\CardCase{z}}.
\end{align*}
\end{Demo}

As previously, proving the thermodynamic limit is done by uniformly bounding the relative entropy.

\begin{Prop}\label{Prop_entropy}
Let $\kappa\in\interoo{0;1}$. Assume
\[
\pMeasure{p}=\prth{\frac{r^2\prth{1-\kappa}}{2\pi\beta}}^{d/2}\int_{p+\case[1]}\exp\prth{-\frac{r^2}{2\beta}\prth{1-\kappa}\norm{x}^{2}}\d x.
\]
Then, there exists $\lambda>0$ such that
\[
\sup_{L\geqslant1}\frac{1}{L^d}\Entropy{\Model{mp}*}{\PoissonRef*}<+\infty
\]
where $\PoissonRef*$ is the Poisson process of intensity measure $\lambda\mpMeasure$ with $\mpMeasure\defequal\Leb[d]\otimes\pMeasure*\otimes\Leb[1]_{\interff{0;1}}\otimes\WienerMeasureNor*$.
\end{Prop}

\begin{Demo}[\Cref{Prop_entropy}]
By definition, we know that
\begin{align*}
\Model{mp}{\d\mpconfig}=\frac{1}{\Zgrandcanonic}\exp\prth{\beta\mu\#\mpconfig-\Ham{mp}{\mpconfig}}\PoissonRef[]{\d\mpconfig}
\end{align*}
Multiplying by $\lambda$ the mass of the intensity measure in $\PoissonRef[]{\d\mpconfig}$ is equivalent to adding an extra $\lambda^{\#\gamma}$ in the density, up to some constant factor:
\[
\PoissonRef{\d\mpconfig}=\E^{(-\lambda+1)L^d}\lambda^{\#\gamma}~\PoissonRef[]{\d\mpconfig}.
\]
Then
\[
\Model{mp}{\d\mpconfig}=\frac{1}{\Zgrandcanonic}\exp\prth{\prth{\lambda-1}L^d+\prth{\beta\mu-\log\prth{\lambda}}\#\mpconfig-\Ham{mp}{\mpconfig}}\PoissonRef{\d\mpconfig}.
\]

We uniformly bound the relative entropy by dealing separately with each non-trivial term of the density
\begin{description}
\item[Step 1] $-\log\prth{\Zgrandcanonic}$
\item[Step 2] $-\sum_{(x,p,u,\brown)\in\mpconfig}\prth{\frac{1}{2\beta}\norm{\Permutationmp{x}-x}^2+\log\prth{\pMeasure{p}}}$
\item[Step 3] $\text{interaction term}=\int_0^{\beta}\d s~-\Interaction{\MathSet{x+\frac{s}{\beta}\brac{\Permutationmp{x}-x}+\sqrt{\beta}~\brown\prth{\frac{s}{\beta}},~(x,p,u,\brown)\in\FKconfig}}$
\end{description}

We proceed in this order and then conclude.
\begin{description}
\item[Step 1] According to inequation (\ref{eq_Z_lower_bound}) from the proof of \Cref{lem_Z},
\[
-\log\prth{\Zgrandcanonic}\leqslant L^d+\beta\InteractionFree{\emptyset}.
\]

\item[Step 2] For any $\mpconfig\in\ConfSpacePerm{mp}$ and $(x,p,u,\brown)\in\mpconfig$,
\[
\norm{\Permutationmp{x}-x}\geqslant r\prth{\norm{p}-\sqrt{d}}.
\]
Given the admitted formula for $\pMeasure*$,
\[
\log\prth{\pMeasure{p}}\geqslant\frac{d}{2}\log\prth{\frac{r^2\prth{1-\kappa}}{2\pi\beta}}-\frac{r^2}{2\beta}\prth{1-\kappa}\prth{\norm{p}+\sqrt{d}}^2.
\]
Thus there exists $C_{\kappa}\in\RR$ such that for any $(x,p,u,\brown)\in\mpconfig$,
\[
\frac{1}{2\beta}\norm{\Permutationmp{x}-x}^2+\log\prth{\pMeasure{p}}\geqslant C_{\kappa}.
\]

\item[Step 3] By definition,
\begin{align*}
&\restrict{\Entropy{\Model{rl}*}{\Poisson{rl}*}}{\mu=0}
+\log\prth{\restrict{\Zgrandcanonic}{\mu=0}}=\int-\Ham{rl}*\d\Model{rl}*.
\end{align*}
By the equivalence of models (mp), (FK) and (rl) from \Cref{Prop_equiv_mp_FK,Prop_equiv_FK_rl}, this equals
\begin{align*}
=&\int\Model{mp}{\d\FKconfig}\int_0^{\beta}\d s~-\Interaction{\MathSet{x+\frac{s}{\beta}\brac{\Permutationmp{x}-x}+\sqrt{\beta}~\brown\prth{\frac{s}{\beta}},~(x,p,u,\brown)\in\FKconfig}}.
\end{align*}
Since the relative (rl) entropy is well defined and finite for $\mu=0$ (see footnote \cpageref{rq_relative_entropy} and the proof of \Cref{th_limit_rl}), this proves the interaction term is integrable under $\Model{mp}*$.

We can then permute the two integral signs
\begin{align*}
&\int\Model{mp}{\d\FKconfig}\int_0^{\beta}\d s~\Interaction{\MathSet{x+\frac{s}{\beta}\brac{\Permutationmp{x}-x}+\sqrt{\beta}~\brown\prth{\frac{s}{\beta}},~(x,p,u,\brown)\in\FKconfig}}
\\
=&
\int_0^{\beta}\d s\int\Model{mp}{\d\FKconfig}~\Interaction{\MathSet{x+\frac{s}{\beta}\brac{\Permutationmp{x}-x}+\sqrt{\beta}~\brown\prth{\frac{s}{\beta}},~(x,p,u,\brown)\in\FKconfig}}
\end{align*}
By the equivalence of models (mp), (FK) and (rl) from \Cref{Prop_equiv_mp_FK,Prop_equiv_FK_rl}, we can write
\begin{align*}
	=&\int_0^{\beta}\d s\int\Model{FK}{\d\FKconfig}~
	\Interaction{\MathSet{\sect(s),~\sect\in\FKconfig}}
	\\
	=&\int_0^{\beta}\d s\int\Model{rl}{\d\rlconfig}~
	\Interaction{\MathSet{\loopath(\beta j+s),~0\leqslant j<\length{\loopath},~\loopath\in\rlconfig}}.
\end{align*}

By time-shift stationarity (see \Cref{Prop_time-shift}), this equals
\begin{align*}
	&\int_0^{\beta}\d s~\int\Model{rl}{\d\rlconfig}~
	\brac{\Interaction{\MathSet{\loopath(\beta j),~0\leqslant j<\length{\loopath},~\loopath\in\cdot}}\circ\TimeTranslation[s]*}(\rlconfig)
	\\
	=&\beta\int\Model{rl}{\d\rlconfig}~
	\Interaction{\MathSet{\loopath(\beta j),~0\leqslant j<\length{\loopath},~\loopath\in\rlconfig}}.
\end{align*}
By equivalence of models again,
\begin{align*}
\int\Model{rl}{\d\rlconfig}~\Interaction{\MathSet{\loopath(\beta j),~0\leqslant j<\length{\loopath},~\loopath\in\rlconfig}}
=\int\Model{mp}{\d\mpconfig}~\Interaction{\SpatialComponent*}.
\end{align*}
Therefore, the interaction term is integrable under $\Model{mp}*$ and equals
\[
\int\d\Model{mp}*~\textnormal{interaction term}
=-\int\Model{mp}{\d\mpconfig}~\Interaction{\SpatialComponent*}.
\]
\end{description}

Now we make use of those results to conclude.

We have seen just above that the interaction term is integrable under $\Model{mp}*$. Thus
\begin{align*}
	&\Entropy{\Model{mp}*}{\PoissonRef*}
	\\
	&=
	\int\Model{mp}{\d\mpconfig}~\prth{-\log\prth{\Zgrandcanonic}+(\lambda-1)L^d+\prth{\beta\mu-\log\prth{\lambda}}\#\mpconfig-\Ham{mp}{\mpconfig}-\textnormal{interaction term}(\mpconfig)}
	\\
	&\quad+\int\Model{mp}{\d\mpconfig}~\textnormal{interaction term}(\mpconfig)
\end{align*}
According to step 3, this equals
\begin{align*}
	=&-\log\prth{\Zgrandcanonic}+(\lambda-1)L^d+\int\Model{mp}{\d\mpconfig}\Bigg\{\prth{\beta\mu-\log\prth{\lambda}-\frac{d}{2}\log\prth{2\pi\beta}}\#\mpconfig
	\\
	&\qquad\qquad\qquad\qquad\qquad\qquad+\sum_{(x,p,u,\brown)\in\mpconfig}-\frac{1}{2\beta}\norm{\Permutationmp{x}-x}^2-\log\prth{\pMeasure{p}}+\log\prth{\CardCase{x+rp}}\Bigg\}
	\\
	&+\int\Model{mp}{\d\mpconfig}~-\Interaction{\SpatialComponent*}.
\end{align*}
Since the function $\MyFunction{}{}{}{\mpconfig}{-\Interaction{\SpatialComponent*}}$ is integrable under $\Model{mp}*$ (see step 3), we can sum the two integrals:
\begin{align*}
=&-\log\prth{\Zgrandcanonic}+(\lambda-1)L^d+\int\Model{mp}{\d\mpconfig}\Bigg\{\prth{\beta\mu-\log\prth{\lambda}-\frac{d}{2}\log\prth{2\pi\beta}}\#\mpconfig-\Interaction{\SpatialComponent*}
\\
&\qquad\qquad\qquad\qquad\qquad\qquad+\sum_{(x,p,u,\brown)\in\mpconfig}-\frac{1}{2\beta}\norm{\Permutationmp{x}-x}^2-\log\prth{\pMeasure{p}}+\log\prth{\CardCase{x+rp}}\Bigg\}.
\end{align*}
According to steps 1 and 2, this can be upper-bounded by
\begin{align*}
\leqslant&\beta\InteractionFree{\emptyset}+\lambda L^d
\\
&+\int\Model{mp}{\d\mpconfig}\prth{\prth{\beta\mu-\log\prth{\lambda}+\frac{d}{2}\log\prth{2\pi\beta}-C_{\kappa}}\#\mpconfig
-\Interaction{\SpatialComponent*}
-\sum_{(x,p,u,\brown)\in\mpconfig}\log\prth{\CardCase{x+rp}}}.
\end{align*}
By superstability of the interaction and according to \Cref{lem_xlogx}, we can upper-bound further by
\begin{align*}
\leqslant\beta\InteractionFree{\emptyset}+\lambda L^d
+\int\Model{mp}{\d\mpconfig}\Bigg\{&\prth{\beta\mu-\log\prth{\lambda}+\frac{d}{2}\log\prth{2\pi\beta}-C_{\kappa}+2^d\log\prth{2^d}+\beta A}\#\mpconfig
\\
&+\sum_{\underset{(z+\case)\cap\window\neq\emptyset}{z\in r\ZZ^d}}2^d\CardCase{z}\log\prth{\CardCase{z}}-\beta B\CardCase{z}^2\Bigg\}.
\end{align*}
There exists $C^{(0)}\in\RR$ such that $2^d x\log(x)-\beta Bx^2\leqslant C^{(0)}$ for any $x\geqslant 0$. We deduce
\begin{align*}
\leqslant&
\beta\InteractionFree{\emptyset}+\lambda L^d+C^{(0)}\prth{\frac{L}{r}+1}^d
\\
&+\int\Model{mp}{\d\mpconfig}\prth{\beta\mu-\log\prth{\lambda}+\frac{d}{2}\log\prth{2\pi\beta}-C_{\kappa}+2^d\log\prth{2^d}+\beta A}\#\mpconfig.
\end{align*}
For $\lambda>0$ large enough, the integrand is non-positive and we conclude
\[
\sup_{L\geqslant1}\frac{1}{L^d}\Entropy{\Model{mp}*}{\PoissonRef*}\leqslant\beta\InteractionFree{\emptyset}+\lambda+C^{(0)}\prth{\frac{1}{r}+1}^d<+\infty.
\]
\end{Demo}

We claimed in \Cref{th_limit_mp} the class of functions which are compatible with this thermodynamic limit includes functions $\MyFunction{f}{\ConfSpace{mp}}{\RR}{}{}$ such that
\[
a\abs{f(\mpconfig)}\leqslant1+\sum_{(x,p,u,\brown)\in\mpconfig\cap\Delta}\abs{\Sausage[\delta]{\MyFunction{}{\interff{0;\beta}}{\RR^d}{s}{\frac{s}{\beta}rp+\sqrt{\beta}~\brown\brac{\frac{s}{\beta}}}}}^{\alpha}
\]
for some $a,\delta>0$, $\alpha<2$ and compact $\Delta$ (see \Cref{def_local_tame_mp}).

If we refer to \Cref{def_local_tame_general} about the general entropic method, this corresponds to choosing
\[
\psi_{\alpha,\delta}\prth{p,u,\brown}=\abs{\Sausage[\delta]{\MyFunction{}{\interff{0;\beta}}{\RR^d}{s}{\frac{s}{\beta}rp+\sqrt{\beta}~\brown\brac{\frac{s}{\beta}}}}}^{\alpha}.
\]
So, to apply the entropic method, we need to prove that any $\psi_{\alpha,\delta}$ have all their exponential moments along the reference measure $\pMeasure*\otimes\Leb[1]_{\interff{0;1}}\otimes\WienerMeasureNor*$ where
\[
\pMeasure{p}=\prth{\frac{r^2\prth{1-\kappa}}{2\pi\beta}}^{d/2}\int_{p+\case[1]}\exp\prth{-\frac{r^2}{2\beta}\prth{1-\kappa}\norm{x}^{2}}\d x
\]
for some $\kappa>0$ small enough.

\begin{Lem}\label{lem_psi}
Let $\alpha\in\interfo{0;2}$ and $\delta>0$. There exists $\kappa\in\interoo{0;1}$ small enough such that for any $\lambda>0$,
\[
\int\exp\prth{\lambda\abs{\Sausage[\delta]{\MyFunction{}{\interff{0;\beta}}{\RR^d}{s}{\frac{s}{\beta}rp+\sqrt{\beta}~\brown\brac{\frac{s}{\beta}}}}}^{\alpha}}\pMeasure{\d p}\WienerMeasureNor{\d\brown}<+\infty
\]
with
\[
\pMeasure{p}=\prth{\frac{r^2\prth{1-\kappa}}{2\pi\beta}}^{d/2}\int_{p+\case[1]}\exp\prth{-\frac{r^2}{2\beta}\prth{1-\kappa}\norm{x}^{2}}\d x.
\]
\end{Lem}

\begin{Rq}
The importance of this lemma is crucial. It justifies our first intuition that loops become very large because of their length, not because each bridge inside them becomes large. This is made even more clear in \Cref{prop_equiint_FK,prop_equiint_cap_FK}.
\end{Rq}

\begin{Demo}[\Cref{lem_psi}]
For any $y\in\RR^d$, we denote as $\ent{y}$ the point of the lattice $\ZZ^d$ closest to $x$.

If a random vector $Y\in\RR^d$ has a Gaussian distribution given by
\[
G_{r^2(1-\kappa)}(\d y)=
\prth{\frac{r^2\prth{1-\kappa}}{2\pi\beta}}^{d/2}\exp\prth{-\frac{r^2}{2\beta}\prth{1-\kappa}\norm{y}^{2}}\d y
\]
then $\ent{Y}$'s distribution is $\pMeasure*$. Therefore
\begin{align*}
&\int\exp\prth{\lambda\abs{\Sausage[\delta]{\MyFunction{}{}{}{s}{\frac{s}{\beta}rp+\sqrt{\beta}~\brown\brac{\frac{s}{\beta}}}}}^{\alpha}}\pMeasure{\d p}\WienerMeasureNor{\d\brown}
\\
&\leqslant\int\exp\prth{\lambda\abs{\Sausage[\delta+r\sqrt{d}]{\MyFunction{}{}{}{s}{\frac{s}{\beta}ry+\sqrt{\beta}~\brown\brac{\frac{s}{\beta}}}}}^{\alpha}}G_{r^2(1-\kappa)}(\d y)\WienerMeasureNor{\d\brown}
\\
&\leqslant\int\exp\prth{\lambda\abs{\Sausage[\delta+r\sqrt{d}]{\MyFunction{}{}{}{s}{\frac{s}{\beta}y+\brown\brac{s}}}}^{\alpha}}G_{1-\kappa}(\d y)\WienerMeasureNor[\beta]{\d\brown}
\\
&\leqslant(1-\kappa)^{d/2}\int\E^{\frac{\kappa}{2\beta}\norm{y}^2}\exp\prth{\lambda\abs{\Sausage[\delta+r\sqrt{d}]{\MyFunction{}{}{}{s}{\frac{s}{\beta}y+\brown\brac{s}}}}^{\alpha}}G_{1}(\d y)\WienerMeasureNor[\beta]{\d\brown}
\end{align*}
If $Y$ is a random vector with Gaussian distribution of variance $\beta$ and $\brown$ is a Brownian bridge from $0$ to $0$ in time $\beta$, then the trajectory $\MyFunction{}{}{}{s}{\frac{s}{\beta}Y+\brown(s)}$ has the distribution of a Brownian motion, up to time $\beta$. So
\begin{align*}
&\int\E^{\frac{\kappa}{2\beta}\norm{y}^2}\exp\prth{\lambda\abs{\Sausage[\delta+r\sqrt{d}]{\MyFunction{}{}{}{s}{\frac{s}{\beta}y+\brown\brac{s}}}}^{\alpha}}G_{1}(\d y)\WienerMeasureNor[\beta]{\d\brown}
\\
\leqslant&\Esp[\brown]{\E^{\frac{\kappa}{2\beta}\norm{\brown(\beta)}^2}\exp\prth{\lambda\abs{\Sausage[\delta+r\sqrt{d}]{\brown}}^{\alpha}}}.
\end{align*}
where the expectancy is taken along the standard Brownian motion.

By \Cref{prop_saucisse}, the cylinder whose axis goes from $0$ to $\brown(\beta)$ with radius $\delta+r\sqrt{d}$ has smaller volume than the Wiener sausage. Then
\[
\norm{\brown(\beta)}\leqslant\frac{1}{c_{d-1}\prth{\delta+r\sqrt{d}}^{d-1}}\abs{\Sausage[\delta+r\sqrt{d}]{\brown}}
\]
where we denote as $c_{d-1}$ the volume of the $d-1$ dimensional ball of radius $1$.

Thus
\begin{align*}
&\Esp[\brown]{\E^{\frac{\kappa}{2\beta}\norm{\brown(\beta)}^2}\exp\prth{\lambda\abs{\Sausage[\delta+r\sqrt{d}]{\brown}}^{\alpha}}}
\\
\leqslant&\Esp[\brown]{\exp\prth{\frac{\kappa}{2\beta c_{d-1}^2\prth{\delta+r\sqrt{d}}^{2d-2}}\abs{\Sausage[\delta+r\sqrt{d}]{\brown}}^2+\lambda\abs{\Sausage[\delta+r\sqrt{d}]{\brown}}^{\alpha}}}.
\end{align*}

There exists $x_0>0$ such that for any $x\geqslant0$
\[
\frac{\kappa}{2\beta c_{d-1}^2\prth{\delta+r\sqrt{d}}^{2d-2}}x^2\leqslant\lambda x^{\alpha}\iff x\leqslant x_0.
\]
If we treat seperately the cases whether $\abs{\Sausage[\delta+r\sqrt{d}]{\brown}}$ is smaller or bigger than $x_0$, we can see it is sufficient to prove
\[
\Esp[\brown]{\exp\prth{\frac{\kappa}{\beta c_{d-1}^2\prth{\delta+r\sqrt{d}}^{2d-2}}\abs{\Sausage[\delta+r\sqrt{d}]{\brown}}^2}}<+\infty
\]
which is true for $\kappa$ small enough, according to \Cref{th_saucisse}.
\end{Demo}

\begin{Demo}[\Cref{th_limit_mp}]\label{demo_th_limit_mp}
In the assumptions of \Cref{Prop_entropy}, the Poisson point process $\PoissonRef[\lambda]*$ has the intensity measure
\[
\Leb[d]\otimes\prth{\lambda\nu\otimes\Leb[1]_{\interff{0;1}}\otimes\WienerMeasureNor*}.
\]
The mark measure has a finite mass $\lambda$. So according to \Cref{lem_entropic_method}, the uniform bound from \Cref{Prop_entropy} on the entropy is enough to prove the existence of the thermodynamic limit $\Model*{mp}*$. More precisely, this is enough to prove tightness of the family $\EmpiricalField{mp}*,~L\geqslant1$, hence the existence of a converging subsequence.

The class of tame functions is justified by \Cref{lem_psi}.
\end{Demo}

\subsection{Permutation in Infinite Volume}\label{sec_permutation}

In \Cref{sec_mp} we constructed the (mp) framework to encode the permutation of the (FK) model into the $p$, $u$ and $\brown$ marks. We just proved the (mp) model to have a thermodynamic limit, but does the encoding pass to the limit? Is it only possible to define an infinite volume permutation with those marks? Thankfully the answer is yes.

\begin{Prop}\label{Prop_confperm_mp}
$\Model*{mp}{\ConfSpacePerm{mp}*}=1$.
\end{Prop}

\begin{Demo}[\Cref{Prop_confperm_mp}]
If we are given a configuration $\mpconfig$, the map $\Permutationmp*{\cdot}$ is a well-defined bijection if and only if
\begin{enumerate}
\setcounter{enumi}{-1}
\item The marked point configuration $\mpconfig$ is simple.
\item $\forall(x,p,u,\brown)\in\mpconfig,~\CardCase*{x+rp}\geqslant 1$ so that the mark is not pointing to an empty region of space. The map $\Permutationmp*{\cdot}$ is then well defined.
\item $\forall(x,p,u,\brown)\in\mpconfig,~\sum_{(x',p',u',\brown')\in\mpconfig}\carac{\Permutationmp*{x'}=x}\leqslant1$ so that the map is injective.
\item The map $\Permutationmp*{\cdot}$ is surjective.
\end{enumerate}
We check that those four properties hold $\Model*{mp}*$ almost surely.

\begin{description}
\item[Step 0] Let us assume $\Model*{mp}{\mpconfig\text{ is not simple}}>0$. So there exists a compact $\Delta\subset\RR^d$ such that
\[
\Model*{mp}{\mpconfig\cap\Delta\text{ is not simple}}>0.
\]
This event is local so there exists $L>0$ such that
\[
\EmpiricalField{mp}{\mpconfig\cap\Delta\text{ is not simple}}=\frac{1}{L^d}\int_{\window[L]}\d v~\Model{mp}{\prth{\mpconfig+v}\cap\Delta\text{ is not simple}}>0
\]
which is wrong.

\item[Step 1]Let us denote as $\prth{x_0^{\mpconfig},p_0^{\mpconfig},u_0^{\mpconfig},\brown_0^{\mpconfig}}$ the closest point to $0$ in the configuration $\mpconfig$. We assume
\[
\Model*{mp}{\CardCase*{x_0^{\mpconfig}+rp_0^{\mpconfig}}=0}>0.
\]
By monotone convergence, there exists $D>0$ such that
\[
\Model*{mp}{\CardCase*{x_0^{\mpconfig}+rp_0^{\mpconfig}}=0\text{ and }\norm{x_0^{\mpconfig}}\leqslant D\text{ and }\norm{p_0^{\mpconfig}}\leqslant D}>0.
\]
This event is local so there exists $L>0$ such that
\[
\EmpiricalField{mp}{\CardCase*{x_0^{\mpconfig}+rp_0^{\mpconfig}}=0\text{ and }\norm{x_0^{\mpconfig}}\leqslant D\text{ and }\norm{p_0^{\mpconfig}}\leqslant D}>0.
\]
By definition of the empirical field,
\begin{align*}
&\EmpiricalField{mp}{\CardCase*{x_0^{\mpconfig}+rp_0^{\mpconfig}}=0\text{ and }\norm{x_0^{\mpconfig}}\leqslant D\text{ and }\norm{p_0^{\mpconfig}}\leqslant D}
\\
&=\frac{1}{L^d}\int_{\window[L]}\d v
~\Model{mp}{\CardCase*{x_{0}^{\mpconfig+v}+rp_{0}^{\mpconfig+v}}[\SpatialComponent*+v]=0\text{ and }\norm{x_{0}^{\mpconfig+v}}\leqslant D\text{ and }\norm{p_{0}^{\mpconfig+v}}\leqslant D}.
\end{align*}

So there exists $v\in\window$ such that
\[
\Model{mp}{\CardCase*{x_{0}^{\mpconfig+v}+rp_{0}^{\mpconfig+v}}[\SpatialComponent*+v]=0\text{ and }\norm{x_{0}^{\mpconfig+v}}\leqslant D\text{ and }\norm{p_{0}^{\mpconfig+v}}\leqslant D}>0
\]
so that
\[
\Model{mp}{\CardCase*{x_{0}^{\mpconfig+v}+rp_{0}^{\mpconfig+v}}[\SpatialComponent*+v]=0}>0.
\]
If we denote as $x_{-v}^{\mpconfig}$ the point closest to $-v$ in the configuration $\mpconfig$, it is clear that $x_0^{\mpconfig+v}=x_{-v}^{\mpconfig}+v$. Then
\[
\CardCase{x_0^{\mpconfig+v}+rp_0^{\mpconfig+v}}[\SpatialComponent*+v]=\CardCase{x_{-v}^{\mpconfig}+rp_{-v}^{\mpconfig}}.
\]
Therefore
\[
\Model{mp}{\CardCase{x_{-v}^{\mpconfig}+rp_{-v}^{\mpconfig}}=0}>0
\]
which is not true because $\Model{mp}{\ConfSpaceAuth}=1$.
\\

Let us change notations. If we denote by $(x_n,p_n,u_n,\brown_n)$ the $n$'th closest point to $0$ in the configuration $\mpconfig$, we have proven
\[
\Model*{mp}{\CardCase*{x_1+rp_1}\geqslant1}=1
\]
and with the same procedure as previously, we can prove this equality for the $n$'th closest point to $0$ in the configuration, for any $n\geqslant1$. Finally
\[
\Model*{mp}{\forall n\geqslant1,~\CardCase*{x_n+rp_n}\geqslant1}=1.
\]

\item[Step 2] According to the previous steps, the map $\Permutationmp{\cdot}$ is well defined $\Model*{mp}*$ almost surely.

We follow a similar proof path in step 2. We assume
\[
\Model*{mp}{\sum_{(x',p',u',\brown')\in\mpconfig}\carac{\Permutationmp*{x'}=x_0^{\mpconfig}}>1}>0.
\]
By monotone convergence, there exists $D>0$ such that
\[
\Model*{mp}{\sum_{\underset{\norm{x'}\leqslant D,~\norm{p'}\leqslant D}{(x',p',u',\brown')\in\mpconfig}}\carac{\Permutationmp*{x'}=x_0^{\mpconfig}}>1\text{ and }\norm{x_0^{\mpconfig}}\leqslant D}>0.
\]
Since this event is local, there exists $L>0$ such that
\[
\EmpiricalField{mp}{\sum_{\underset{\norm{x'}\leqslant D,~\norm{p'}\leqslant D}{(x',p',u',\brown')\in\mpconfig}}\carac{\Permutationmp*{x'}=x_0^{\mpconfig}}>1\text{ and }\norm{x_0^{\mpconfig}}\leqslant D}>0.
\]
By definition of the empirical field,
\begin{align*}
&\EmpiricalField{mp}{\sum_{\underset{\norm{x'}\leqslant D,~\norm{p'}\leqslant D}{(x',p',u',\brown')\in\mpconfig}}\carac{\Permutationmp*{x'}=x_0^{\mpconfig}}>1\text{ and }\norm{x_0^{\mpconfig}}\leqslant D}
\\
&=\frac{1}{L^d}\int_{\window}\d v~\Model{mp}{\sum_{\underset{\norm{x'}\leqslant D,~\norm{p'}\leqslant D}{(x',p',u',\brown')\in\mpconfig+v}}\carac{\Permutationmp*{x'}[\mpconfig+v]=x_0^{\mpconfig+v}}>1\text{ and }\norm{x_0^{\mpconfig+v}}\leqslant D}
\end{align*}
So there exists $v\in\window$ such that
\[
\Model{mp}{\sum_{\underset{\norm{x'}\leqslant D,~\norm{p'}\leqslant D}{(x',p',u',\brown')\in\mpconfig+v}}\carac{\Permutationmp*{x'}[\mpconfig+v]=x_0^{\mpconfig+v}}>1\text{ and }\norm{x_0^{\mpconfig+v}}\leqslant D}>0
\]
so that
\[
\Model{mp}{\sum_{(x',p',u',\brown')\in\mpconfig+v}\carac{\Permutationmp{x'}[\FKconfig+v]=x_{0}^{\mpconfig+v}}>1}>0.
\]
Like in step $1$, it is clear that $x_0^{\mpconfig+v}=x_{-v}^{\mpconfig}+v$. Then
\[
\Model{mp}{\sum_{(x',p',u',\brown')\in\mpconfig}\carac{\Permutationmp{x'}=x_{-v}^{\mpconfig}}>1}>0.
\]
which is not true because $\Permutationmp{\cdot}$ is $\Model{mp}*$ almost surely injective.

Once again, we conclude by generalizing the procedure for any $n$'th closest point to $0$ of the configuration.

\item[Step 3] We prove a seemingly weaker, yet sufficient result :
\[
\forall k\in\ZZ^d,~\sum_{(x,p,u,\brown)\in\mpconfig}\carac{\Permutationmp*{x}\in rk+\case}=\#\prth{\SpatialComponent*\cap\prth{rk+\case}}.
\]
By stationarity, it is enough to prove it for $k=0$.

By injectivity, we already know
\[
\sum_{(x,p,u,\brown)\in\mpconfig}\carac{\Permutationmp*{x}\in\case}\leqslant\#\prth{\SpatialComponent*\cap\case}
\]
To prove the equality more easily, we introduce a discretized version of the configuration
\[
\MyFunction{}{\mpconfig}{\ent{\mpconfig}}{(x,p,u,\brown)}{\brac{\ent{x},\ent{\Permutationmp*{x}}-\ent{x}}}
\]
where, for $x\in\RR^d$, the point $\ent{x}$ is the closest point of $r\ZZ^d$ to $x$.

Then
\begin{align*}
\int\Model*{mp}{\d\mpconfig}\sum_{(x,p,u,\brown)\in\mpconfig}\carac{\Permutationmp*{x}\in\case}
&=\int\Model*{mp}{\d\mpconfig}\sum_{(z,k)\in\ent{\mpconfig}}\carac{z+k=0}
\\
&=\sum_{j\in\ZZ^d}\int\Model*{mp}{\d\mpconfig}\sum_{(z,k)\in\ent{\mpconfig}}\carac{z=rj}\carac{k=-rj}.
\end{align*}
The probability measure $\Model*{mp}*$ is stationary so $\ent{\mpconfig}$'s distribution is $r\ZZ^d$-stationary. Thus
\begin{align*}
&=\sum_{j\in\ZZ^d}\int\Model*{mp}{\d\mpconfig}\sum_{(z,k)\in\ent{\mpconfig}}\carac{z=0}\carac{k=-rj}
\\
&=\int\Model*{mp}{\d\mpconfig}\sum_{(z,k)\in\ent{\mpconfig}}\carac{z=0}
\\
&=\int\Model*{mp}{\d\mpconfig}~\#\prth{\SpatialComponent*\cap\case}.
\end{align*}
We conclude $\sum_{(x,p,u,\brown)\in\mpconfig}\carac{\Permutationmp*{x}\in\case}=\#\prth{\SpatialComponent*\cap\case}$ is true $\Model*{mp}*$ almost surely.

We know each cell of the lattice receives the good number of marks pointing to it. By injectivity of the map $\Permutationmp*{\cdot}$, it guarantees all points in the cell are reached.
\end{description}
\end{Demo}

\subsection{Thermodynamic Limit}\label{sec_thermo_limit}

Section \ref{sec_entropy} was dedicated to the thermodynamic limits of (mp) and (rl) models. We then proved (\Cref{sec_permutation}) the marks of the (mp) infinite volume model to still have meaning and still encode a (FK) representation.

\begin{Def}\label{def_FK}
We define the probability measure $\Model*{FK}*$ over $\ConfSpacePerm{FK}$ by
\[
\Model*{FK}{E}\defequal\Model*{mp}{\prth{\ViewChange*{FK}{mp}*}^{-1}\prth{E}}
\]
for any event $E\in\ConfAlgebraPerm{FK}$.
\end{Def}

But is this new (FK) model the thermodynamic limit of the finite volume (FK) model we presented in \Cref{lemdef_FK}? We claimed this is true in our main thermodynamic limit result (\Cref{th_limit_FK}) and the current section proves this statement.

\begin{Cor}\label{cor_equiint_mp}
For any $\delta,\varepsilon>0$, $\alpha\in\interfo{0;2}$ and compact $\Delta$, there exists $m\geqslant1$ such that for any $L>0$,
\[
\int\Thresh[m]{\sum_{(x,p,u,\brown)\in\mpconfig,~x\in\Delta}\abs{\Sausage{\MyFunction{}{}{}{s}{\frac{s}{\beta}rp+\sqrt{\beta}~\brown\prth{\frac{s}{\beta}}}}}^{\alpha}}~\EmpiricalField{mp}{\d\mpconfig}\leqslant\varepsilon.
\]
Furthermore
\[
\int\sum_{(x,p,u,\brown)\in\mpconfig,~x\in\Delta}\abs{\Sausage{\MyFunction{}{}{}{s}{\frac{s}{\beta}rp+\sqrt{\beta}~\brown\prth{\frac{s}{\beta}}}}}^{\alpha}~\Model*{mp}{\d\mpconfig}<+\infty.
\]
\end{Cor}

\begin{Demo}[\Cref{cor_equiint_mp}]
This is immediate from \Cref{lem_entropic_method,lem_psi} (see \Cref{sec_entropic_method}).
\end{Demo}

We establish a result analogous to \Cref{cor_equiint_mp} for the (FK) models.

\begin{Prop}\label{prop_equiint_FK}
For any $V,\delta,\varepsilon>0$, $\alpha\in\interfo{0;2}$ and $K\in\NN$, there exists $m>0$ such that for any $L>0$,

for any compact $\Delta$ of volume $V$, if $\Delta\subseteq\window[KL]$ then
\[
\int\Thresh[m]{\sum_{\sect\in\FKconfig,~\sect(0)\in\Delta}\abs{\Sausage{\sect}}^{\alpha}}~\EmpiricalField{FK}{\d\FKconfig}\leqslant\varepsilon.
\]
Furthermore for any compact $\Delta$ of volume $V$,
\[
\int\sum_{\sect\in\FKconfig,~\sect(0)\in\Delta}\abs{\Sausage{\sect}}^{\alpha}~\Model*{FK}{\d\FKconfig}<+\infty.
\]
\end{Prop}

\begin{Demo}[\Cref{prop_equiint_FK}]
By definition,
\begin{align*}
&\int\Thresh[m]{\sum_{\sect\in\FKconfig,~\sect(0)\in\Delta}\abs{\Sausage{\sect}}^{\alpha}}~\EmpiricalField{FK}{\d\FKconfig}
\\
=&\frac{1}{L^d}\int_{\window[L]}\d v\int
\Thresh[m]{\sum_{\sect\in\FKconfig,~\sect(0)\in\Delta-v}\abs{\Sausage{\sect}}^{\alpha}}~\Model{FK}{\d\FKconfig}
\end{align*}

According to \Cref{Prop_equiv_mp_FK} and the definition of the map $\ViewChange{FK}{mp}*$, this equals
\begin{align*}
\frac{1}{L^d}\int_{\window[L]}\d v\int
\Thresh[m]{\sum_{\underset{x\in\Delta-v}{(x,p,u,\brown)\in\mpconfig}}\abs{\Sausage{\MyFunction{}{}{}{s}{\frac{s}{\beta}\brac{\Permutationmp{x}-x}+\sqrt{\beta}~\brown\prth{\frac{s}{\beta}}}}}^{\alpha}}~\Model{mp}{\d\mpconfig}.
\end{align*}

By definition of the map $\Permutationmp{\cdot}$, we know
\[
\norm[\infty]{\Permutationmp{x}-\prth{x+rp}}\leqslant r.
\]
So we can approximate the trajectory of any bridge by $\MyFunction{}{}{}{s}{\frac{s}{\beta}rp+\sqrt{\beta}~\brown\prth{\frac{s}{\beta}}}$ with a $\norm[2]{\cdot}$ error of at most $r\sqrt{d}$. Embiggening the sausage by an extra $r\sqrt{d}$ thickness yields the following bound
\begin{align*}
\leqslant&
\frac{1}{L^d}\int_{\window}\d v\int
\Thresh[m]{\sum_{\underset{x\in\Delta-v}{(x,p,u,\brown)\in\mpconfig}}\abs{\Sausage[\delta+r\sqrt{d}]{\MyFunction{}{}{}{s}{\frac{s}{\beta}rp+\sqrt{\beta}~\brown\prth{\frac{s}{\beta}}}}}^{\alpha}}~\Model{mp}{\d\mpconfig}
\\
\leqslant&
\int\Thresh[m]{\sum_{\underset{x\in\Delta}{(x,p,u,\brown)\in\mpconfig}}\abs{\Sausage[\delta+r\sqrt{d}]{\MyFunction{}{}{}{s}{\frac{s}{\beta}rp+\sqrt{\beta}~\brown\prth{\frac{s}{\beta}}}}}^{\alpha}}~\EmpiricalField{mp}{\d\mpconfig}.
\end{align*}
Finally we just need \Cref{cor_equiint_mp} to conclude.

With a very analog procedure as in finite volume, we can show that
\begin{align*}
&\int\sum_{\sect\in\FKconfig,~\sect(0)\in\Delta}\abs{\Sausage{\sect}}^{\alpha}~\Model*{FK}{\d\FKconfig}
\\
\leqslant&\int\sum_{(x,p,u,\brown)\in\mpconfig,~x\in\Delta}\abs{\Sausage[\delta+r\sqrt{d}]{\MyFunction{}{}{}{s}{\frac{s}{\beta}rp+\sqrt{\beta}~\brown\prth{\frac{s}{\beta}}}}}^{\alpha}~\Model*{mp}{\d\mpconfig}.
\end{align*}
which is finite by \Cref{cor_equiint_mp}.
\end{Demo}

We can now prove the first part of \Cref{th_limit_FK}, that is, for any $\MyFunction{f}{\ConfSpacePerm{FK}}{\RR}{}{}$ $\in$-local and $\in$-tame,
\[
\lim_{L\to+\infty}\int f\d\EmpiricalField{FK}*=\int f\d\Model*{FK}*.
\]

\begin{Demo}[\Cref{th_limit_FK}, 1/2]\label{demo_th_limit_FK_1}
Let $\varepsilon>0$ and $\MyFunction{f}{\ConfSpacePerm{FK}}{\RR}{}{}$ $\in$-local such that
\[
\forall\FKconfig\in\ConfSpacePerm{FK},~
a\abs{f(\FKconfig)}\leqslant1+\sum_{\sect\in\Proj{\in}{\FKconfig}}\abs{\Sausage[\delta]{\sect}}^{\alpha}.
\]
for some $a,\delta>0$ and $\alpha<2$. We assume without any loss of generality that $\delta>r\sqrt{d}$.

For any $m\geqslant1$, we define $\MyFunction{f_m}{\ConfSpacePerm{FK}}{\RR}{}{}$ by
\[
f_m(\FKconfig)\defequal
\begin{cases}
f(\FKconfig) & \text{ if }\sum_{\sect\in\Proj{\in}{\FKconfig}}\abs{\Sausage{\sect}}^{\alpha}<\prth{c_{d-1}\prth{\delta-r\sqrt{d}}^{d-1}\frac{m}{2}}^{\alpha}\\
0 & \text{ otherwise.}
\end{cases}
\]

It remains to prove that for $m\geqslant1$ large enough, the error between $f$ and $f_m$ is arbitrarily small. Our goal is to write the integrals of the $f_m$ as integrals of local bounded $\MyFunction{h_m}{\ConfSpace{mp}}{\RR}{}{}$. By using the thermodynamic limit of the (mp) model, this is enough to conclude.

The functions $f_m$, $m\geqslant1$, are uniformly dominated by the $\in$-tame bound of $f$, which is integrable under $\Model*{FK}*$ (see \Cref{prop_equiint_FK}). So by dominated convergence theorem there exists $m_0\geqslant 1$ large enough such that
\[
\forall m\geqslant m_0,~
\int\abs{f_{m}-f}\d\Model*{FK}*\leqslant\varepsilon.
\]
Furthermore, according to \Cref{prop_equiint_FK}, there exists $m_1\geqslant m_0$ such that
\[
\forall L>0,~
\int\abs{f_{m_1}-f}\d\EmpiricalField{FK}*\leqslant\varepsilon.
\]
Then for all $L\geqslant1$, the error term is
\[
\abs{\int f\d\Model*{FK}*-\int f\d\EmpiricalField{FK}*}\leqslant2\varepsilon+\abs{\int f_{m_1}\d\Model*{FK}*-\int f_{m_1}\d\EmpiricalField{FK}*}
\]

We define the function $\MyFunction{g_{m_1}}{\ConfSpace{FK}}{\RR}{}{}$ by
\[
g_{m_1}(\FKconfig)=
\begin{cases}
f_{m_1}\prth{\FKconfig_0} & \text{ if }\exists\FKconfig_0\in\ConfSpacePerm{FK},~\FKconfig=\Proj{\in}{\FKconfig_0}\\
0 & \text{ otherwise.}
\end{cases}
\]
This is well defined thanks to $\in$-locality of $f_{m_1}$ and it is clear that
\[
\forall\FKconfig\in\ConfSpacePerm{FK},~\prth{g_{m_1}\circ\Proj{\in}*}(\FKconfig)=f_{m_1}(\FKconfig).
\]
Since the map $\ViewChange{FK}{mp}*$ commutes with any translation $\MyFunction{}{}{}{\FKconfig}{\FKconfig+v}$, we deduce
\begin{align*}
\int f_{m_1}\d\EmpiricalField{FK}*
&=\frac{1}{L^d}\int_{\window}\d v\int \prth{g_{m_1}\circ\Proj{\in}*\circ\ViewChange{FK}{mp}*}(\FKconfig+v)~\Model{mp}{\d\FKconfig}
\\
&=\int g_{m_1}\circ\Proj{\in}*\circ\ViewChange{FK}{mp}*\d\EmpiricalField{mp}*
\end{align*}

Our goal is to define a local function $\MyFunction{h_{m_1}}{\ConfSpace{mp}}{\RR}{}{}$ such that
\[
\forall\mpconfig\in\ConfSpacePerm{mp},~
h_{m_1}(\mpconfig)=\prth{g_{m_1}\circ\Proj{\in}*\circ\ViewChange{FK}{mp}*}(\mpconfig).
\]
We make a case by case study of the map $\Proj{\in}*\circ\ViewChange{FK}{mp}*$ to identify the desired $h_{m_1}$.

For any $\mpconfig\in\ConfSpacePerm{mp}$,
\begin{align*}
&\prth{\Proj{\in}*\circ\ViewChange{FK}{mp}*}(\mpconfig)
\\
&=\MathSet{\MyFunction{}{\interff{0;\beta}}{\RR^d}{s}{x+\frac{s}{\beta}\prth{\Permutationmp{x}-x}+\sqrt{\beta}~\brown\prth{\frac{s}{\beta}}},
\begin{array}{c}
(x,p,u,\brown)\in\mpconfig\\
\text{s.t. }x\in\Delta
\end{array}
}
\end{align*}
In the following, we make a slight abuse of notations by writing $(x,p,u,\brown)\in\mpconfig\cap\Delta$.

\begin{description}
\item[Case 1] We assume the condition
\[
\sum_{(x,p,u,\brown)\in\mpconfig\cap\Delta}\abs{\Sausage[\delta-r\sqrt{d}]{\MyFunction{}{}{}{s}{\frac{s}{\beta}rp+\sqrt{\beta}~\brown\prth{\frac{s}{\beta}}}}}^{\alpha}>\prth{c_{d-1}\prth{\delta-r\sqrt{d}}^{d-1}\frac{m_1}{2}}^{\alpha}.
\]
The map $\Proj{\in}*\circ\ViewChange{FK}{mp}*$ induces a bijection between the set of marked points $(x,p,u,\brown)\in\mpconfig\cap\Delta$ and the image set of bridges $\sect\in\prth{\Proj{\in}*\circ\ViewChange{FK}{mp}*}\prth{\mpconfig}$. We use a similar reasoning as in the proof of \Cref{prop_equiint_FK}: approximating the Brownian bridge by $\MyFunction{}{}{}{s}{\frac{s}{\beta}rp+\sqrt{\beta}~\brown\prth{\frac{s}{\beta}}}$ and getting from it a bound of the sausage's size
\begin{align*}
\sum_{\sect\in\prth{\Proj{\in}*\circ\ViewChange{FK}{mp}*}\prth{\mpconfig}}\abs{\Sausage{\sect}}^{\alpha}
&\geqslant
\sum_{(x,p,u,\brown)\in\mpconfig\cap\Delta}\abs{\Sausage[\delta-r\sqrt{d}]{\MyFunction{}{}{}{s}{\frac{s}{\beta}rp+\sqrt{\beta}~\brown\prth{\frac{s}{\beta}}}}}^{\alpha}
\\
&>\prth{c_{d-1}\prth{\delta-r\sqrt{d}}^{d-1}\frac{m_1}{2}}^{\alpha}.
\end{align*}
Thus
\[
\prth{g_{m_1}\circ\Proj{\in}*\circ\ViewChange{FK}{mp}*}(\mpconfig)=f_{m_1}\prth{\ViewChange{FK}{mp}{\mpconfig}}=0.
\]

\item[Case 2] We assume
\[
\sum_{(x,p,u,\brown)\in\mpconfig\cap\Delta}\abs{\Sausage[\delta-r\sqrt{d}]{\MyFunction{}{}{}{s}{\frac{s}{\beta}rp+\sqrt{\beta}~\brown\prth{\frac{s}{\beta}}}}}^{\alpha}\leqslant\prth{c_{d-1}\prth{\delta-r\sqrt{d}}^{d-1}\frac{m_1}{2}}^{\alpha}.
\]
By \Cref{prop_saucisse}, the cylinder whose axis goes from $0$ ro $rp$ with diameter $\delta-r\sqrt{d}$ has a smaller volume than the Wiener sausage:
\[
\begin{array}{c}
\forall(x,p,u,\brown)\in\mpconfig\\
x\in\Delta
\end{array}
\!,~r\norm{p}\leqslant\frac{1}{c_{d-1}\prth{\delta-r\sqrt{d}}^{d-1}}\abs{\Sausage[\delta-r\sqrt{d}]{\MyFunction{}{}{}{s}{\frac{s}{\beta}rp+\sqrt{\beta}~\brown\prth{\frac{s}{\beta}}}}}
\leqslant\frac{m_1}{2}.
\]
We can assume without any loss of generality that $\Delta+\case\subseteq\window[m_1]$. Then, for all $(x,p,u,\brown)\in\mpconfig\cap\Delta$, the expression $\Permutationmp*{x}$ only depends on $\mpconfig\cap\prth{x+rp+\case}\subseteq\mpconfig\cap\window[2m_1]$. We slightly abuse notations by writing
\begin{align*}
&\prth{\Proj{\in}*\circ\ViewChange{FK}{mp}*}(\mpconfig)
\\
&=\MathSet{\MyFunction{}{\interff{0;\beta}}{\RR^d}{s}{x+\frac{s}{\beta}\prth{\Permutationmp*{x}[\mpconfig\cap\window[2m_1]]-x}+\sqrt{\beta}~\brown\prth{\frac{s}{\beta}}},~(x,p,u,\brown)\in\mpconfig\cap\Delta}.
\end{align*}
\end{description}

In all cases, we denote
\[
\ViewChangeDelta{\mpconfig}\defequal
\MathSet{\!\MyFunction{}{\interff{0;\beta}}{\RR^d}{s}{x+\frac{s}{\beta}\prth{\Permutationmp*{x}[\mpconfig\cap\window[2m_1]]-x}+\sqrt{\beta}~\brown\prth{\frac{s}{\beta}}}\!,~
\begin{array}{c}
(x,p,u,\brown)\in\mpconfig\cap\Delta\text{ s.t.}\\
\CardCase*{x+rp}[\SpatialComponent*\cap\window[2m_1]]\geqslant 1
\end{array}}.
\]
We define $\MyFunction{h_{m_1}}{\ConfSpace{mp}}{\RR}{}{}$ by
\[
h_{m_1}(\mpconfig)\defequal
\begin{cases}
\prth{g_{m_1}\circ\ViewChangeDelta*}(\mpconfig)
& \text{if }
\dps\sum_{(x,p,u,\brown)\in\mpconfig\cap\Delta}\abs{\Sausage[\delta-r\sqrt{d}]{\MyFunction{}{}{}{s}{\frac{s}{\beta}rp+\sqrt{\beta}~\brown\prth{\frac{s}{\beta}}}}}^{\alpha}\leqslant\prth{c_{d-1}\prth{\delta-r\sqrt{d}}^{d-1}\frac{m_1}{2}}^{\alpha}
\\
0 & \text{otherwise}.
\end{cases}
\]
In all cases, $h_{m_1}$ equals $g_{m_1}\circ\Proj{\in}*\circ\ViewChange{FK}{mp}*$ on $\ConfSpacePerm{mp}$. Therefore
\begin{align*}
\int f_{m_1}\d\EmpiricalField{FK}*
=\int g_{m_1}\circ\Proj{\in}*\circ\ViewChange{FK}{mp}*\d\EmpiricalField{mp}*
=\int h_{m_1}\d\EmpiricalField{mp}*.
\end{align*}
Furthermore,
\begin{align*}
\int f_{m_1}\d\Model*{FK}*
&=\int g_{m_1}\circ\Proj{\in}*\d\Model*{FK}*
\\
&=\int g_{m_1}\circ\Proj{\in}*\circ\ViewChange{FK}{mp}*
\d\Model*{mp}*
\\
&=\int h_{m_1}\d\Model*{mp}*.
\end{align*}

Since the function $\MyFunction{h_{m_1}}{\ConfSpace{mp}}{\RR}{}{}$ is bounded and local, we have
\[
\lim_{L\to+\infty}\int h_{m_1}\d\EmpiricalField{mp}*
=\int h_{m_1}\d\Model*{mp}*.
\]
Thus
\[
\lim_{L\to+\infty}\int f_{m_1}\d\EmpiricalField{mp}*
=\int f_{m_1}\d\Model*{mp}*.
\]
\end{Demo}

\subsection{Extension of Locality}

Locality in the sense of $\Proj{\in}*$ is natural but it does not preserve the cycle structure from $\PermutationFK{\cdot}$. Cycles of a configuration $\FKconfig\in\ConfSpacePerm{FK}$ are split open by the projection map $\Proj{\in}*$ and only bridges starting in the compact $\Delta$ are preserved. We loose the information on the length of the cycle any bridge $\sect\in\Proj{\in}{\FKconfig}$ was part of in $\FKconfig$. The family of projection maps $\Proj{\cap^n}*$ ($n\geqslant0$) is more compatible with the cycle structure of $\PermutationFK{\cdot}$ because if a cycle intersects $\Delta$, then it is completely preserved in $\Proj{\cap^n}{\FKconfig}$, provided its length is not too long. The second part of \Cref{th_limit_FK} is adapted to functions which are $\cap^n$-local.

In \Cref{sec_thermo_limit}, we made use of \Cref{prop_equiint_FK} in the proof of the first part of the thermodynamic limit. Similarly, we need the following proposition.

\begin{Prop}\label{prop_equiint_cap_FK}
For any $D,\delta,\varepsilon>0$ and $\alpha\in\interfo{0;1}$, there exists $m>0$ such that for any $L\geqslant4\max\prth{\delta,D,\frac{1}{4}}$ and compact $\Delta$ such that $\sup_{x\in\Delta}\norm{x}\leqslant D$,
\[
\int\sum_{\sect\in\Proj{\cap}{\FKconfig}}\carac{\sect(0)\notin\window[m]}\abs{\Sausage{\sect}}^{\alpha}~\EmpiricalField{FK}{\d\FKconfig}
\leqslant\varepsilon.
\]
Furthermore,
\[
\int\sum_{\sect\in\Proj{\cap}{\FKconfig}}\abs{\Sausage{\sect}}^{\alpha}~\Model*{FK}{\d\FKconfig}<+\infty.
\]
\end{Prop}

\begin{Rq}
Accessing the bridges that only intersect $\Delta$ made us lose a power in the volume of the sausage. This can be seen at the end of the proof below: an exponent $1+\alpha$ appears.
\end{Rq}

\begin{Demo}[\Cref{prop_equiint_cap_FK}]
For any $L\geqslant1$,
\begin{align*}
&\int\EmpiricalField{FK}{\d\FKconfig} \sum_{\sect\in\FKconfig}\carac{\sect\cap\Delta\neq\emptyset}\carac{\sect(0)\notin\window[m]}\abs{\Sausage{\sect}}^{\alpha}
\\
\leqslant
&\sum_{i\in4\delta\ZZ^d\cap\window[m-4\delta]\complementary}\int\EmpiricalField{FK}{\d\FKconfig}\sum_{\sect\in\FKconfig}\carac{\sect\cap\Delta\neq\emptyset}\carac{\sect(0)\in i+\case[4\delta]}\abs{\Sausage{\sect}}^{\alpha}
\\
\leqslant
&\sum_{i\in4\delta\ZZ^d\cap\window[m-4\delta]\complementary}\int\EmpiricalField{FK}{\d\FKconfig}\sum_{\sect\in\FKconfig}\carac{0\in\Sausage[D]{\sect}}\carac{\sect(0)\in i+\case[4\delta]}\abs{\Sausage{\sect}}^{\alpha}.
\end{align*}
We assume without any loss of generality $\delta\geqslant D$. Then
\begin{equation*}
\leqslant
\sum_{i\in4\delta\ZZ^d\cap\window[m-4\delta]\complementary}\int\EmpiricalField{FK}{\d\FKconfig}\sum_{\sect\in\FKconfig}\carac{0\in\Sausage{\sect}}\carac{\sect(0)\in i+\case[4\delta]}\abs{\Sausage{\sect}}^{\alpha}.
\end{equation*}
We directly bound this expression by replacing the configuration by its periodized version
\[
\leqslant
\sum_{i\in4\delta\ZZ^d\cap\window[m-4\delta]\complementary}\int\EmpiricalField{FK}[dir]{\d\FKconfig}\sum_{\sect\in\FKconfig_{(\textnormal{per}),L}}\carac{0\in\Sausage{\sect}}\carac{\sect(0)\in i+\case[4\delta]}\abs{\Sausage{\sect}}^{\alpha}
\]
where we denote $\FKconfig_{(\textnormal{per}),L}=\bigcup_{k\in\ZZ^d}\prth{\FKconfig+Lk}$.

If $\FKconfig$ is distributed along $\EmpiricalField{FK}[dir]*$ then the configuration $\FKconfig_{(\textnormal{per}),L}$ has a stationary distribution. Thus
\begin{align*}
&\leqslant
\sum_{i\in4\delta\ZZ^d\cap\window[m-4\delta]\complementary}\int\EmpiricalField{FK}[dir]{\d\FKconfig}\sum_{\sect\in\FKconfig_{(\textnormal{per}),L}}\carac{-i\in\Sausage{\sect}}\carac{\sect(0)\in \case[4\delta]}\abs{\Sausage{\sect}}^{\alpha}
\\
&\leqslant
\sum_{i\in4\delta\ZZ^d\cap\window[m-4\delta]\complementary}\int\EmpiricalField{FK}[dir]{\d\FKconfig}\sum_{k\in\ZZ^d}\sum_{\sect\in\FKconfig}\carac{-i\in\Sausage{\sect}+Lk}\carac{\sect(0)+Lk\in \case[4\delta]}\abs{\Sausage{\sect}}^{\alpha}.
\end{align*}
Under $\EmpiricalField{FK}[dir]*$, the set $\MathSet{\sect(0),~\sect\in\FKconfig}$ is included inside $\window[2L]$. Plus, we know $\window[4\delta]\subseteq\window$. So the condition $\sect(0)+Lk\in\window[4\delta]$ allows us to restrict the sum to $k\in\{-1;0;1\}^d$
\begin{align*}
&\leqslant
\sum_{i\in4\delta\ZZ^d\cap\window[m-4\delta]\complementary}\int\EmpiricalField{FK}[dir]{\d\FKconfig}\sum_{k\in\{-1;0;1\}^d}\sum_{\sect\in\FKconfig}\carac{-i\in\Sausage{\sect}+Lk}\carac{\sect(0)+Lk\in \case[4\delta]}\abs{\Sausage{\sect}}^{\alpha}.
\end{align*}

If $-i\in\Sausage{\sect}+Lk$ then
\[
\abs{(-i+\case[4\delta])\cap\prth{\Sausage{\sect}+Lk}}\geqslant c_d\delta^d
\]
where $c_d$ is the volume of the $d$-dimensional unit ball. We deduce
\begin{align*}
\sum_{i\in4\delta\ZZ^d\cap\window[m-4\delta]\complementary}\carac{-i\in\Sausage{\sect}+Lk}
&\leqslant\frac{1}{c_d\delta^d}\sum_{i\in4\delta\ZZ^d\cap\window[m-4\delta]\complementary}\abs{(-i+\case[4\delta])\cap\prth{\Sausage{\sect}+Lk}}
\\
&\leqslant\frac{1}{c_d\delta^d}\abs{\window[m-8\delta]\complementary\cap\prth{\Sausage{\sect}+Lk}}.
\end{align*}

Therefore
\begin{align*}
&\int\EmpiricalField{FK}{\d\FKconfig} \sum_{\sect\in\FKconfig}\carac{\sect\cap\Delta\neq\emptyset}\carac{\sect(0)\notin\window[m]}\abs{\Sausage{\sect}}^{\alpha}
\\
\leqslant&\frac{1}{c_d\delta^d}\sum_{k\in\{-1;0;1\}^d}\int\EmpiricalField{FK}{\d\FKconfig}\sum_{\sect\in\Proj{\in}[\case[4\delta]-Lk]{\FKconfig}}\abs{\window[m-8\delta]\complementary\cap\prth{\Sausage{\sect}+Lk}}\cdot\abs{\Sausage{\sect}}^{\alpha}
\\
\leqslant&\frac{1}{c_d\delta^d}\sum_{k\in\{-1;0;1\}^d}\int\EmpiricalField{FK}{\d\FKconfig}\sum_{\sect\in\Proj{\in}[\case[4\delta]-Lk]{\FKconfig}}\carac{\prth{\window[m-8\delta]\complementary-Lk}\cap\Sausage{\sect}\neq\emptyset}\cdot\abs{\Sausage{\sect}}^{1+\alpha}.
\end{align*}

Let a Brownian bridge $\sect$ have its starting point $\sect(0)$ in $\window[4\delta]$. We assume $\Sausage{\sect}\cap\window[m-8\delta]\complementary\neq\emptyset$. This means there exists $s\in\interff{0;\beta}$ such that $\sect(s)$ is at a distance less than $\frac{\delta}{2}$ from $\window[m-8\delta]\complementary$. So the point $\sect(s)$ is at a distance \emph{at least} $\frac{m-13\delta}{2}$ from $\window[4\delta]$. In particular,\label{reasoning_sausage}
\[
\norm{\sect(0)-\sect(s)}\geqslant\frac{m-13\delta}{2}.
\]
By \Cref{prop_saucisse}, a cylinder whose axis goes from $\sect(0)$ to $\sect(s)$ with radius $\delta$ has a smaller volume than the sausage. Therefore
\[
\abs{\Sausage{\sect}}\geqslant c_{d-1}\delta^{d-1}\frac{m-13\delta}{2}
\]
where $c_{d-1}$ is the volume of the $d-1$ dimensional unit ball. A translation by $Lk$ does not change the picture.

We conclude with the following upper bound
\begin{align*}
\leqslant&\frac{1}{c_d\delta^d}\sum_{k\in\{-1;0;1\}^d}\int\EmpiricalField{FK}{\d\FKconfig}\sum_{\sect\in\Proj{\in}[\case[4\delta]-Lk]{\FKconfig}}\carac{\abs{\Sausage{\sect}}\geqslant\frac{1}{2}c_{d-1}\delta^{d-1}\prth{m-13\delta}}\cdot\abs{\Sausage{\sect}}^{1+\alpha}.
\end{align*}
Thanks to \Cref{prop_equiint_FK}, this is enough to conclude on the first part of the proposition. Infinite volume is managed in a similar way, without any need for a periodization.
\end{Demo}

We state the following intuitive proposition which allows us to treat similarly outgoing and ingoing bridges in the proof of the second part of \Cref{th_limit_FK}.

\begin{Def}\label{def_proj_subset}
	Let $\Delta\subset\RR^d$ be a compact. We define the projection $\MyFunction{\Proj{\subset}*}{\ConfSpace{FK}}{\ConfSpace{FK}}{}{}$ by
	\begin{align*}
		\Proj{\subset}{\FKconfig}\defequal\MathSet{\sect\in\FKconfig}[\sect\subset\Delta]
	\end{align*}
	if we accept the abuse of notation \enquote{$\sect\subset\Delta$} to mean $\sect\prth{\interff{0;\beta}}\subset\Delta$.\\
	
	A function $\MyFunction{f}{}{}{}{}$ defined over $\ConfSpacePerm{FK}$ is said to be \emph{$\subset$-local} if there exists a compact $\Delta\subset\RR^d$ such that
	\[
	\Proj{\in}{\FKconfig}=\Proj{\in}{\FKconfig'}\implies f(\FKconfig)=f(\FKconfig').
	\]
\end{Def}

\begin{Prop}\label{prop_time_reversal}
The (FK) model is time-reversal invariant. More precisely, for any measurable $\MyFunction{f}{\ConfSpacePerm{FK}}{\RR^+}{}{}$ and $L\in\interof{0;+\infty}$,
\[
\int f\circ\TimeReversal{FK}*\d\Model{FK}*=\int f\d\Model{FK}*
\]
where we define the time-reversal operator as
\[
\MyFunction{\TimeReversal{FK}*}{\ConfSpace{FK}}{\ConfSpace{FK}}{\FKconfig}{
\MathSet{\MyFunction{}{\interff{0;\beta}}{\RR^d}{s}{\sect(\beta-s)},~\sect\in\FKconfig}.}
\]
\end{Prop}

\begin{Demo}[\Cref{prop_time_reversal}]
In finite volume, one just needs to replace $\PermutationSymbol\in\PermutationSpace{\finiteconfig}$ by $\PermutationSymbol^{-1}$ in the definition of $\Model{FK}*$.

In infinite volume, we can already obtain the property for the indicators of $\subset$-local events by thermodynamic limit (first part of \Cref{th_limit_FK}). Since $\in$-local events are a ring of sets, by Carathéodory's extension theorem, the two measures $\Model*{FK}*$ and $\Model*{FK}*\circ\TimeReversal{FK}*$ coincide on the generated $\sigma$-algebra. It remains to prove this $\sigma$-algebra is the whole $\ConfAlgebraPerm{FK}$.

Let $A$ be an $\in$-local event (relatively to some compact $\Delta$). We define the sequence of $\subset$-local events
\begin{equation*}
	B_n=\MathSet{\gamma\in A}[\forall\eta\in\ConfSpacePerm{FK},~\Proj{\subset}[\window[n]]{\eta}=\Proj{\subset}[\window[n]]{\gamma}\implies\eta\in A].
\end{equation*}
The inclusion $\bigcup_{n\geqslant0}B_n\subseteq A$ is trivial. Conversely, for any $\gamma\in A$, there exists $n(\gamma)\geqslant1$ large enough such that
\begin{equation*}
	\Proj{\in}{\gamma}\subseteq\Proj{\subset}[\window[n(\gamma)]]{\gamma}.
\end{equation*}
Then $\gamma\in B_{n(\gamma)}$. This proves $A=\bigcup_{n\geqslant0}B_n$ and more generally that $\in$-local events are part of the $\sigma$-algebra generated by $\subset$-local events.
\end{Demo}

Our next and final lemma is the last specific result we need for the sake of our theorem.

\begin{Lem}\label{lem_inter_n_lipschitz}
	Let $\MyFunction{f}{\ConfSpacePerm{FK}}{\RR}{}{}$ be $\cap^n$-Lipschitz for some $n\geqslant0$. Then there exists an extension $\MyFunction{g}{\ConfSpace{FK}}{\RR}{}{}$ of $f$ such that for any $\FKconfig\in\ConfSpacePerm{FK}$ and $\FKconfig'\subseteq\FKconfig$,
	\begin{align*}
		a\abs{g(\FKconfig)-g\prth{\FKconfig\setminus\FKconfig'}}
		\leqslant&
		\sum_{\sect\in\Proj{\in}{\FKconfig}\cap\FKconfig'}\abs{\Sausage{\sect}}^{1+\alpha}+\sum_{\sect\in\Proj{\cap}{\FKconfig}\cap\FKconfig'}\abs{\Sausage{\sect}}^{\alpha}
		\\
		&+\sum_{\sect\in\Proj{\cap}{\FKconfig}}\carac{\exists k\in\intgff{-n;n},~\prth{\PermutationFK{\cdot}}^{k}(\sect)\in\FKconfig'}
	\end{align*}
	where the compact $\Delta$ and the constants $a$, $\alpha$ and $\delta$ are the same as in the Lipschitz property of $f$ (see \Cref{def_tame}).
	
	Furthermore, for any compact $\Delta'$, there exists $b_{\Delta'}>0$ such that for any $\FKconfig\in\ConfSpace{FK}$,
	\begin{align*}
		\prth{\forall\sect\in\FKconfig,~\sect\subset\Delta'}
		\implies
		b_{\Delta'}\abs{g(\FKconfig)}\leqslant1+\sum_{\sect\in\Proj{\in}{\FKconfig}}\abs{\Sausage{\sect}}^{1+\alpha}+\sum_{\sect\in\FKconfig}\abs{\Sausage{\sect}}^{\alpha}
	\end{align*}
	where the slight abuse of notation $\sect\subset\Delta'$ means $\sect\prth{\interff{0;\beta}}\subset\Delta'$.
\end{Lem}

\begin{Demo}[\Cref{lem_inter_n_lipschitz}]
	For any $\FKconfig\in\ConfSpacePerm{FK}$ and $\rlconfig\in\ConfSpace{FK}$, we denote
	\begin{align*}
		\bound{\FKconfig,\rlconfig}=
		&
		\sum_{\sect\in\Proj{\in}{\FKconfig}\setminus\rlconfig}\abs{\Sausage{\sect}}^{1+\alpha}+\sum_{\sect\in\Proj{\cap}{\FKconfig}\setminus\rlconfig}\abs{\Sausage{\sect}}^{\alpha}
		\\
		&+\sum_{\sect\in\Proj{\cap}{\FKconfig}}\carac{\exists k\in\intgff{-n;n},~\prth{\PermutationFK{\cdot}}^{k}(\sect)\notin\rlconfig}.
	\end{align*}
	
	We prove that for any $\FKconfig\in\ConfSpace{FK}$ if there exists $\rlconfig\in\ConfSpacePerm{FK}$ such that $\FKconfig\subseteq\rlconfig$ then
	\begin{equation}\label{eq_sup_inf}
		\sup_{\FKconfig\subseteq\rlconfig}
		\prth{f(\rlconfig)-\frac{1}{a}\bound{\rlconfig,\FKconfig}}
		\leqslant
		\inf_{\FKconfig\subseteq\rlconfig}
		\prth{f(\rlconfig)+\frac{1}{a}\bound{\rlconfig,\FKconfig}}
	\end{equation}
	if we agree to take the $\inf$ and $\sup$ over permutation-wise $\rlconfig$. All the subsequent $\inf$ and $\sup$ are also implicitly defined over the set of permutation-wise configurations.
	
	Let $\FKconfig\in\ConfSpace{FK}$ and $\rlconfig,\rlconfig'\in\ConfSpacePerm{FK}$ such that $\FKconfig\subseteq\rlconfig\cap\rlconfig'$. By $\cap^n$-Lipschitz property,
	\[
	f(\rlconfig)-f(\rlconfig')\leqslant\frac{1}{a}\bound{\rlconfig,\rlconfig\cap\rlconfig'}+\frac{1}{a}\bound{\rlconfig',\rlconfig\cap\rlconfig'}.
	\]
	Then
	\begin{align*}
		f(\rlconfig)-\frac{1}{a}\bound{\rlconfig,\FKconfig}
		&\leqslant
		f(\rlconfig)-\frac{1}{a}\bound{\rlconfig,\rlconfig\cap\rlconfig'}
		\\
		&\leqslant
		f(\rlconfig')+\frac{1}{a}\bound{\rlconfig',\rlconfig\cap\rlconfig'}
		\\
		&\leqslant
		f(\rlconfig')+\frac{1}{a}\bound{\rlconfig',\FKconfig}.
	\end{align*}
	We conclude by taking the $\sup$ over all such $\rlconfig$ and the $\inf$ over all such $\rlconfig'$.

	We define $\MyFunction{g}{\ConfSpace{FK}}{\RR}{}{}$ by setting
	\[
	g(\FKconfig)\defequal
	\sup_{\FKconfig\subseteq\rlconfig}\prth{f(\rlconfig)-\frac{1}{a}\bound{\rlconfig,\FKconfig}}
	\]
	if the set defining this $\sup$ is non-empty (we remind it is restricted to permutation-wise configurations). We set $g(\FKconfig)=0$ otherwise.
	
	First, we know by definition that for any $\FKconfig\in\ConfSpacePerm{FK}$,
	\[
	g(\FKconfig)\geqslant f(\FKconfig)-\frac{1}{a}\bound{\FKconfig,\FKconfig}=f(\FKconfig).
	\]
	Furthermore by inequality (\ref{eq_sup_inf}),
	\begin{align*}
		g(\FKconfig)\leqslant
		\inf_{\FKconfig\subseteq\rlconfig}
		\prth{f(\rlconfig)+\frac{1}{a}\bound{\rlconfig,\FKconfig}}
		\leqslant
		f(\FKconfig)+\frac{1}{a}\bound{\FKconfig,\FKconfig}=f(\FKconfig).
	\end{align*}
	So the function $g$ does coincide with $f$ on $\ConfSpacePerm{FK}$.

	Let $\FKconfig\in\ConfSpacePerm{FK}$ and $\FKconfig'\subseteq\FKconfig$. Then
	\begin{align*}
		g(\FKconfig)-g(\FKconfig\setminus\FKconfig')
		&=
		f(\FKconfig)-\sup_{\FKconfig\setminus\FKconfig'\subseteq\rlconfig}\prth{f(\rlconfig)-\frac{1}{a}\bound{\rlconfig,\FKconfig\setminus\FKconfig'}}
		\\
		&\leqslant
		f(\FKconfig)-\prth{f(\FKconfig)-\frac{1}{a}\bound{\FKconfig,\FKconfig\setminus\FKconfig'}}
		\\
		&\leqslant
		\frac{1}{a}\bound{\FKconfig,\FKconfig\setminus\FKconfig'}.
	\end{align*}
	Similarly according to inequality (\ref{eq_sup_inf}),
	\begin{align*}
		g(\FKconfig\setminus\FKconfig')-g(\FKconfig)
		&\leqslant
		\inf_{\FKconfig\setminus\FKconfig'\subseteq\rlconfig}\prth{f(\rlconfig)+\frac{1}{a}\bound{\rlconfig,\FKconfig\setminus\FKconfig'}}
		-f(\FKconfig)
		\\
		&\leqslant
		\prth{f(\FKconfig)+\frac{1}{a}\bound{\FKconfig,\FKconfig\setminus\FKconfig'}}-f(\FKconfig)
		\\
		&\leqslant
		\frac{1}{a}\bound{\FKconfig,\FKconfig\setminus\FKconfig'}.
	\end{align*}
	Therefore,
	\begin{align*}
		a\abs{g(\rlconfig)-g(\rlconfig\setminus\rlconfig')}
		\leqslant
		\bound{\rlconfig,\rlconfig\setminus\rlconfig'}.
	\end{align*}

All that remains is to prove the final part of the lemma.

Let $\Delta'$ be a compact. There exists $R>0$ large enough so that $\Delta\cup\Delta'\subseteq B_R$. Let $\FKconfig\in\ConfSpace{FK}$ such that $\forall\sect\in\FKconfig,~\sect\subset\Delta'$.

If there is no $\rlconfig\in\ConfSpacePerm{FK}$ such that $\FKconfig\subseteq\rlconfig$ then by definition $g(\FKconfig)=0$. In the following we assume the existence of such an $\rlconfig$. It is then clear
\begin{align*}
	\forall\sect\in\FKconfig,
	\begin{array}{l}
		\#\MathSet{\sect'\in\FKconfig}[\sect'(0)=\sect(\beta)]\leqslant1\\
		\#\MathSet{\sect'\in\FKconfig}[\sect'(\beta)=\sect(0)]\leqslant1.
	\end{array}
\end{align*}
We use this fact to build a configuration $\widehat{\FKconfig}\in\ConfSpacePerm{FK}$ containing $\FKconfig$. We denote the set of points lacking a follow-up bridge as
\begin{align*}
	\partial^{\textnormal{fol}}\FKconfig\defequal
	\MathSet{x\in\Delta'}[
	\begin{array}{l}
		\exists\sect\in\FKconfig,~x=\sect(\beta)\\
		\forall\sect\in\FKconfig,~x\neq\sect(0)
	\end{array}
	]
\end{align*}
and the set of points lacking a precursor bridge as
\begin{align*}
	\partial^{\textnormal{pre}}\FKconfig\defequal
	\MathSet{y\in\Delta'}[
	\begin{array}{l}
		\exists\sect\in\FKconfig,~y=\sect(0)\\
		\forall\sect\in\FKconfig,~y\neq\sect(\beta)
	\end{array}
	].
\end{align*}
For each $x\in\partial^{\textnormal{fol}}\FKconfig$, there exists a $\sect_{x,0}^{\textnormal{fol}}\in\WienerSpace[\beta]$ from $\sect_{x,0}^{\textnormal{fol}}(0)=x$ to $\sect_{x,0}^{\textnormal{fol}}(\beta)\notin B_R$ such that
\[
\abs{\Sausage{\sect_{x,0}^{\textnormal{fol}}}}\leqslant c_{d-1}\delta^{d-1}R+c_d\delta^d+1
\]
where $c_{d-1}$ and $c_d$ are the respective volumes of $d-1$ and $d$ dimensional unit balls. We just need $\sect_{x,0}^{\textnormal{fol}}$ to be a straight line barely poking out of $B_R$.

Similarly, for each $y\in\partial^{\textnormal{pre}}\FKconfig$, there exists a $\sect_{y,0}^{\textnormal{pre}}\in\WienerSpace[\beta]$ from $\sect_{y,0}^{\textnormal{pre}}(0)\notin B_R$ to $\sect_{y,0}^{\textnormal{pre}}(\beta)=y$ such that
\[
\abs{\Sausage{\sect_{y,0}^{\textnormal{pre}}}}\leqslant c_{d-1}\delta^{d-1}R+c_d\delta^d+1.
\]
Of course it is possible to choose all these bridges so that the vectors $\sect_{x,0}^{\textnormal{fol}}(\beta),~x\in\partial^{\textnormal{fol}}\FKconfig$ and $\sect_{y,0}^{\textnormal{pre}}(\beta),~y\in\partial^{\textnormal{pre}}\FKconfig$ are all distinct from one another (neither collinear).

For each $x\in\partial^{\textnormal{fol}}\FKconfig$, we recursively define the family of bridges $\sect_{x,n}^{\textnormal{fol}}\in\WienerSpace[\beta],~n\geqslant1$ by
\begin{align*}
\sect_{x,n}^{\textnormal{fol}}(s)\defequal\sect_{x,n-1}^{\textnormal{fol}}(\beta)+\frac{s}{\beta}\sect_{x,0}^{\textnormal{fol}}(\beta).
\end{align*}
Similarly, for each $y\in\partial^{\textnormal{pre}}\FKconfig$, we define recursively
\begin{align*}
	\sect_{y,n}^{\textnormal{pre}}(s)\defequal\sect_{y,n-1}^{\textnormal{pre}}(0)+\frac{\beta-s}{\beta}\sect_{y,0}^{\textnormal{pre}}(0).
\end{align*}
It turns out the (FK) configuration
\begin{align*}
	\widehat{\FKconfig}\defequal\FKconfig\cup\bigcup_{x\in\partial^{\textnormal{fol}}\FKconfig}\MathSet{\sect_{x,n}^{\textnormal{fol}},~n\geqslant0}\cup\bigcup_{y\in\partial^{\textnormal{pre}}\FKconfig}\MathSet{\sect_{y,n}^{\textnormal{pre}},~n\geqslant0}
\end{align*}
is permutation-wise.

\begin{figure}[h]
\centering
\begin{tikzpicture}[scale=.95]
	\tikzstyle{fleche_int} = [rounded corners,->,>=latex]
	\tikzstyle{fleche_ext} = [->,>=latex,color=blue,dashed]
	\tikzstyle{fleche_pre} = [->,>=latex,color=red,dotted,thick]
	\draw (1.5,3.5) rectangle (3,4.7);
	\draw (1.3,3.3) node {$\Delta$};
	\draw (0,0) rectangle (4,4);
	\draw (-0.3,1) node {$\Delta'$};
	\draw (2,2) circle (3);
	\draw (-1,0) node {$B_R$};
	\draw (3.3,1.2) node (B) {} ;
	\draw (0.5,2.3) node (J) {};
	\draw (1.5,0.9) node (K) {};
	\draw (-1.5,2) node (M) {};
	\draw (-3.5,2) node (N) {};
	\draw (5.5,1) node (O) {};
	\draw[fleche_int] (3.3,1.2) to[bend left, looseness=0.5] (K);
	\draw[fleche_int] (1.5,0.9) to[bend right, looseness=1.4] (J);
	\draw[fleche_ext] (0.5,2.3) to (M);
	\draw[fleche_ext] (-1.5,2) to (N);
	\draw[fleche_pre] (5.5,1) to (B);
	\draw[fleche_pre] (9,0) to (O);
	\draw (3.3,1.2) node {\tiny$\blacksquare$};
	\draw (0.5,2.3) node {\small$\blacktriangle$};
	%
	\draw[fleche_int] (6,4) to (6.5,4);
	\draw (7,4) node{$\FKconfig$};
	\draw[fleche_pre] (6,3) to (6.5,3);
	\draw (8,3) node{$\MathSet{\sect_{{\blacksquare},n}^{\textnormal{pre}},~n\geqslant0}$};
	\draw[fleche_ext] (6,2) to (6.5,2);
	\draw (8,2) node{$\MathSet{\sect_{{\blacktriangle},n}^{\textnormal{fol}},~n\geqslant0}$};
	\draw (11,3) node{{\tiny$\blacksquare$} $~\partial^{\textnormal{pre}}\FKconfig$};
	\draw (11,2) node{{\small$\blacktriangle$} $~\partial^{\textnormal{fol}}\FKconfig$};
\end{tikzpicture}
\label{fig_extension_locality}
\caption{Illustration of $\widehat{\FKconfig}$}
\end{figure}

Let us bound the value $\abs{g(\FKconfig)}$
\begin{align*}
g(\FKconfig)&=\sup_{\FKconfig\subseteq\rlconfig}\prth{f(\rlconfig)-\frac{1}{a}\bound{\rlconfig,\FKconfig}}
\geqslant f(\widehat{\FKconfig})-\frac{1}{a}\bound{\widehat{\FKconfig},\FKconfig}
\geqslant -\abs{f(\widehat{\FKconfig})}-\frac{1}{a}\bound{\widehat{\FKconfig},\FKconfig}.
\end{align*}
By inequality (\ref{eq_sup_inf}),
\begin{align*}
g(\FKconfig)&\leqslant\inf_{\FKconfig\subseteq\rlconfig}\prth{f(\rlconfig)+\frac{1}{a}\bound{\rlconfig,\FKconfig}}
\leqslant f\prth{\widehat{\FKconfig}}+\frac{1}{a}\bound{\widehat{\FKconfig},\FKconfig}
\leqslant \abs{f\prth{\widehat{\FKconfig}}}+\frac{1}{a}\bound{\widehat{\FKconfig},\FKconfig}.
\end{align*}
So we have
\[
\abs{g(\FKconfig)}\leqslant\abs{f\prth{\widehat{\FKconfig}}}+\frac{1}{a}\bound{\widehat{\FKconfig},\FKconfig}.
\]
Since $f$ is $\cap^n$-Lipschitz, according to \Cref{rq_inter_n_lipschitz}, it also is $\cap$-tame relatively to the same compact. Then
\begin{align*}
a\abs{f\prth{\widehat{\FKconfig}}}+\bound{\widehat{\FKconfig},\FKconfig}
\leqslant&~1+\sum_{\sect\in\Proj{\in}{\widehat{\FKconfig}}}\abs{\Sausage[\delta]{\sect}}^{1+\alpha}+\sum_{\sect\in\Proj{\cap}{\widehat{\FKconfig}}}\abs{\Sausage[\delta]{\sect}}^{\alpha}
\\
&+\sum_{\sect\in\Proj{\in}{\widehat{\FKconfig}}\setminus\FKconfig}\abs{\Sausage{\sect}}^{1+\alpha}+\sum_{\sect\in\Proj{\cap}{\widehat{\FKconfig}}\setminus\FKconfig}\abs{\Sausage{\sect}}^{\alpha}
\\
&+\sum_{\sect\in\Proj{\cap}{\widehat{\FKconfig}}}\carac{\exists k\in\intgff{-n;n},~\prth{\PermutationFK{\cdot}}^k(\sect)\notin\FKconfig'}.
\end{align*}
Since
\begin{align*}
	&\Proj{\in}{\widehat{\FKconfig}}\subseteq\Proj{\in}{\FKconfig}\cup\MathSet{\sect_{x,0}^{\textnormal{fol}},~x\in\partial^{\textnormal{fol}}\FKconfig}
	\\
	&\Proj{\cap}{\widehat{\FKconfig}}\subseteq\FKconfig\cup\MathSet{\sect_{x,0}^{\textnormal{fol}},~x\in\partial^{\textnormal{fol}}\FKconfig}\cup\MathSet{\sect_{y,0}^{\textnormal{pre}},~y\in\partial^{\textnormal{pre}}\FKconfig}
\end{align*}
we deduce
\begin{align*}
\leqslant&~1+\sum_{\sect\in\Proj{\in}{\FKconfig}}\abs{\Sausage[\delta]{\sect}}^{1+\alpha}+\sum_{\sect\in\FKconfig}\abs{\Sausage[\delta]{\sect}}^{\alpha}
\\
&+2\sum_{x\in\partial^{\textnormal{fol}}\FKconfig}\prth{\abs{\Sausage{\sect_{x,0}^{\textnormal{fol}}}}^{1+\alpha}+\abs{\Sausage{\sect_{x,0}^{\textnormal{fol}}}}^{\alpha}}
+2\sum_{y\in\partial^{\textnormal{pre}}\FKconfig}\abs{\Sausage{\sect_{y,0}^{\textnormal{pre}}}}^{\alpha}
\\
&+\prth{\#\FKconfig+\#\partial^{\textnormal{fol}}\FKconfig+\#\partial^{\textnormal{pre}}\FKconfig}.
\end{align*}
Given the definition of $\widehat{\FKconfig}$, we can further bound by
\begin{align*}
\leqslant&~1+\sum_{\sect\in\Proj{\in}{{\FKconfig}}}\abs{\Sausage[\delta]{\sect}}^{1+\alpha}+\sum_{\sect\in\FKconfig}\prth{1+\abs{\Sausage[\delta]{\sect}}^{\alpha}}
\\
&+\#\FKconfig+\prth{1+2\prth{c_{d-1}\delta^{d-1}R+c_d\delta^d+1}^{1+\alpha}+2\prth{c_{d-1}\delta^{d-1}R+c_d\delta^d+1}^{\alpha}}\#\partial^{\textnormal{fol}}\FKconfig
\\
&\qquad~~+\prth{1+2\prth{c_{d-1}\delta^{d-1}R+c_d\delta^d+1}^{\alpha}}\#\partial^{\textnormal{pre}}\FKconfig.
\end{align*}
The number of points lacking a follow-up or precursor bridge is at most $\#\FKconfig$. Therefore
\[
a\abs{g(\FKconfig)}\leqslant1+\sum_{\sect\in\Proj{\in}{{\FKconfig}}}\abs{\Sausage[\delta]{\sect}}^{1+\alpha}+\sum_{\sect\in\FKconfig}\prth{C+\abs{\Sausage[\delta]{\sect}}^{\alpha}}
\]
where $C=3+2\prth{c_{d-1}\delta^{d-1}R+c_d\delta^d+1}^{1+\alpha}+4\prth{c_{d-1}\delta^{d-1}R+c_d\delta^d+1}^{\alpha}$.

The volume of a sausage is bounded from below by $c_d\delta^d$ so we can conclude
\[
b\abs{g(\FKconfig)}\leqslant1+\sum_{\sect\in\Proj{\in}{{\FKconfig}}}\abs{\Sausage[\delta]{\sect}}^{1+\alpha}+\sum_{\sect\in\FKconfig}\abs{\Sausage[\delta]{\sect}}^{\alpha}
\]
where $b=a\prth{1+\frac{C}{c_{d}^{\alpha}\delta^{d\alpha}}}^{-1}$.
\end{Demo}

We can now prove the second part of \Cref{th_limit_FK}, that is, for any $\cap^n$-Lipschitz function $\MyFunction{f}{\ConfSpacePerm{FK}}{\RR}{}{}$,
\[
\lim_{L\to+\infty}\int f\d\EmpiricalField{FK}*=\int f\d\Model*{FK}*.
\]

\begin{Demo}[\Cref{th_limit_FK}, 2/2]\label{demo_th_limit_FK_2}
The class of $\cap^n$-Lipschitz functions is increasing in $n$, so we can assume without any loss of generality that $n\geqslant1$.

Let $\varepsilon,a,\delta>0$, $\alpha\in\interfo{0;1}$ and a compact $\Delta$. We denote $D=\sup_{x\in\Delta}\norm{x}$. According to \Cref{prop_equiint_cap_FK}, there exists $m_*\geqslant 1$ such that for any $L\geqslant4\max\prth{D,\delta,\frac{1}{4}}$,
\begin{align*}
&\int\EmpiricalField{FK}{\d\FKconfig}
\sum_{\sect\in\Proj{\cap}{\FKconfig}}
\carac{\sect(0)\notin\window[m_*]}\abs{\Sausage{\sect}}^{\alpha}\leqslant\varepsilon
\\
&\int\Model*{FK}{\d\FKconfig}
\sum_{\sect\in\Proj{\cap}{\FKconfig}}
\carac{\sect(0)\notin\window[m_*]}\abs{\Sausage{\sect}}^{\alpha}\leqslant\varepsilon.
\end{align*}
We can assume without any loss of generality that $\Delta\subseteq\window[m_*]$ and $\delta$ is large enough so that $\abs{B_{\delta}}^{\alpha}\geqslant1$.

According to \Cref{prop_equiint_FK}, there also exists $m_0\geqslant1$ such that for any $L>0$
\begin{align*}
	&\int\EmpiricalField{FK}{\d\FKconfig}~\Thresh[\prth{c_{d-1}\delta^{d-1}\prth{m_0-m_*}/2}^{\alpha+1}]{\sum_{\sect\in\FKconfig,~\sect(0)\in\window[m_*]}\abs{\Sausage{\sect}}^{\alpha+1}}\leqslant\frac{\varepsilon}{2}
	\\
	&\int\EmpiricalField{FK}{\d\FKconfig}~\Thresh[\prth{c_{d-1}\delta^{d-1}\prth{m_0-m_*}/2}^{\alpha}]{\sum_{\sect\in\FKconfig,~\sect(0)\in\window[m_*]}\abs{\Sausage{\sect}}^{\alpha}}\leqslant\frac{\varepsilon}{2}
	\\
	&\int\Model*{FK}{\d\FKconfig}~\Thresh[\prth{c_{d-1}\delta^{d-1}\prth{m_0-m_*}/2}^{\alpha+1}]{\sum_{\sect\in\FKconfig,~\sect(0)\in\window[m_*]}\abs{\Sausage{\sect}}^{\alpha+1}}\leqslant\frac{\varepsilon}{2}
	\\
	&\int\Model*{FK}{\d\FKconfig}~\Thresh[\prth{c_{d-1}\delta^{d-1}\prth{m_0-m_*}/2}^{\alpha}]{\sum_{\sect\in\FKconfig,~\sect(0)\in\window[m_*]}\abs{\Sausage{\sect}}^{\alpha}}\leqslant\frac{\varepsilon}{2}.
\end{align*}
For any $x_1\dots x_n\geqslant0$, it is clear that for any $b>0$
\[
\sum_{i=1}^n\carac{x_i\geqslant b}~x_i\leqslant\Thresh[b]{\sum_{i=1}^{n}x_i}.
\]
Then
\begin{align*}
	&\int\EmpiricalField{FK}{\d\FKconfig}
	\sum_{\sect\in\FKconfig,~\sect(0)\in\window[m_*]}\carac{\abs{\Sausage{\sect}}\geqslant c_{d-1}\delta^{d-1}\prth{m_0-m_*}/2}\prth{\abs{\Sausage{\sect}}^{\alpha+1}+\abs{\Sausage{\sect}}^{\alpha}}\leqslant\varepsilon
	\\
	&\int\Model*{FK}{\d\FKconfig}
	\sum_{\sect\in\FKconfig,~\sect(0)\in\window[m_*]}\carac{\abs{\Sausage{\sect}}\geqslant c_{d-1}\delta^{d-1}\prth{m_0-m_*}/2}\prth{\abs{\Sausage{\sect}}^{\alpha+1}+\abs{\Sausage{\sect}}^{\alpha}}\leqslant\varepsilon.
\end{align*}

Recursively, according to \Cref{prop_equiint_FK} again, there exists a family of positive $m_i,~1\leqslant i\leqslant n$ such that for any $i$ and $L>0$,
\begin{align*}
&\int\EmpiricalField{FK}{\d\FKconfig}
\sum_{\sect\in\FKconfig,~\sect(0)\in\window[m_0+\dots+m_{i-1}]}\carac{\abs{\Sausage{\sect}}\geqslant c_{d-1}\delta^{d-1}m_i/2}\leqslant\varepsilon
\\
&\int\Model*{FK}{\d\FKconfig}
\sum_{\sect\in\FKconfig,~\sect(0)\in\window[m_0+\dots+m_{i-1}]}\carac{\abs{\Sausage{\sect}}\geqslant c_{d-1}\delta^{d-1}m_i/2}\leqslant\varepsilon.
\end{align*}

For any $\FKconfig\in\ConfSpacePerm{FK}$, we define
\begin{align*}
\FKconfig_m^0&\defequal\MathSet{\sect\in\Proj{\cap}{\FKconfig}}[\sect\subset\window[m_0]]
\\
\FKconfig_m^+&\defequal\MathSet{\sect\in\Proj{\cap^n}{\FKconfig}}[
\exists0<q\leqslant n,~
\begin{array}{ll}
&\PermutationFK{\cdot}^{-q}(\sect)\cap\Delta\neq\emptyset\\
\forall0\leqslant i<q,~&\PermutationFK{\cdot}^{-i}(\sect)\cap\Delta=\emptyset\\
\forall0\leqslant i\leqslant q,~&\PermutationFK{\cdot}^{-q+i}(\sect)\subset\window[m_0+\dots+m_i]
\end{array}
]
\\
\FKconfig_m^-&\defequal\MathSet{\sect\in\Proj{\cap^n}{\FKconfig}}[
\exists0<q\leqslant n,~
\begin{array}{ll}
	&\PermutationFK{\cdot}^{q}(\sect)\cap\Delta\neq\emptyset\\
	\forall0\leqslant i<q,~&\PermutationFK{\cdot}^{i}(\sect)\cap\Delta=\emptyset\\
	\forall0\leqslant i\leqslant q,~&\PermutationFK{\cdot}^{q-i}(\sect)\subset\window[m_0+\dots+m_i]
\end{array}
]
\\
\FKconfig_m&\defequal\FKconfig_m^0\cup\FKconfig_m^+\cup\FKconfig_m^-.
\end{align*}

Let us denote $\MyFunction{g}{\ConfSpace{FK}}{\RR}{}{}$ the extension of $f$ described in \Cref{lem_inter_n_lipschitz}. We then define $\MyFunction{h}{\ConfSpacePerm{FK}}{\RR}{}{}$ by
\begin{equation*}
h\prth{\FKconfig}\defequal g\prth{\FKconfig_m}.
\end{equation*}
This makes $h$ $\in$-local relatively to the compact $\window[m_0+\dots+m_n]$ and $\in$-tame relatively to the same compact (see the second part of \Cref{lem_inter_n_lipschitz}). Therefore, according to the first part of \Cref{th_limit_FK},
\[
\lim_{L\to+\infty}\int h\d\EmpiricalField{FK}*=
\int h\d\Model*{FK}*.
\]
It remains to prove $\int\abs{f-h}$ is uniformly small under $\Model*{FK}*$ and the $\EmpiricalField{FK}*$ ($L>0$).

Let $\FKconfig\in\ConfSpacePerm{FK}$. According to \Cref{lem_inter_n_lipschitz},
\begin{align*}
a\abs{f(\FKconfig)-h(\FKconfig)}=&~a\abs{g(\FKconfig)-g(\FKconfig_m)}
\\
\leqslant&
\sum_{\sect\in\Proj{\in}{\FKconfig}}\carac{\sect\notin\FKconfig_m}\abs{\Sausage{\sect}}^{\alpha+1}
+\sum_{\sect\in\Proj{\cap}{\FKconfig}}\carac{\sect\notin\FKconfig_m}\abs{\Sausage{\sect}}^{\alpha}
\\
&+\sum_{\sect\in\Proj{\cap}{\FKconfig}}\carac{\exists k\in\intgff{-n;n},~\prth{\PermutationFK{\cdot}}^k(\sect)\notin\FKconfig_m}
\\
\\
\leqslant&
\sum_{\sect\in\Proj{\in}{\FKconfig}}\carac{\sect\notin\FKconfig_m}\abs{\Sausage{\sect}}^{\alpha+1}
+2\sum_{\sect\in\Proj{\cap}{\FKconfig}}\carac{\sect\notin\FKconfig_m}\abs{\Sausage{\sect}}^{\alpha}
\\
&+\sum_{\sect\in\Proj{\cap}{\FKconfig}}\sum_{k=1}^{n}\carac{\prth{\PermutationFK{\cdot}}^k(\sect)\notin\FKconfig_m\text{ and }\prth{\PermutationFK{\cdot}}^{k-1}(\sect)\in\FKconfig_m}
\\
&+\sum_{\sect\in\Proj{\cap}{\FKconfig}}\sum_{k=1}^{n}\carac{\prth{\PermutationFK{\cdot}}^{-k}(\sect)\notin\FKconfig_m\text{ and }\prth{\PermutationFK{\cdot}}^{-k+1}(\sect)\in\FKconfig_m}.
\end{align*}

Thanks to \Cref{prop_time_reversal}, we know the probability measures $\Model*{FK}*$ and $\EmpiricalField{FK}*$ ($L>0$) are time-reversal invariant. Yet for any probability measure $P$ on $\ConfSpacePerm{FK}$ which has this property,
\begin{align*}
&\int P(\d\FKconfig)~\sum_{\sect\in\Proj{\cap}{\FKconfig}}\sum_{k=1}^{n}\carac{\prth{\PermutationFK{\cdot}}^{-k}(\sect)\notin\FKconfig_m\text{ and }\prth{\PermutationFK{\cdot}}^{-k+1}(\sect)\in\FKconfig_m }
\\
=&\int P(\d\FKconfig)~\sum_{\sect\in\Proj{\cap}{\FKconfig}}\sum_{k=1}^{n}\carac{\prth{\PermutationFK{\cdot}}^k(\sect)\notin\FKconfig_m\text{ and }\prth{\PermutationFK{\cdot}}^{k-1}(\sect)\in\FKconfig_m}.
\end{align*}
So from now on we are studying the following bound
\begin{align}\label{eq_ext_bound}
&\sum_{\sect\in\Proj{\in}{\FKconfig}}\carac{\sect\notin\FKconfig_m}\abs{\Sausage{\sect}}^{\alpha+1}
+2\sum_{\sect\in\Proj{\cap}{\FKconfig}}\carac{\sect\notin\FKconfig_m}\abs{\Sausage{\sect}}^{\alpha}\nonumber
\\
&+2\sum_{\sect\in\Proj{\cap}{\FKconfig}}\sum_{k=1}^{n}\carac{\prth{\PermutationFK{\cdot}}^k(\sect)\notin\FKconfig_m\text{ and }\prth{\PermutationFK{\cdot}}^{k-1}(\sect)\in\FKconfig_m}.
\end{align}

Let $\sect\in\Proj{\cap}{\FKconfig}$ and $k\in\intgff{1;n}$ be such that $\prth{\PermutationFK{\cdot}}^{k}(\sect)\notin\FKconfig_m$ and $\prth{\PermutationFK{\cdot}}^{k-1}(\sect)\in\FKconfig_m$.
\begin{description}
\item[Case 1] We assume $\prth{\PermutationFK{\cdot}}^{k}(\sect)\cap\Delta\neq\emptyset$.

By the change of variables $\sect'\defequal\prth{\PermutationFK{\cdot}}^k(\sect)$, we can rewrite
\begin{align*}
&\sum_{\sect\in\Proj{\cap}{\FKconfig}}~\sum_{k=1}^{n}\carac{\prth{\PermutationFK{\cdot}}^k(\sect)\notin\FKconfig_m
\text{ and }
\prth{\PermutationFK{\cdot}}^{k-1}(\sect)\in\FKconfig_m
\text{ and }
\prth{\PermutationFK{\cdot}}^{k}(\sect)\cap\Delta\neq\emptyset
}
\\
\leqslant&~\sum_{k=1}^{n}~\sum_{\sect'\in\Proj{\cap}{\FKconfig}}
\carac{\sect'\notin\FKconfig_m\text{ and }\PermutationFK{\sect'}\in\FKconfig_m\text{ and }\prth{\PermutationFK{\cdot}}^{-k}(\sect')\cap\Delta\neq\emptyset}
\\
\leqslant&~n\sum_{\sect'\in\Proj{\cap}{\FKconfig}}
\carac{\sect'\notin\FKconfig_m}.
\end{align*}

\item[Case 2] We assume $\prth{\PermutationFK{\cdot}}^{k}(\sect)\cap\Delta=\emptyset$.
\begin{description}
	\item[Subcase 2a] We assume $\prth{\PermutationFK{\cdot}}^{k-1}(\sect)\in\FKconfig_m^-$. By definition, this means there exists $1\leqslant q\leqslant n$ such that
	\begin{align*}
		&\prth{\PermutationFK{\cdot}^{q}\circ\PermutationFK{\cdot}^{k-1}}(\sect)\cap\Delta\neq\emptyset
		\\
		\forall 0\leqslant i<q,~
		&\prth{\PermutationFK{\cdot}^{i}\circ\PermutationFK{\cdot}^{k-1}}(\sect)\cap\Delta=\emptyset
		\\
		\forall 0\leqslant i\leqslant q,~
		&\prth{\PermutationFK{\cdot}^{q-i}\circ\PermutationFK{\cdot}^{k-1}}(\sect)\subset\window[m_0+\dots+m_i].
	\end{align*}
	Since $\prth{\PermutationFK{\cdot}}^k(\sect)\cap\Delta=\emptyset$, we know from the first line that $q\geqslant2$. Then,
	\begin{align*}
		&\prth{\PermutationFK{\cdot}^{q-1}\circ\PermutationFK{\cdot}^{k}}(\sect)\cap\Delta\neq\emptyset
		\\
		\forall 0\leqslant i<q-1,~
		&\prth{\PermutationFK{\cdot}^{i}\circ\PermutationFK{\cdot}^{k}}(\sect)\cap\Delta=\emptyset
		\\
		\forall 0\leqslant i\leqslant q-1,~
		&\prth{\PermutationFK{\cdot}^{q-1-i}\circ\PermutationFK{\cdot}^{k}}(\sect)\subset\window[m_0+\dots+m_i]
	\end{align*}
	which contradicts the fact $\prth{\PermutationFK{\cdot}}^{k}(\sect)\notin\FKconfig_m$ so this subcase is impossible.
	
	\item[Subcase 2b] We assume $\prth{\PermutationFK{\cdot}}^{k-1}(\sect)\in\FKconfig_m^0$. Then
	\begin{align*}
		&\prth{\PermutationFK{\cdot}^{-1}\circ\PermutationFK{\cdot}^k}(\sect)\cap\Delta\neq\emptyset
		\\
		&\prth{\PermutationFK{\cdot}}^k(\sect)\cap\Delta=\emptyset
		\\
		&\prth{\PermutationFK{\cdot}^{-1}\circ\PermutationFK{\cdot}^k}(\sect)\subset\window[m_0]
	\end{align*}
	The fact $\prth{\PermutationFK{\cdot}}^k(\sect)\notin\FKconfig_m$ implies $\prth{\PermutationFK{\cdot}}^k(\sect)\not\subset\window[m_0+m_1]$. So in subcase 2b we can state
	\begin{align*}
		&\sum_{\sect\in\Proj{\cap}{\FKconfig}}\sum_{k=1}^{n}\carac{\prth{\PermutationFK{\cdot}}^k(\sect)\notin\FKconfig_m
			\text{ and }
			\prth{\PermutationFK{\cdot}}^{k-1}(\sect)\in\FKconfig_m^0
			\text{ and }
			\prth{\PermutationFK{\cdot}}^k(\sect)\cap\Delta=\emptyset
		}
		\\
		\leqslant&
		\sum_{\sect\in\Proj{\cap}{\FKconfig}}\sum_{k=1}^{n}\carac{\prth{\PermutationFK{\cdot}}^{k-1}(\sect)\subset\window[m_0]\text{ and }\prth{\PermutationFK{\cdot}}^{k}(\sect)\not\subset\window[m_0+m_{1}]}.
	\end{align*}
	By the change of variables $\sect'\defequal\prth{\PermutationFK{\cdot}}^{k}(\sect)$,
	\begin{align*}
		\leqslant&
		~\sum_{k=1}^{n}\sum_{\underset{\prth{\PermutationFK{\cdot}}^{-1}(\sect')\subset\window[m_0]}{\sect'\in\FKconfig}}\carac{\sect'\not\subset\window[m_0+m_{1}]\text{ and }\prth{\PermutationFK{\cdot}}^{-k}(\sect')\cap\Delta\neq\emptyset}
		\\
		\leqslant&
		~n\sum_{\underset{\prth{\PermutationFK{\cdot}}^{-1}(\sect')\subset\window[m_0]}{\sect'\in\FKconfig}}\carac{\sect'\not\subset\window[m_0+m_{1}]}
		\\
		\leqslant&
		~n\sum_{\underset{\sect'(0)\in\window[m_0]}{\sect'\in\FKconfig}}\carac{\sect'\not\subset\window[m_0+m_{1}]}.
	\end{align*}
	
	\item[Subcase 2c] We assume $\prth{\PermutationFK{\cdot}}^{k-1}(\sect)\in\FKconfig_m^+$. Then there exists $1\leqslant q\leqslant n$ such that
	\begin{align*}
		&\prth{\PermutationFK{\cdot}^{-q}\circ\PermutationFK{\cdot}^{k-1}}(\sect)\cap\Delta\neq\emptyset
		\\
		\forall 0\leqslant i<q,~
		&\prth{\PermutationFK{\cdot}^{-i}\circ\PermutationFK{\cdot}^{k-1}}(\sect)\cap\Delta=\emptyset
		\\
		\forall 0\leqslant i\leqslant q,~
		&\prth{\PermutationFK{\cdot}^{i-q}\circ\PermutationFK{\cdot}^{k-1}}(\sect)\subset\window[m_0+\dots+m_i].
	\end{align*}
	Since $q$ is the smallest integer satisfying the first inequality above and $\sect\cap\Delta\neq\emptyset$, we know $q\leqslant k-1$. Hence $q\leqslant n-1$. And since $\prth{\PermutationFK{\cdot}}^k(\sect)\cap\Delta=\emptyset$, we deduce
	\begin{align*}
		&\prth{\PermutationFK{\cdot}^{-(q+1)}\circ\PermutationFK{\cdot}^{k}}(\sect)\cap\Delta\neq\emptyset
		\\
		\forall 0\leqslant i<q+1,~
		&\prth{\PermutationFK{\cdot}^{-i}\circ\PermutationFK{\cdot}^{k}}(\sect)\cap\Delta=\emptyset
		\\
		\forall 0\leqslant i\leqslant q,~
		&\prth{\PermutationFK{\cdot}^{i-(q+1)}\circ\PermutationFK{\cdot}^k}(\sect)\subset\window[m_0+\dots+m_i].
	\end{align*}
	The fact $\prth{\PermutationFK{\cdot}}^k(\sect)\notin\FKconfig_m$ implies then $\prth{\PermutationFK{\cdot}}^k(\sect)\not\subset\window[m_0+\dots+m_{q+1}]$. So in subcase 2c we can state
	\begin{align*}
		&\sum_{\sect\in\Proj{\cap}{\FKconfig}}\sum_{k=1}^{n}\carac{\prth{\PermutationFK{\cdot}}^k(\sect)\notin\FKconfig_m
			\text{ and }
			\prth{\PermutationFK{\cdot}}^{k-1}(\sect)\in\FKconfig_m^+
			\text{ and }
			\prth{\PermutationFK{\cdot}}^k(\sect)\cap\Delta=\emptyset
		}
		\\
		\leqslant&
		\sum_{\sect\in\Proj{\cap}{\FKconfig}}\sum_{k=1}^{n}\sum_{q=1}^{n-1}\carac{\prth{\PermutationFK{\cdot}}^{k-1}(\sect)\subset\window[m_0+\dots+m_{q}]\text{ and }\prth{\PermutationFK{\cdot}}^{k}(\sect)\not\subset\window[m_0+\dots+m_{q+1}]}.
	\end{align*}
	By the change of variables $\sect'\defequal\prth{\PermutationFK{\cdot}}^{k}(\sect)$,
	\begin{align*}
		\leqslant&
		\sum_{k=1}^{n}\sum_{q=1}^{n-1}\sum_{\underset{\prth{\PermutationFK{\cdot}}^{-1}(\sect')\subset\window[m_0+\dots+m_{q}]}{\sect'\in\FKconfig}}\carac{\sect'\not\subset\window[m_0+\dots+m_{q+1}]\text{ and }\prth{\PermutationFK{\cdot}}^{-k}(\sect')\cap\Delta\neq\emptyset}
		\\
		\leqslant&
		~n\sum_{q=1}^{n-1}\sum_{\underset{\sect'(0)\in\window[m_0+\dots+m_{q}]}{\sect'\in\FKconfig}}\carac{\sect'\not\subset\window[m_0+\dots+m_{q+1}]}.
	\end{align*}
\end{description}
\end{description}

Therefore, in all cases we know
\begin{align}\label{eq_ext_1}
	&\sum_{\sect\in\Proj{\cap}{\FKconfig}}\sum_{k=1}^{n}\carac{\prth{\PermutationFK{\cdot}}^k(\sect)\notin\FKconfig_m
		\text{ and }
		\prth{\PermutationFK{\cdot}}^{k-1}(\sect)\in\FKconfig_m}\nonumber
	\\
	\leqslant&
	~n\sum_{\sect'\in\Proj{\cap}{\FKconfig}}
	\carac{\sect'\notin\FKconfig_m}
	+n\sum_{q=0}^{n-1}\sum_{\underset{\sect'(0)\in\window[m_0+\dots+m_{q}]}{\sect'\in\FKconfig}}\carac{\sect'\not\subset\window[m_0+\dots+m_{q+1}]}.
\end{align}

Let $\sect\in\Proj{\cap}{\FKconfig}$. If $\sect\notin\FKconfig_m$ then $\sect\not\subset\window[m_0]$, and either $\sect(0)$ is in $\window[m_*]$ either it is not:
\begin{align}\label{eq_ext_2}
\sum_{\sect\in\Proj{\cap}{\FKconfig}}\carac{\sect\notin\FKconfig_m}\abs{\Sausage{\sect}}^{\alpha}
\leqslant
&~\sum_{\sect\in\Proj{\in}[\window[m_*]]{\FKconfig}}\carac{\sect\not\subset\window[m_0]}\abs{\Sausage{\sect}}^{\alpha}\nonumber
\\
&+\sum_{\sect\in\Proj{\cap}{\FKconfig}}\carac{\sect(0)\notin\window[m_*]}\abs{\Sausage{\sect}}^{\alpha}.
\end{align}
By combining inequations (\ref{eq_ext_1}) and (\ref{eq_ext_2}) with (\ref{eq_ext_bound}), our bound becomes
\begin{align*}
	a\int\abs{f-h}\d P\leqslant\int P(\d\FKconfig)~
	&~(2n+3)\sum_{\sect\in\Proj{\in}[\window[m_*]]{\FKconfig}}\carac{\sect\not\subset\window[m_0]}\prth{\abs{\Sausage{\sect}}^{1+\alpha}+\abs{\Sausage{\sect}}^{\alpha}}
	\\
	&+(2n+2)\sum_{\sect\in\Proj{\cap}{\FKconfig}}\carac{\sect(0)\notin\window[m_*]}\abs{\Sausage{\sect}}^{\alpha}
	\\
	&+2n\sum_{q=0}^{n-1}\sum_{\underset{\sect'(0)\in\window[m_0+\dots+m_{q}]}{\sect'\in\FKconfig}}\carac{\sect'\not\subset\window[m_0+\dots+m_{q+1}]}.
\end{align*}
for any $P$ among $\EmpiricalField{FK}*$ ($L>0$) or $\Model*{FK}*$.

Let $\sect\in\Proj{\in}[\window[m_*]]{\FKconfig}$. If $\sect\not\subset\window[m_0]$ then there exists $s\in\interff{0;\beta}$ such that $\norm{\sect(0)-\sect(s)}\geqslant\frac{m_0-m_*}{2}$. Therefore\footnote{This is a reasoning similar to \cpageref{reasoning_sausage} in the proof of \Cref{prop_equiint_cap_FK}.}
\[
\abs{\Sausage{\sect}}\geqslant\abs{\Sausage{\restrict{\sect}{\interff{0;s}}}}\geqslant c_{d-1}\delta^{d-1}\frac{m_0-m_*}{2}.
\]
So the conditions in the indicator functions can be weakened as
\begin{align*}
\leqslant\int P(\d\FKconfig)~&
~(2n+3)\sum_{\underset{\sect(0)\in\window[m_*]}{\sect\in\FKconfig}}\carac{\abs{\Sausage{\sect}}\geqslant c_{d-1}\delta^{d-1}(m_0-m_*)/2}\prth{\abs{\Sausage{\sect}}^{\alpha+1}+\abs{\Sausage{\sect}}^{\alpha}}
\\
&+(2n+2)\sum_{\sect\in\Proj{\cap}{\FKconfig}}\carac{\sect(0)\notin\window[m_*]}\abs{\Sausage{\sect}}^{\alpha}
\\
&+n\sum_{q=0}^{n-1}\sum_{\underset{\sect(0)\in\window[m_0+\dots+m_{q}]}{\sect\in\FKconfig}}\carac{\abs{\Sausage{\sect}}\geqslant c_{d-1}\delta^{d-1}m_{q+1}/2}.
\end{align*}

Given the definitions of $m_*,m_0\dots m_n$, all these terms are uniformly small under $\EmpiricalField{FK}*$ (for $L$ large enough) and $\Model*{FK}*$. Thus
\[
\int\abs{f-h}\d\EmpiricalField{FK}*\leqslant\frac{\varepsilon}{a}\prth{\prth{2n+3}+\prth{2n+2}+n^2}
\]
and
\[
\int\abs{f-h}\d\Model*{FK}*\leqslant\frac{\varepsilon}{a}\prth{\prth{2n+3}+\prth{2n+2}+n^2}
\]
for $L$ large enough. So finally
\[
\abs{\int f\d\EmpiricalField{FK}*-
\int f\d\Model*{FK}*}\underset{L\to+\infty}{\leqslant} \frac{2\varepsilon}{a}\prth{n^2+4n+5}+o(1).
\]
\end{Demo}

\subsection{Technical Lemmas}

We now prove any limiting model $\Model*{FK}*$ to be Gibbs, that is to say, to be solution of DLR equations. From now on, we are assuming all hypotheses I to VI.

Most of the technicalities in this section are due to the fact we are stating DLR equations over trajectories rather than points. Indeed, cutting down the configuration into interior and exterior configurations respectively to some compact $\Delta$ is not as straightforward as for points.

Firstly, we make use below of \Cref{hyp_finite_range,hyp_boundedly_attractive} to justify $\Hamconditional*$ is well defined and write the Hamiltonian $\Ham{FK}*$ as a sum of two distinct terms. Concerning stationarity, its necessity is linked to the empirical field. The probability $\EmpiricalField{FK}*$ can be seen as sampling a configuration along $\Model{FK}*$ with the extra randomness of not knowing where is the origin in $\window$. To write DLR equations in infinite volume, we need some coherence between the distribution of the empirical field and the original model, and stationarity ensures that sampling a configuration inside a compact $\Delta$ is the same as sampling a configuration inside $\Delta+v$, then shifting it by $-v$. This is intuitive, but not guaranteed without stationarity.

\begin{Demo}[\Cref{rq_ham_local}]\label{demo_rq}
Let $\finiteconfig\in\ConfFinite$. We denote $N=\#\prth{\finiteconfig\cap\prth{\Delta+B_R}}$.
\begin{description}
\item[Case 1] $\forall\finiteconfig'\in\ConfFinite,~\finiteconfig\cap\prth{\Delta+B_R}=\finiteconfig'\cap\prth{\Delta+B_R}\implies\InteractionFree{\finiteconfig'\cap\Delta\complementary}=+\infty$

Then the equality defining $\InteractionLocal*$ from \Cref{hyp_finite_range} becomes $+\infty=+\infty$. In other words, we can assume without any loss of generality that $\InteractionLocal{\finiteconfig\cap\prth{\Delta+B_R}}=C_{\Delta,N}$.

\item[Case 2]
$\exists\finiteconfig'\in\ConfFinite,~\finiteconfig\cap\prth{\Delta+B_R}=\finiteconfig'\cap\prth{\Delta+B_R}\text{ and } \InteractionFree{\finiteconfig'\cap\Delta\complementary}<+\infty$

Then for any such $\finiteconfig'$, the equality defining $\InteractionLocal*$ from \Cref{hyp_finite_range} becomes
\[
\InteractionLocal{\finiteconfig\cap\prth{\Delta+B_R}}=\InteractionFree{\finiteconfig'}-\InteractionFree{\finiteconfig'\cap\Delta\complementary}\geqslant C_{\Delta,N}
\]
according to \Cref{hyp_boundedly_attractive}.
\end{description}
\end{Demo}

Then, we need to be able to cut down the permutation part of $\Model{FK}*$ into the shuffling of interior points and exterior points respectively.

\begin{Lem}\label{lem_permutation}
Let $X$ be a finite set and $Y\subseteq X$. For any function $\MyFunction{f}{\PermutationSpace{X}}{\RR}{}{}$,
\[
\sum_{\PermutationSymbol\in\PermutationSpace{X}}f(\PermutationSymbol)
=
\sum_{Z_1\subseteq Y}~\sum_{Z_2\subseteq X\setminus Y}~
\sum_{\PermutationSymbol^{\textnormal{int}}\in\PermutationSpace{Y\to Z_2\cup (Y\setminus Z_1)}}~
\sum_{\PermutationSymbol^{\textnormal{ext}}\in\PermutationSpace{X\setminus Y\to Z_1\cup X\setminus\prth{Z_2\cup Y}}}
f\prth{\PermutationSymbol^{\textnormal{int}}\cup\PermutationSymbol^{\textnormal{ext}}}
\]
where $\PermutationSymbol^{\textnormal{int}}\cup\PermutationSymbol^{\textnormal{ext}}$ is a notation for the permutation induced on the whole set $X$ by the \emph{interior} (relatively to $Y$) bijection $\PermutationSymbol^{\textnormal{int}}$ and the \emph{exterior} bijection $\PermutationSymbol^{\textnormal{ext}}$.
\end{Lem}

\begin{Demo}[\Cref{lem_permutation}]
For any $Z_1,Z_2\subseteq X\setminus Y$, we denote
\[
\SS_{X,Y,Z_1,Z_2}\defequal
\MathSet{\PermutationSymbol\in\PermutationSpace{X}}[
\begin{array}{l}
Z_1=\PermutationSymbol(X\setminus Y)\cap Y\\
Z_2=\PermutationSymbol(Y)\setminus Y
\end{array}
].
\]
Then we can define the natural map
\[
\MyFunction{\varphi_{X,Y,Z_1,Z_2}}{\SS_{X,Y,Z_1,Z_2}}{
\PermutationSpace{Y\to Z_2\cup \prth{Y\setminus Z_1}}
\times
\PermutationSpace{X\setminus Y\to Z_1\cup X\setminus\prth{Z_2\cup Y}}
}{\PermutationSymbol}{
\prth{\PermutationSymbol^{\textnormal{int}},\PermutationSymbol^{\textnormal{ext}}
}}
\]
by
\[
\MyFunction{\PermutationSymbol^{\textnormal{int}}}{Y}{Z_2\cup (Y\setminus Z_1)}{x}{\PermutationSymbol(x)}
\quad\text{ and }\quad
\MyFunction{\PermutationSymbol^{\textnormal{ext}}}{X\setminus Y}{Z_1\cup X\setminus(Z_2\cup Y)}{x}{\PermutationSymbol(x).}
\]
This map turns out to be bijective. Furthermore, the sets $\SS_{X,Y,Z_1,Z_2}$ for $Z_1\subseteq Y$ and $Z_2\subseteq X\setminus Y$ form a partition of $\PermutationSpace{X}$.
\end{Demo}

We need one last easy result about the sampling of subsets from a Poisson point process.

\begin{Lem}\label{lem_uniform_subset}
For any measurable $\MyFunction{f}{\ConfFinite^2}{\RR^+}{}{}$,
\[
\int\PoissonLeb[\Delta]{\d\finiteconfig}\sum_{\zeta\subseteq\finiteconfig}f\prth{\zeta,\finiteconfig\setminus\zeta}=\E^{\abs{\Delta}}\int\PoissonLeb[\Delta]{\d\finiteconfig_1}~\PoissonLeb[\Delta]{\d\finiteconfig_2}~f\prth{\finiteconfig_1,\finiteconfig_2}.
\]
\end{Lem}

\begin{Demo}[\Cref{lem_uniform_subset}]
For any $\finiteconfig\in\ConfFinite$,
\begin{align*}
\sum_{\zeta\subseteq\finiteconfig}f\prth{\zeta,\finiteconfig\setminus\zeta}
&=
\sum_{n=0}^{+\infty}~\sum_{\zeta\subseteq\finiteconfig,~\#\zeta=n}f\prth{\zeta,\finiteconfig\setminus\zeta}
\\
&=f\prth{\emptyset,\finiteconfig}+\sum_{n=1}^{+\infty}\frac{1}{n!}\sum_{x_1\in\finiteconfig}\dots\sum_{x_n\in\finiteconfig\setminus\{x_1\dots x_{n-1}\}}f\prth{\{x_1\dots x_n\},\finiteconfig\setminus\{x_1\dots x_n\}}.
\end{align*}
By Mecke formula (see Theorem 4.1 p. 27 in \autocite{LP17}),
\begin{align*}
\int\PoissonLeb[\Delta]{\d\finiteconfig}\sum_{\zeta\subseteq\finiteconfig}f\prth{\zeta,\finiteconfig\setminus\zeta}
&=
\int\PoissonLeb[\Delta]{\d\finiteconfig}\sum_{n=0}^{+\infty}\frac{1}{n!}\int_{\Delta^n}\d x^{\otimes n}f\prth{\{x_1\dots x_n\},\finiteconfig}
\\
&=\E^{\abs{\Delta}}\int\PoissonLeb[\Delta]{\d\finiteconfig}~\PoissonLeb[\Delta]{\d\zeta}~f\prth{\zeta,\finiteconfig}.
\end{align*}
\end{Demo}

\subsection{DLR Equations}

Now, we make use of the previous lemmas to introduce an auxiliary measure which takes care of all the exterior sampling (relatively to a compact $\Delta$).

\begin{Def}\label{def_Q}
Let $L>0$ and $\Delta$ be a compact.

For any $x,y\in\RR^d$, we define the measure
\[
\WienerMeasureXYTBC[x,y][\beta,\not\subset\Delta]*
\defequal
\WienerMeasureXYTBC[x,y][\beta]*
-\WienerMeasureXYT[x,y][\beta,\subset\Delta]*.
\]

We also define for any finite $\FKconfig\in\ConfSpace{FK}$ the exterior Hamiltonian as
\[
\Hamexterior{\FKconfig}\defequal\int_{0}^{\beta}\Interaction{\MathSet{\sect\prth{s},~\sect\in\FKconfig}\cap\Delta\complementary}.
\]

Finally, we define the measure $\QQ_{\Delta,L}$ over $\ConfSpace{FK}$ by
\begin{align*}
\QQ_{\Delta,L}(\d\FKconfig)\defequal
~\E^{2\abs{\Delta}}\int&\PoissonLeb[\window]{\d\finiteconfig}~\PoissonLeb[\Delta]{\d\zeta_1}~\E^{\beta\mu\#\prth{\finiteconfig\cup\zeta_1}}
\\
&\sum_{\zeta_2\subseteq\finiteconfig}~\sum_{\PermutationSymbol\exterior\in\PermutationSpace{\finiteconfig\to\zeta_1\cup(\finiteconfig\setminus\zeta_2)}}\prth{\bigotimes_{x\in\finiteconfig}\WienerMeasureXYTBC[x,\PermutationSymbol\exterior(x)][\beta,\not\subset\Delta]*}(\d\FKconfig)~\E^{-\Hamexterior{\FKconfig}}.
\end{align*}
\end{Def}

In the following proposition, we decompose the formula of $\Model{FK}*$ into an exterior and an interior sampling.

\begin{Prop}[non-normalized DLR equations]\label{Prop_QL}
Let $\Delta\subset\RR^d$ be a compact and $\MyFunction{f}{\ConfSpace{FK}}{\RR}{}{}$. Then
\begin{align*}
&\int\PoissonLeb{\d\finiteconfig}~\E^{\beta\mu\#\finiteconfig}\sum_{\PermutationSymbol\in\PermutationSpace{\finiteconfig}}\int\prth{\bigotimes_{x\in\finiteconfig}\WienerMeasureXYTBC[x,\PermutationSymbol(x)][\beta]*}(\d\FKconfig)~\E^{-\Ham{FK}{\FKconfig}}~f\prth{\FKconfig}
\\
=&\int\QQ_{\Delta,L}(\d\FKconfig)~\int\PoissonLeb[\Delta]{\d\zeta}~\E^{\beta\mu\#\zeta}~\sum_{\PermutationSymbol\interior\in\PermutationSpace{\inward\cup\zeta\to\outward\cup\zeta}}
\\
&\qquad\int\prth{\bigotimes_{x\in\inward\cup\zeta}\WienerMeasureXYT[x,\PermutationSymbol\interior(x)][\beta,\subset\Delta]*}\prth{\d\eta}~\E^{-\Hamconditional{\eta\cup\FKconfig\exterior}}~f\prth{\eta\cup\FKconfig\exterior}.
\end{align*}
\end{Prop}

\begin{Demo}[\Cref{Prop_QL}]
We rewrite the measure over Brownian bridges as
\begin{align*}
&\int\PoissonLeb{\d\finiteconfig}~\E^{\beta\mu\#\finiteconfig}\sum_{\PermutationSymbol\in\PermutationSpace{\finiteconfig}}\int\prth{\bigotimes_{x\in\finiteconfig}\WienerMeasureXYTBC[x,\PermutationSymbol(x)][\beta]*}(\d\FKconfig)~\E^{-\Ham{FK}{\FKconfig}}~f\prth{\FKconfig}
\\
=&\int\PoissonLeb{\d\finiteconfig}~\E^{\beta\mu\#\finiteconfig}\sum_{\PermutationSymbol\in\PermutationSpace{\finiteconfig}}\int\sum_{\zeta\subseteq\finiteconfig}\prth{\bigotimes_{x\in\zeta}\WienerMeasureXYT[x,\PermutationSymbol(x)][\beta,\subset\Delta]*}\prth{\d\eta}~\prth{\bigotimes_{x\in\finiteconfig\setminus\zeta}\WienerMeasureXYTBC[x,\PermutationSymbol(x)][\beta,\not\subset\Delta]*}\prth{\d\FKconfig}~\E^{-\Ham{FK}{\eta\cup\FKconfig}}~f\prth{\eta\cup\FKconfig}.
\end{align*}
We can restrict the summation over subsets of $\finiteconfig$ to subsets of $\finiteconfig_{\Delta}=\finiteconfig\cap\Delta$
\begin{align*}
=&\int\PoissonLeb[\window\setminus\Delta]{\d\finiteconfig_{\Delta\complementary}}~\PoissonLeb[\Delta]{\d\finiteconfig_{\Delta}}~\E^{\beta\mu\#\prth{\finiteconfig_{\Delta\complementary}\cup\finiteconfig_{\Delta}}}\sum_{\zeta\subseteq\finiteconfig_{\Delta}}~\sum_{\PermutationSymbol\in\PermutationSpace{\finiteconfig_{\Delta\complementary}\cup\finiteconfig_{\Delta}}}
\\
&\int\prth{\bigotimes_{x\in\zeta}\WienerMeasureXYT[x,\PermutationSymbol(x)][\beta,\subset\Delta]*}\prth{\d\eta}~\prth{\bigotimes_{x\in\finiteconfig_{\Delta\complementary}\cup\finiteconfig_{\Delta}\setminus\zeta}\WienerMeasureXYTBC[x,\PermutationSymbol(x)][\beta,\not\subset\Delta]*}\prth{\d\FKconfig}~\E^{-\Ham{FK}{\eta\cup\FKconfig}}~f\prth{\eta\cup\FKconfig}.
\end{align*}
According to \Cref{lem_uniform_subset} this equals
\begin{align*}
=&~\E^{\abs{\Delta}}\int\PoissonLeb[\window\setminus\Delta]{\d\finiteconfig_{\Delta\complementary}}~\PoissonLeb[\Delta]{\d\finiteconfig'_{\Delta}}~\PoissonLeb[\Delta]{\d\zeta}~\E^{\beta\mu\#\prth{\finiteconfig_{\Delta\complementary}\cup\finiteconfig'_{\Delta}\cup\zeta}}~\sum_{\PermutationSymbol\in\PermutationSpace{\finiteconfig_{\Delta\complementary}\cup\finiteconfig'_{\Delta}\cup\zeta}}
\\
&\int\prth{\bigotimes_{x\in\zeta}\WienerMeasureXYT[x,\PermutationSymbol(x)][\beta,\subset\Delta]*}\prth{\d\eta}~\prth{\bigotimes_{x\in\finiteconfig_{\Delta\complementary}\cup\finiteconfig'_{\Delta}}\WienerMeasureXYTBC[x,\PermutationSymbol(x)][\beta,\not\subset\Delta]*}\prth{\d\FKconfig}~\E^{-\Ham{FK}{\eta\cup\FKconfig}}~f\prth{\eta\cup\FKconfig}
\\
\\
=&~\E^{\abs{\Delta}}\int\PoissonLeb[\window]{\d\finiteconfig}~\PoissonLeb[\Delta]{\d\zeta}~\E^{\beta\mu\#\prth{\finiteconfig\cup\zeta}}~\sum_{\PermutationSymbol\in\PermutationSpace{\finiteconfig\cup\zeta}}
\\
&\int\prth{\bigotimes_{x\in\zeta}\WienerMeasureXYT[x,\PermutationSymbol(x)][\beta,\subset\Delta]*}\prth{\d\eta}~\prth{\bigotimes_{x\in\finiteconfig}\WienerMeasureXYTBC[x,\PermutationSymbol(x)][\beta,\not\subset\Delta]*}\prth{\d\FKconfig}~\E^{-\Ham{FK}{\eta\cup\FKconfig}}~f\prth{\eta\cup\FKconfig}
\end{align*}
According to \Cref{lem_permutation} this equals
\begin{align*}
=&~\E^{\abs{\Delta}}\int\PoissonLeb[\window]{\d\finiteconfig}~\PoissonLeb[\Delta]{\d\zeta}~\E^{\beta\mu\#\prth{\finiteconfig\cup\zeta}}~\sum_{\zeta_1\subseteq\zeta}~\sum_{\zeta_2\subseteq\finiteconfig}~\sum_{\PermutationSymbol\interior\in\PermutationSpace{\zeta\to\zeta_2\cup\zeta\setminus\zeta_1}}~\sum_{\PermutationSymbol\exterior\in\PermutationSpace{\finiteconfig\to\zeta_1\cup\finiteconfig\setminus\zeta_2}}
\\
&\int\prth{\bigotimes_{x\in\zeta}\WienerMeasureXYT[x,\PermutationSymbol\interior(x)][\beta,\subset\Delta]*}\prth{\d\eta}~\prth{\bigotimes_{x\in\finiteconfig}\WienerMeasureXYTBC[x,\PermutationSymbol\exterior(x)][\beta,\not\subset\Delta]*}\prth{\d\FKconfig}~\E^{-\Ham{FK}{\eta\cup\FKconfig}}~f\prth{\eta\cup\FKconfig}
\end{align*}
According to \Cref{lem_uniform_subset} again, this equals
\begin{align}
=&~\E^{2\abs{\Delta}}\int\PoissonLeb[\window]{\d\finiteconfig}~\PoissonLeb[\Delta]{\d\zeta'}~\PoissonLeb[\Delta]{\d\zeta_1}~\E^{\beta\mu\#\prth{\finiteconfig\cup\zeta'\cup\zeta_1}}~\sum_{\zeta_2\subseteq\finiteconfig}~\sum_{\PermutationSymbol\interior\in\PermutationSpace{\zeta_1\cup\zeta'\to\zeta_2\cup\zeta'}}~\sum_{\PermutationSymbol\exterior\in\PermutationSpace{\finiteconfig\to\zeta_1\cup(\finiteconfig\setminus\zeta_2)}}
\nonumber
\\
&\int\prth{\bigotimes_{x\in\zeta_1\cup\zeta'}\WienerMeasureXYT[x,\PermutationSymbol\interior(x)][\beta,\subset\Delta]*}\prth{\d\eta}~\prth{\bigotimes_{x\in\finiteconfig}\WienerMeasureXYTBC[x,\PermutationSymbol\exterior(x)][\beta,\not\subset\Delta]*}\prth{\d\FKconfig}~\E^{-\Ham{FK}{\eta\cup\FKconfig}}~f\prth{\eta\cup\FKconfig}.\label{eq_DLR_0}
\end{align}

Let $\eta,\FKconfig\in\ConfSpace{FK}$ be finite configurations. We assume $\eta$ is only made up of bridges included in $\Delta$. Then it is clear from the definition of $\Hamexterior*$ (see \Cref{def_Q}) that,
\begin{align*}
\Ham{FK}{\eta\cup\FKconfig}=\Hamconditional{\eta\cup\FKconfig}+\Hamexterior{\FKconfig}.
\end{align*}
Then equation (\ref{eq_DLR_0}) becomes
\begin{align*}
=&~\E^{2\abs{\Delta}}\int\PoissonLeb[\window]{\d\finiteconfig}~\PoissonLeb[\Delta]{\d\zeta'}~\PoissonLeb[\Delta]{\d\zeta_1}~\E^{\beta\mu\#\prth{\finiteconfig\cup\zeta'\cup\zeta_1}}~\sum_{\zeta_2\subseteq\finiteconfig}~\sum_{\PermutationSymbol\interior\in\PermutationSpace{\zeta_1\cup\zeta'\to\zeta_2\cup\zeta'}}~\sum_{\PermutationSymbol\exterior\in\PermutationSpace{\finiteconfig\to\zeta_1\cup(\finiteconfig\setminus\zeta_2)}}
\\
&\int\prth{\bigotimes_{x\in\zeta_1\cup\zeta'}\WienerMeasureXYT[x,\PermutationSymbol\interior(x)][\beta,\subset\Delta]*}\prth{\d\eta}~\prth{\bigotimes_{x\in\finiteconfig}\WienerMeasureXYTBC[x,\PermutationSymbol\exterior(x)][\beta,\not\subset\Delta]*}\prth{\d\FKconfig}~\E^{-\Hamconditional{\eta\cup\FKconfig}-\Hamexterior{\FKconfig}}~f\prth{\eta\cup\FKconfig}.
\end{align*}
By definition of the measure $\QQ_{\Delta,L}$, this equals
\begin{align*}
=&\int\QQ_{\Delta,L}(\d\FKconfig)~\int\PoissonLeb[\Delta]{\d\zeta'}~\E^{\beta\mu\#\zeta'}~\sum_{\PermutationSymbol\interior\in\PermutationSpace{\inward\cup\zeta'\to\outward\cup\zeta'}}
\\
&\int\prth{\bigotimes_{x\in\inward\cup\zeta'}\WienerMeasureXYT[x,\PermutationSymbol\interior(x)][\beta,\subset\Delta]*}\prth{\d\eta}~\E^{-\Hamconditional{\eta\cup\FKconfig}}~f\prth{\eta\cup\FKconfig}.
\end{align*}
\end{Demo}

We make use of \Cref{Prop_QL} to prove the probability $\ConditionnalModel{~\cdot~}$ is well defined $\Model{FK}*$ almost surely. The infinite volume part is proved later.

\begin{Demo}[\Cref{lemdef_DLR}, 1/2]\label{demo_lemdef_DLR_1}
It is clear from their definitions that measures $\QQ_{\Delta,L}$ and $\Model{FK}*$ are limited to finite (FK) configurations. So the quantity $\sup_{s\in\interff{0;\beta}}\#\prth{\MathSet{\sect\prth{s},~\sect\in\FKconfig}\cap\prth{\Delta+B_R}}$ is finite $\QQ_{\Delta,L}$ almost everywhere and $\Model{FK}*$ almost surely. Therefore the quantity $\Zconditional{\FKconfig}$ is well defined $\QQ_{\Delta,L}$ almost everywhere.

We know from \Cref{Prop_QL} that
\[
\Zgrandcanonic=\int\Zconditional{\FKconfig}~\QQ_{\Delta,L}(\d\FKconfig).
\]
We also know from \Cref{lem_Z} that $\Zgrandcanonic<+\infty$. So $\Zconditional{\FKconfig}$ is finite $\QQ_{\Delta,L}$ almost everywhere.
Then
\begin{align*}
\Model{FK}{\Zconditional{\FKconfig}=+\infty}&=\frac{1}{\Zgrandcanonic}\int\carac{\Zconditional{\FKconfig}=+\infty}~\Zconditional{\FKconfig}~\QQ_{\Delta,L}(\d\FKconfig)
\\
&=+\infty\cdot\QQ_{\Delta,L}\prth{\Zconditional{\FKconfig}=+\infty}=0.
\end{align*}
Similarly
\[
\Model{FK}{\Zconditional{\FKconfig}=0}=\frac{1}{\Zgrandcanonic}\int\carac{\Zconditional{\FKconfig}=0}~\Zconditional{\FKconfig}~\QQ_{\Delta,L}(\d\FKconfig)=0.
\]
\end{Demo}

We establish the DLR equations in the following proof, mainly based of \Cref{Prop_QL,th_limit_FK}.

\begin{Demo}[\Cref{th_DLR}]\label{demo_th_DLR}
Let $\MyFunction{f}{\ConfSpace{FK}}{\RR}{}{}$ be bounded and $\in$-local relatively to some compact $\Delta_f$.

We know $\Zconditional*$ is well defined, positive and finite $\QQ_{\Delta,L}$ almost everywhere (see the proof of \Cref{lemdef_DLR}). So according to \Cref{Prop_QL},  
\begin{align*}
\Zgrandcanonic\int f(\FKconfig)~\Model{FK}{\d\FKconfig}
=\int\QQ_{\Delta,L}(\d\FKconfig)~\Zconditional{\FKconfig}\int\ConditionnalModel{\d\eta}~f\prth{\eta\cup\FKconfig\exterior}.
\end{align*}
According to \Cref{Prop_QL} again, if $\FKconfig_1$ and $\FKconfig_2$'s respective distributions are given by $\Model{FK}*$ and $\frac{\Zconditional*}{\Zgrandcanonic}\d\QQ_{\Delta,L}$ then $(\FKconfig_1)\exterior$ and $(\FKconfig_2)\exterior$ have the same distribution. So
\[
\int f(\FKconfig)~\Model{FK}{\d\FKconfig}=\int\Model{FK}{\d\FKconfig}\int\ConditionnalModel{\d\eta}~f\prth{\eta\cup\FKconfig\exterior}.
\]
We call this equality the DLR equations in finite volume.

We establish the DLR equations for the empirical field
\begin{align*}
&\int\EmpiricalField{FK}{\d\FKconfig}\int\ConditionnalModel{\d\eta}~f\prth{\eta\cup\FKconfig\exterior}
\\
=&\frac{1}{L^d}\int_{\window}\d v\int\Model{FK}{\d\FKconfig}\int\ConditionnalModel{\d\eta}[\FKconfig+v]~f\prth{\eta\cup\prth{\FKconfig+v}\exterior}.
\end{align*}
By stationarity of the interaction, if $\eta$ is distributed along $\ConditionnalModel[\Delta]{~\cdot~}[\FKconfig+v]$ then $\eta-v$ is distributed along $\ConditionnalModel[\Delta-v]{~\cdot~}[\FKconfig]$. Therefore
\begin{align*}
&=\frac{1}{L^d}\int_{\window}\d v\int\Model{FK}{\d\FKconfig}\int\ConditionnalModel[\Delta-v]{\d\eta}~f\prth{\prth{\eta+v}\cup\prth{\FKconfig+v}\exterior}
\\
&=\frac{1}{L^d}\int_{\window}\d v\int\Model{FK}{\d\FKconfig}\int\ConditionnalModel[\Delta-v]{\d\eta}~f\prth{\prth{\eta\cup\FKconfig\exterior[\Delta-v]}+v}.
\end{align*}
By the DLR equations in finite volume for the compact $\Delta-v$ and the function $f(\cdot+v)$, this equals
\begin{align*}
=\frac{1}{L^d}\int_{\window}\d v\int\Model{FK}{\d\FKconfig}~f\prth{\FKconfig+v}=\int f\d\EmpiricalField{FK}*
\end{align*}

Since the function $f$ is $\in$-local and bounded, according to the first half of \Cref{th_limit_FK},
\begin{equation}\label{eq_DLR_1}
\lim_{L\to+\infty}\int f(\FKconfig)~\EmpiricalField{FK}{\d\FKconfig}=
\int f(\FKconfig)~\Model*{FK}{\d\FKconfig}.
\end{equation}
We denote
\[
f_{\Delta}(\FKconfig)\defequal\int\ConditionnalModel{\d\eta}~f\prth{\eta\cup\FKconfig\exterior}.
\]
Given the definition of the measure $\ConditionnalModel{~\cdot~}$, $\FKconfig\in\ConfSpacePerm{FK}$, the function $f_{\Delta}$ is $\cap^0$-local relatively to the compact $\Delta_f\cup\prth{\Delta+B_R}$ and bounded. Thus, according to the second part of \Cref{th_limit_FK} (and \Cref{rq_inter_n_lipschitz}),
\begin{equation}\label{eq_DLR_2}
\lim_{L\to+\infty}\int f_{\Delta}(\FKconfig)~\EmpiricalField{FK}{\d\FKconfig}=
\int f_{\Delta}(\FKconfig)~\Model*{FK}{\d\FKconfig}.
\end{equation}
According to the DLR equations for the empirical field and equations (\ref{eq_DLR_1}) and (\ref{eq_DLR_2}),
\[
\int f(\FKconfig)~\Model*{FK}{\d\FKconfig}=\int f_{\Delta}(\FKconfig)~\Model*{FK}{\d\FKconfig}.
\]

We just proved the equality of measures $\Model*{FK}*$ and $\MyFunction{}{}{}{E}{\int\Model*{FK}{\d\FKconfig}\int\ConditionnalModel{\d\eta}~\carac{E}\prth{\eta\cup\FKconfig\exterior}}$ over the ring of sets of $\in$-local events. So by Carathéodory's extension theorem, the two measures coincide.
\end{Demo}

\begin{Demo}[\Cref{lemdef_DLR}, 2/2]\label{demo_lemdef_DLR_2}
According to the DLR equations,
\begin{align*}
&\int\carac{\Zconditional*\textnormal{ is not well defined, positive and finite}}\d\Model*{FK}*
\\
=&\int\Model*{FK}{\d\FKconfig}\int\ConditionnalModel{\d\rlconfig}\carac{\Zconditional{\rlconfig\cup\FKconfig\exterior}\textnormal{ is not well defined, positive and finite}}.
\end{align*}
Except the quantity $\Zconditional{\FKconfig}$ only depends on $\FKconfig\exterior$. So this equals
\begin{align*}
=\int\Model*{FK}{\d\FKconfig}\int\ConditionnalModel{\d\rlconfig}\carac{\Zconditional{\FKconfig}\textnormal{ is not well defined, positive and finite}}=0
\end{align*}
by definition of the measure $\ConditionnalModel{~\cdot~}$.

Therefore, the quantity $\Zconditional*$ is well defined, positive and finite $\Model*{FK}*$ almost surely.
\end{Demo}

\appendix

\section{Wiener Sausage}

\begin{Def}
Let $\MyFunction{\sect}{\interff{0;T}}{\RR^d}{}{}$ be a $d$ dimensional continuous trajectory. We define its Wiener sausage of thickness $\delta>0$ as the set
\[
\Sausage[\delta]{\sect}\defequal
\ens{x\in\RR^d}[\exists t\in\interff{0;T},~\norm{x-\brown_t}\leqslant\delta]
\]
and its volume as $\abs{\Sausage[\delta]{\brown}}$.
\end{Def}

\begin{Prop}\label{prop_saucisse}
Let $\MyFunction{\sect}{\interff{0;T}}{\RR^d}{}{}$ be continuous. For any $0\leqslant t<t'\leqslant\beta$,
\begin{equation}
	\abs{\Sausage{\sect}}\geqslant c_{d-1}\delta^{d-1}\norm{\sect(t')-\sect(t)}
\end{equation}
where $c_{d-1}$ is the surface of the $d-1$ dimensional unit disk.
\end{Prop}

In other words, a cylinder whose axis goes from $\sect(t)$ to $\sect(t')$ with radius $\delta$ has a smaller volume than $\Sausage{\sect}$.
\\

\begin{Demo}[\Cref{prop_saucisse}]
We denote $x=\sect(t)$, $y=\sect(t')$, $\ell_{x,y}$ the straight line going from $x$ to $y$ and $\HH^{\perp}(z)$ the hyperplane orthogonal to that straight line in a given $z\in\ell_{x,y}$. Then
\begin{align*}
\Leb[d]\prth{\Sausage{\sect}}&=\int_{\ell_{x,y}}\Leb[d-1]\prth{\Sausage{\sect}\cap\HH^{\perp}(z)}\d z
\\
&\geqslant\int_{x}^y\Leb[d-1]\prth{\Sausage{\sect}\cap\HH^{\perp}(z)}\d z
\end{align*}
By continuity of the trajectory $\MyFunction{\sect}{\interff{0;T}}{\RR^d}{}{}$, for any $z\in\ell_{x,y}$ between $x$ and $y$, there exists $s_z$ between $t$ and $t'$ such that $\sect(s_z)\in\HH^{\perp}(z)$. Therefore a disk in $\HH^{\perp}(z)$ of center $\sect(s_z)$ and radius $\delta$ is included in $\Sausage{\sect}$. We conclude
\begin{equation*}
	\Leb[d]\prth{\Sausage{\sect}}
	\geqslant\int_{x}^y c_{d-1}\delta^{d-1}\d z
\end{equation*}
\end{Demo}

\begin{Th}\label{th_saucisse}
Let $\delta>0$. Let $\brown=\prth{\brown_t}_{0\leqslant t\leqslant T}$ be a $d$ dimensional Brownian motion. There exists $\varepsilon>0$ such that
\[
\Esp{\E^{\varepsilon\abs{\Sausage[\delta]{\brown}}^{2}}}<+\infty.
\]
\end{Th}

The proof of this theorem is heavily based on the proof from \autocite{Szn87} of finite exponential moments for the Wiener sausage.

\begin{Demo}[\Cref{th_saucisse}]
We define the sequence of stopping times $\prth{\stoptime}_{i\in\NN}$ by
$
\begin{cases}
\stoptime[0]\defequal0\\
\stoptime[i+1]\defequal\inf\MathSet{t\geqslant0}[\norm[\infty]{\brown_{\stoptime+t}-\brown_{\stoptime}}>\delta]
\end{cases}
$
and the random variable $N_T\defequal\inf\MathSet{n\in\NN}[\sum_{i=1}^{n}\stoptime[i]>T]$.

It is clear that
\begin{align*}
\Sausage[\delta]{\brown}&=\bigcup_{t=0}^T\BB\prth{\brown_t,\delta}
\subseteq\bigcup_{i=0}^{N_T-1}\prth{\bigcup_{t=\stoptime[0]+\dots+\stoptime}^{\stoptime[0]+\dots+\stoptime[i+1]}\BB\prth{\brown_t,\delta}}
\subseteq\bigcup_{i=0}^{N_T-1}\prth{\bigcup_{t=\stoptime[0]+\dots+\stoptime}^{\stoptime[0]+\dots+\stoptime[i+1]}\prth{\brown_t+\case[2\delta]}}
\subseteq\bigcup_{i=0}^{N_T-1}\prth{\brown_t+\case[4\delta]}
\end{align*}
thus
\begin{align*}
\abs{\Sausage[\delta]{\brown}}^2\leqslant N_T^2\prth{4\delta}^{2d}.
\end{align*}

For any $k\geqslant 0$ and $\lambda>0$, by Markov inequality,
\begin{align*}
\Pb{N_T\geqslant k}=\Pb{\sum_{i=1}^k\stoptime[i]\leqslant T}
\leqslant\E^{\lambda T}\Esp{\E^{-\lambda\sum_{i=1}^{k}\stoptime}}
\end{align*}
Since the stopping times $\stoptime,~i\geqslant 1$ are \iid, we deduce
\[
\Pb{N_T\geqslant k}\leqslant\E^{\lambda T}\Esp{\E^{-\lambda\stoptime[1]}}^k.
\]
For any non-negative random variable $X$ and $\lambda>0$,
\[
\Esp{\E^{-\lambda X}}=\lambda\int_0^{+\infty}\E^{-\lambda x}\prth{1-\Pb{X>x}}\d x.
\]
Furthermore, if we denote as $\brown^{(i)},~1\leqslant i\leqslant d$ the $d$ spatial components of the Brownian motion $\brown$, then we can write
\[
\stoptime[1]=\min_{1\leqslant i\leqslant d}\stoptime[1]*[i]
\quad\text{ where }\quad
\stoptime[1]*[i]\defequal\inf\MathSet{t\geqslant 0}[\abs{\brown_t^{(i)}}>\delta].
\]
Therefore
\begin{align*}
\Esp{\E^{-\lambda\stoptime[1]}}
=\lambda\int_0^{+\infty}\E^{-\lambda x}\prth{1-\Pb{\stoptime[1]>x}}\d x
=\lambda\int_0^{+\infty}\E^{-\lambda x}\prth{1-\Pb{\stoptime[1]*>x}^d}\d x.
\end{align*}
By concavity, it is clear that $\forall p\geqslant 0,~1-p^d\leqslant d(1-p)$ so
\begin{align*}
\Esp{\E^{-\lambda\stoptime[1]}}
\leqslant
d\lambda\int_0^{+\infty}\E^{-\lambda x}\prth{1-\Pb{\stoptime[1]*>x}}\d x
\leqslant
d\Esp{\E^{-\lambda\stoptime[1]*}}.
\end{align*}
According to the \citetitle{BS02} \autocite{BS02} (3.0.1 p218),
\begin{align*}
\Esp{\E^{-\lambda\stoptime[1]*}}=\frac{1}{\cosh\prth{\delta\sqrt{2\lambda}}}
\leqslant 2\E^{-\delta\sqrt{2\lambda}}.
\end{align*}
For $\lambda=\varepsilon k^2$, we get
\[
\Pb{N_T\geqslant k}\leqslant\E^{\varepsilon Tk^2}(2d)^k\E^{-\sqrt{2\varepsilon}\delta k^2}.
\]

We conclude
\begin{align*}
\Esp{\E^{\varepsilon\abs{\Sausage[\delta]{\brown}}^2}}
&\leqslant\Esp{\E^{\varepsilon(4\delta)^{2d}N_T^2}}
\leqslant\sum_{k=1}^{+\infty}\E^{\varepsilon(4\delta)^{2d}k^2}\Pb{N_T\geqslant k}
\leqslant\sum_{k=1}^{+\infty}\E^{\varepsilon(4\delta)^{2d}k^2}\E^{\varepsilon Tk^2}(2d)^k\E^{-\sqrt{2\varepsilon}\delta k^2}
\end{align*}
which converges for $\varepsilon>0$ small enough.
\end{Demo}

\section{Notation Table}\label{sec_table}

We order symbols in an approximate alphabetical order.

\begingroup
\renewcommand\arraystretch{1.5}
\definecolor{lightgray}{gray}{0.9}
\rowcolors{1}{}{lightgray}
\begin{longtable}{clc}
Symbol & Definition  & Page\\
\toprule
\endfirsthead
Symbol & Definition  & Page\\
\toprule
\endhead
\rowcolor{white}
$\dots$ & $\dots$ & $\dots$ \\
\endfoot
\bottomrule
\endlastfoot
$A$ & Superstability constant (see \Cref{hyp_superstable}) & \pageref{hyp_superstable} \\
$B$ & Superstability constant (see \Cref{hyp_superstable}) & \pageref{hyp_superstable} \\
$B_R$ & Closed ball of radius $R$ & \pageref{hyp_finite_range} \\
$\beta$ & Inverse temperature & - \\
$\ConfAlgebra*{}$ & $\sigma$-algebra over $\ConfFinite$ (see \Cref{def_confFinite}) & \pageref{def_confFinite} \\
$\ConfAlgebra{FK}$ & $\sigma$-algebra over $\ConfSpace{FK}$ (see \Cref{def_confspace_FK}) & \pageref{def_confspace_FK} \\
$\ConfAlgebraPerm{FK}$ & $\sigma$-algebra over $\ConfSpacePerm{FK}$ (see \Cref{def_confspaceperm_FK}) & \pageref{def_confspaceperm_FK} \\
$\ConfAlgebra{mp}$ & $\sigma$-algebra over $\ConfSpace{mp}$ (see \Cref{def_confspace_mp}) & \pageref{def_confspace_mp} \\
$\ConfAlgebra{rl}$ & $\sigma$-algebra over $\ConfSpace{rl}$ (see \Cref{def_confspace_rl}) & \pageref{def_confspace_rl} \\
$\ConfFinite$ & $\MathSet{\finiteconfig\subset\RR^d}[\#\finiteconfig<+\infty]$ & \pageref{def_confFinite} \\
$\ConfSpace{FK}$ & $\ens{\FKconfig\subset\WienerSpace[\beta]}[\FKconfig\text{ is locally finite for }\WienerTopology[\beta]]$ & \pageref{def_confspace_FK} \\
$\ConfSpace{mp}$ & $\MathSet{\mpconfig=\prth{x,p_x,u_x,\brown_x}_{x\in\finiteconfig}\subset\RR^d\times\ZZ^d\times\interff{0;1}\times\WienerSpace[1]}[\finiteconfig\subset\RR^d\text{ loc. finite}]$ & \pageref{def_confspace_mp} \\
$\ConfSpace{rl}*$ & $\ens{\rlconfig\subset\WienerSpaceLoop{rl}}[\rlconfig\text{ is locally finite in }\WienerTopologyLoop{rl}]$ & \pageref{def_confspace_rl} \\
$\ConfSpaceAuth$ & $\MathSet{\mpconfig\in\ConfSpace{mp}}[\SpatialComponent*\text{ simple and } \forall(x,p,u,\brown)\in\mpconfig,~\CardCase{x+rp}\geqslant1]$ & \pageref{def_confspace_mp} \\
$\ConfSpacePerm{FK}$ & $\MathSet{\FKconfig\in\ConfSpace{FK}}[\FKconfig\text{ is permutation-wise}]$ & \pageref{def_confspace_FK} \\
$\ConfSpacePerm{mp}$ & $\MathSet{\mpconfig\in\ConfSpaceAuth}[\forall y\in\SpatialComponent*,~\exists!x\in\SpatialComponent*,~\Permutationmp{x}=y]$ & \pageref{def_confspace_mp} \\
$\inward$ & See \Cref{lemdef_DLR} & \pageref{lemdef_DLR} \\
$\outward$ & See \Cref{lemdef_DLR} & \pageref{lemdef_DLR} \\
$\SpatialComponent*$ & $\MathSet{x,~(x,p,u,\brown)\in\mpconfig}$ & \pageref{def_confspace_mp} \\
$\FKconfig\exterior$ & $\MathSet{\sect\in\FKconfig}[\sect\not\subset\Delta]$ & \pageref{lemdef_DLR} \\
$\Ham{FK}*$ & $\MyFunction{}{}{}{\FKconfig}{\int_0^{\beta}\Interaction{\ens{\sect\prth{s},~\sect\in\FKconfig}}\d s}$ & \pageref{def_ham_FK} \\
$\Ham{mp}*$ & Hamiltonian in (mp) framework (see \Cref{lemdef_mp}) & \pageref{lemdef_mp} \\
$\Ham{rl}*$ & $\MyFunction{}{}{}{\rlconfig}{\int_0^{\beta}\Interaction{\ens{\loopath\prth{\beta j+s},~0\leqslant j<\length{\loopath},~\loopath\in\rlconfig}}\d s}$ & \pageref{def_ham_rl} \\
$\Hamconditional*$ & $\MyFunction{}{}{}{\FKconfig}{\int_0^{\beta}\InteractionLocal{\MathSet{\sect\prth{s},~\sect\in\FKconfig}}\d s}$ & \pageref{lemdef_DLR} \\
$\Hamexterior*$ & $\MyFunction{}{}{}{\FKconfig}{\int_0^{\beta}\Interaction{\MathSet{\sect\prth{s},~\sect\in\FKconfig}\cap\Delta\complementary}\d s}$ & \pageref{def_Q} \\
$\Entropy{\cdot}{\cdot}$ & See \Cref{def_entropy} & \pageref{def_entropy} \\
$\length*$ & Length of a loop (see \Cref{def_LoopRoot}) & \pageref{def_LoopRoot} \\
$\window$ & $\interfo{-L/2;L/2}^d$ & \pageref{hyp_superstable} \\
$\Leb[d]$ & $d$ dimensional Lebesgue measure & \pageref{def_Poisson_mp} \\
$\mu$ & Chemical potential & - \\
$\CardCase{\cdot}[\finiteconfig]$ & $\MyFunction{}{}{}{z}{\#\prth{\finiteconfig\cap\prth{z+\case}}}$ & \pageref{def_confspace_mp} \\
$\pMeasure*$ & Probability measure over $\ZZ^d$ such that $\forall p\in\ZZ^d,~\pMeasure{p}>0$ & \pageref{def_Poisson_mp} \\
$\WienerSpace[t]$ & Set of continuous trajectories $\MyFunction{}{\interff{0;t}}{\RR^d}{}{}$ & \pageref{def_Wiener} \\
$\WienerSpaceExtended{t}$ & Subset of continuous trajectories $\MyFunction{}{\RR}{\RR^d}{}{}$ (see \Cref{def_Loop_t}) & \pageref{def_Loop_t} \\
$\WienerSpaceLoop{rl}$ & Space of rooted loops (see \Cref{def_LoopRoot}) & \pageref{def_LoopRoot} \\
$\ViewChange{FK}{rl}*$ & Transition map from (rl) to (FK) (see \Cref{def_confspace_rl}) & \pageref{def_confspace_rl} \\
$\ViewChange{FK}{mp}*$ & Transition map from (mp) to (FK) (see \Cref{def_vc_mp_FK}) & \pageref{def_vc_mp_FK} \\
$\PoissonLeb*$ & Poisson point process over $\window$ with intensity $1$ & \pageref{def_poisson} \\
$\Poisson{mp}*$ & Poisson process over $\window\times\ZZ^d\times\interff{0;1}\times\WienerSpace[1]$ (see \Cref{def_Poisson_mp}) & \pageref{def_Poisson_mp} \\
$\Poisson{rl}*$ & Poisson process with intensity measure $\WienerMeasureRootBC*$ & \pageref{def_W_root} \\
$\Model{FK}*$ & Probability measure over $\ConfSpace{FK}$ (see \Cref{lemdef_FK}) & \pageref{lemdef_FK} \\
$\Model{mp}*$ & Probability measure over $\ConfSpace{mp}$ (see \Cref{lemdef_mp}) & \pageref{lemdef_mp} \\
$\Model{rl}*$ & Probability measure over $\ConfSpace{rl}$ (see \Cref{lemdef_rl}) & \pageref{lemdef_rl} \\
$\EmpiricalField{xx}*$ & 
$\int f\d\EmpiricalField{xx}*\defequal\frac{1}{L^d}\int_{\window}\d v\int f(\FKconfig+v)~\Model{xx}{\d\FKconfig}$ & - \\
$\Model*{xx}*$ & Infinite volume (xx) model (Def. \ref{def_FK} and Thms. \ref{th_limit_rl} and \ref{th_limit_mp}) & - \\
$\ConditionnalModel{\cdot}$ & Conditional measure over $\ConfSpace{FK}$ (see \Cref{lemdef_DLR}) & \pageref{lemdef_DLR} \\
$\Proj{?}*$ & Projection map over $\ConfSpace{FK}$, for $?$ being $\in$, $\cap$, ... (see Def. \ref{def_proj}) & \pageref{def_proj} \\
$\QQ_{\Delta,L}$ & Measure over $\ConfSpace{FK}$ (see \Cref{def_Q}) & \pageref{def_Q} \\
$r$ & Superstability constant (see \Cref{hyp_superstable}) & \pageref{hyp_superstable} \\
$R$ & Range of the interaction $\InteractionFree*$ (see \Cref{hyp_finite_range}) & \pageref{hyp_finite_range} \\
$\TimeReversal{FK}*$ & $\MyFunction{}{}{}{\FKconfig}{\MathSet{\MyFunction{}{}{}{s}{\sect(\beta-s)},~\sect\in\FKconfig}}$ & \pageref{prop_time_reversal} \\
$\PermutationSpace{X}$ & Set of permutations over $X$ & \pageref{lemdef_FK} \\
$\PermutationSpace{X\to Y}$ & Set of bijections from $X$ to $Y$ & \pageref{lemdef_DLR} \\
$\PermutationFK{\cdot}$ & Natural permutation over $\FKconfig\in\ConfSpacePerm{FK}$ (see \Cref{def_confspaceperm_FK}) & \pageref{def_confspaceperm_FK} \\
$\Permutationmp{\cdot}$ & Encoded map over $\SpatialComponent*$ (see \Cref{def_confspace_mp}) & \pageref{def_confspace_mp} \\
$\TimeTranslation*$ & $\MyFunction{}{}{}{\rlconfig}{\MathSet{\loopath\prth{\cdot+s},~\loopath\in\rlconfig}}$ & \pageref{Prop_time-shift} \\
$\Thresh[m]*$ & $\MyFunction{}{}{}{x}{x\cdot\carac{x>m}}$ & \pageref{lem_entropic_method} \\
$\InteractionFree*$ & $\MyFunction{}{\ConfFinite}{\RR\cup\{+\infty\}}{}{}$ & \pageref{lab_InteractionFree} \\
$\Interaction*$ & Interaction with Dirichlet boundary condition (see \Cref{def_interaction}) & \pageref{def_interaction} \\
$\InteractionLocal*$ & Local interaction (see \Cref{hyp_finite_range}) & \pageref{hyp_finite_range} \\
$\WienerMeasureXYT[x,y][t]*$ & Wiener measure over Brownian bridges (see \Cref{def_Wiener}) & \pageref{def_Wiener} \\
$\WienerMeasureXYT[x,y][\beta,\subset\Delta]*$ & $\WienerMeasureXYT[x,y][\beta,\subset\Delta]{\d\sect}\defequal\carac{\sect\subset\Delta}\WienerMeasureXYT[x,y][\beta]{\d\sect}$ & \pageref{lemdef_DLR}\\
$\WienerMeasureXYTBC[x,\PermutationSymbol(x)][\beta,\not\subset\Delta]*$ & $\WienerMeasureXYTBC[x,\PermutationSymbol(x)][\beta]*
-\WienerMeasureXYT[x,\PermutationSymbol(x)][\beta,\subset\Delta]*$ & \pageref{def_Q} \\
$\WienerMeasureRootBC*$ & $\int_{\window}\d x\sum_{j\geqslant 1}\frac{1}{j}\WienerMeasureXYTBC[x,x][\beta j]*$ & \pageref{def_W_root} \\
$\WienerMeasureNor[t][x,y]*$ & $\WienerMeasureXYT[x,y][t]*\big/\WienerMeasureXYT[x,y][t]{\WienerSpace[t]}$ & \pageref{def_Poisson_mp} \\
$\WienerAlgebra[t]$ & Borel $\sigma$-algebra over $\WienerSpace[t]$ associated to the uniform norm & \pageref{def_Wiener} \\
$\WienerAlgebraExtended{t}$ & $\sigma$-algebra over $\WienerSpaceExtended{t}$ (see \Cref{def_Loop_t}) & \pageref{def_Loop_t} \\
$\WienerAlgebraLoop{rl}$ & $\sigma$-algebra over $\WienerSpaceLoop{rl}$ (see \Cref{lemdef_LoopRoot_alg_top}) & \pageref{lemdef_LoopRoot_alg_top} \\
$\WienerTopology[t]$ & Topology over $\WienerSpace[t]$ associated to the uniform norm & \pageref{def_Wiener} \\
$\WienerTopologyExtended{t}$ & Topology over $\WienerSpaceExtended{t}$ (see \Cref{def_Loop_t}) & \pageref{def_Loop_t} \\
$\WienerTopologyLoop{rl}$ & Topology over $\WienerSpaceLoop{rl}$ (see \Cref{lemdef_LoopRoot_alg_top}) & \pageref{lemdef_LoopRoot_alg_top} \\
$\Zgrandcanonic$ & Partition function (see \Cref{lemdef_FK}) & \pageref{lemdef_FK} \\
$\Zconditional{\FKconfig}$ & Local partition function (see \Cref{lemdef_DLR}) & \pageref{lemdef_DLR}
\end{longtable}
\endgroup

\section*{Thanks}\addcontentsline{toc}{section}{Thanks}

We acknowledge support from the Labex CEMPI (ANR-11-LABX-0007-01). We also acknowledge the support of the CDP C2EMPI, as well as the French State under the France-2030 programme, the University of Lille, the Initiative of Excellence of the University of Lille, the European Metropolis of Lille for their funding and support of the R-CDP-24-004-C2EMPI project.

We thank Quirin Vogel for the insightful discussions and Benoît Henry for the help in \Cref{th_saucisse}'s computations.

We would like to acknowledge the importance of private exchanges with H.-O. Georgii where the idea of encoding the Bose gas as a marked point process was discussed.

\addcontentsline{toc}{section}{References}
\printbibliography

\end{document}